\documentclass[10pt,preprint]{aastex}
\usepackage{amsmath}
\usepackage{ulem}

\begin{document}


\title{The hydrodynamical models of the cometary compact H II region}


\author{Feng-Yao Zhu\altaffilmark{1,2}, Qing-Feng Zhu\altaffilmark{1,2}, Juan Li\altaffilmark{3}, Jiang-Shui Zhang\altaffilmark{4} and Jun-Zhi Wang\altaffilmark{5}}
\altaffiltext{1}{Astronomy Department, University of Science and Technology of China,
    Hefei, China 230026, zhuqf@ustc.edu.cn }
\altaffiltext{2}{Key Laboratory for Research in Galaxies and Cosmology, University of Science and Technology of China, Chinese Academy of Sciences, Hefei, China, 230026}

\altaffiltext{3}{Shanghai Astronomical Observatory, Chinese Academy of Sciences, Shanghai, China}
\altaffiltext{4}{Center for Astrophysics, Guangzhou University, Guangzhou, China}
\altaffiltext{5}{Shanghai Astronomical Observatory, Chinese Academy of Sciences, Shanghai, China}






\begin{abstract}
We \textrm{have} developed a full numerical method to study the gas dynamics of cometary ultracompact (UC) H II regions, and associated photodissociation regions (PDRs). The \textrm{bow-shock} and \textrm{champagne-flow} models with a $40.9/21.9~M_\odot$ star are simulated. In the \textrm{bow-shock} models, the massive star is assumed to move through dense ($n=8000~cm^{-3}$) molecular material with a stellar velocity of $15~km~s^{-1}$. In the \textrm{champagne-flow} models, an exponential distribution of density with a scale height of $0.2~pc$ is assumed. The profiles of the [Ne II] $12.81 \mu m$ and $H_2~S(2)$ lines from the ionized regions and PDRs are compared for two \textrm{sets} of models. In \textrm{champagne-flow} models, emission lines from the ionized gas \textrm{clearly show} the effect of acceleration along the direction toward the tail due to the density gradient. The kinematics of the molecular gas inside the dense shell is mainly due to the expansion of the H II region.
However, in \textrm{bow-shock} models the ionized gas mainly moves in the same direction as the stellar motion. The kinematics of the molecular gas inside the dense shell simply reflects the motion of the dense shell with \textrm{respect to} the star. These differences can be used to distinguish two \textrm{sets} of models.


\end{abstract}





\section{Introduction}

Massive stars are always formed in dense molecular clouds, and the feedback processes of these stars can affect the surrounding interstellar medium (ISM) significantly. When a massive star is born, the surrounding neutral ISM begins to be ionized by stellar EUV ($h\upsilon\geq13.6~eV$) radiation and an H II region is formed. The \textrm{over-pressurized} ionized gas in the H II region will expand and compress the nearby neutral gas. The materials around the H II region are \textrm{swept up} by the expanding H II region to form a dense shell containing mainly neutral gases \citep{ten79,mac91}. Outside of the photoionized region, the molecular gas is dissociated by FUV \textrm{radiation} ($11.26~eV\leq h\upsilon<13.6~eV$) to form a photodissociation region (PDR). The temperature of the neutral gas in PDR ranges from $\sim3000~K$ (close to the ionized region) to $\sim100~K$ \citep{spi78,rog92}. Besides the stellar radiation, \textrm{stellar winds also have a significant impact} on the ISM. The high-velocity outflowing wind from massive stars is embedded in the photoionized region and interacts with the ionized gas. The photoionized gas decelerates the stellar wind so that the kinetic energy of the stellar wind is \textrm{converted into thermal energy}. A hot low-density bubble composed of stellar wind material ($T\geq10^6~K,~n\leq5~cm^{-3}$) with a surrounding shell of swept-up ionized gas is produced. The kinetic and thermal energy of the \textrm{stellar-wind} bubble can increase the expansion of the H II region \citep{com97}.

If the expansion of H II regions are  envisaged in a uniform environment, a spherical H II region \textrm{results} and its evolution is well described by the Str\"{o}mgren sphere model \citep{str39}. However, a large fraction ($\sim 20\%$) of young H II regions are not spherical, but have a cometary morphology \citep{woo89,kur94,wal98}. The study of the formation of such a morphology can improve our understanding of the interaction between massive stars and \textrm{the} interstellar medium. In order to explain this morphology, two kinds of models, \textrm{champagne-flow} models and \textrm{bow-shock} models, are suggested. The \textrm{champagne-flow} model was first suggested by \citet{ten79}, and further improved by including stellar wind, various ambient density distributions and magnetic fields \citep{yor83,com97,art06,gen12}. The main \textrm{characteristic} of these models is that density gradients in the molecular cloud result in the anisotropic expansion of \textrm{the} H II region and \textrm{thence} the cometary morphology. The Orion Nebula is the archetype of this kind of H II \textrm{region} \citep{isr78,ten79}. \textrm{Observations} of the nebula show that there is a bright ionization front on the surface of a molecular cloud and ionized gas flows away from the cloud in the opposite direction. There are also a series of \textrm{bow-shock} models \citep{mac91,bur92,com98,art06,mac15}. The key point of these models is that the supersonic movement of a \textrm{wind-blowing} massive star relative to the molecular cloud causes the ionization front to be trapped in a dense cometary shell which is behind the bow shock ahead of the star.

In order for the \textrm{champagne-flow} models to work, a density gradient must exist around the newly formed star. \textrm{Density gradients seem} to be unavoidable in the realistic environment of the molecular cloud \citep{kle77,sch78,arq84}. Observations show that the molecular clouds are centrally condensed. The density gradients are approximated by power laws $\rho(r)\propto r^{-\alpha}$ in some previous works \citep{arq85,hat03,fra00}. \textrm{Evidence of various power indices from $1.6$ to $2.5$ is found} by different authors \citep{pir09,san10}. Some evidences suggest a steeper index in the outer parts of dense cores (eg., $3.9$ in \citet{hun10}). Since the density does not go to infinity at the center, a simple power law is combined with an exponential law in \citet{fra07}. It is very likely that \textrm{stars form} at the center of the cores and is surrounded by non-uniform halos. In the current work, an exponential density distribution with a scale height of $0.2~pc$ is used in \textrm{our champagne-flow} models. Such a distribution is not very different from a power law with \textrm{an} index of $3$.

For \textrm{bow-shock} models, a UV photon emitting star moving supersonically with respect to \textrm{a molecular cloud} is needed. High velocity massive stars could be present in massive \textrm{star-forming} regions \citep{jon88}. Stars could have a modest initial velocity ($2-5~km~s^{-1}$) due to the turbulent dynamics of their parental clouds \citep{mck07,pet10}. Binary interactions allow particular massive stars to obtain more substantial velocities ($\sim30~km~s^{-1}$) \citep{hoo01}. The temperatures in H II regions are typically $T_i\approx10^4~K$, and the corresponding sound speeds are $a_i\approx13~km~s^{-1}$. If the stellar velocity is $v_\ast > a_i$, there will be a stellar wind bow shock formed in the H II region. And for \textrm{a subsonic stellar velocity} with respect to the H II region, no stellar wind bow shock will form in the photoionized region \citep{bar71,mac15}.

The PDR is important to the kinematics and radiation transfer of the entire region. Most of the high-energy photons emitted from massive stars are absorbed in PDRs and converted into low-energy photons. Moreover, the dynamical structure and expansion of the H II region are also influenced by the distribution of the neutral gas. At the same time, the development of H II regions also \textrm{affects} the kinematics and morphology of the neutral gas. These should be reflected \textrm{in} the line emission. The study of the dynamical structure of the neutral gas is therefore an important complement to the study of H II regions. To complete our picture about massive star formation, a hydrodynamical model including details both in ionized regions and PDRs is necessary. Although many \textrm{champagne-flow} and \textrm{bow-shock} models have been studied, there are only a few models including detailed thermal and dynamical processes both in the ionized region and the PDR.
In most models, the heating by dissociating photons \textrm{is} neglected. When calculating the cooling in the PDR, it is assumed that hydrogen and oxygen are atomic meanwhile carbon is maintained in \textrm{the} singly ionized stage by FUV photons. This assumption is appropriate if the ISM is low-density because the timescales for \textrm{the dissociation} of $H_2$ and $CO$ are very short ($<100~yr$), and the low density prevents the molecules from reformation. However, if the density is as high as $n\sim10^5~cm^{-3}$ in the shell of H II regions, the timescales for dissociations and reformations of $H_2$ and $CO$ could be both $\sim10,000~yr$, and this method will overestimate cooling rates because the real fractional abundances of $H$, $O$ and $C^+$ are lower than assumed. In order to get more accurate cooling rates, the accurate fractional abundances of different components in the PDR should be calculated. The heating processes in the PDR also need to be included.

In the current work, we try to develop a new model including detailed processes both in the H II region and the PDR to obtain a more complete picture \textrm{of} cometary compact H II regions. The dynamics of the gas in the PDR is also included in the hydrodynamic simulation. The heating and cooling processes in the PDR are treated \textrm{once} the fractional abundances are calculated. After completing the hydrodynamic simulation, the [Ne II] $12.81~\mu m$ line and the $H_2~S(2)$ $J=4-2~12.28~\mu m$ rotational line are selected to depict the gas motions in the H II region and the PDR, respectively. These two lines are both in the mid-infrared and can be accessed by Stratospheric Observatory for Infrared Astronomy (SOFIA). We also investigate the possibilities of using these mid-infrared lines to discriminate between bow shocks and champagne flows. For this purpose, the profiles and position-velocity diagrams of these lines from \textrm{bow-shock and champagne-flow} models are compared.



The organization of the current work is as follows: In \S \ref{sect:method} we explain our numerical method used in simulations. We present the results of models and the discussions in \S \ref{sect:result}. Our conclusions are presented in \S \ref{sect:conclusion}.

\section{Numerical Mehthod}
\label{sect:method}

We investigate the evolution of the cometary H II region around a newly formed massive star together with the surrounding PDR. In our simulations, the coupled hydrodynamics and the radiative transfer equations are solved at the same time. \textrm{Gravity and magnetic fields are neglected}.

\subsection{Hydrodynamics}

The dynamical evolution of the H II region and the PDR is treated with a 2D explicit Eulerian method in cylindrical coordinates. The models in this paper are all computed on a $250\times500$ grid. Due to the various sizes of the cometary H II regions in different models, the cell size of the grids in \textrm{champagne-flow} models is chosen to be $0.005~pc$, and that in \textrm{bow-shock} models is $0.0025~pc$. The equations of hydrodynamics are as follows,

\begin{equation}
        \frac{\partial \rho}{\partial t}=-\frac{\partial \rho v_r}{\partial r}-\frac{\partial \rho v_z}{\partial z}-\frac{\rho v_r}{r}
\end{equation}

\begin{equation}
        \frac{\partial \rho v_r}{\partial t}=-\frac{\partial (\rho v_r^2+p)}{\partial r}-\frac{\partial \rho v_rv_z}{\partial z}-\frac{\rho v_r^2}{r}
\end{equation}

\begin{equation}
        \frac{\partial \rho v_z}{\partial t}=-\frac{\partial \rho v_rv_z}{\partial r}-\frac{\partial (\rho v_z^2+p)}{\partial z}-\frac{\rho v_rv_z}{r}
\end{equation}

\begin{equation}
        \frac{\partial E_t}{\partial t}=-\frac{\partial(E_t+p)v_r}{\partial r}-\frac{\partial (E_t+p)v_z}{\partial z}-\frac{(E_t+p)v_r}{r}
\end{equation}

where $r$ and $z$ are the radial and axial coordinates, respectively; \textrm{$\rho$ is the density;} $v_r$ and $v_z$ are the radial and axial components of velocity; $p$ is the pressure, and $E_t$ is the total energy density as the sum of the kinetic and thermal energy densities.

These hydrodynamical equations are solved by using a hybrid scheme combining a HLLC Riemann solver \citep{miy05} and a NND-1 scheme \citep{zha03,zhu96}. \textrm{The NND-1 scheme smoothes the oscillations.} The stellar wind with a \textrm{mass-loss} rate $\dot{M}$ and a terminal velocity $v_w$ is treated as a volume-distributed source of mass and energy \citep{roz85,com97,art06}. The mass-loss rates and terminal velocities of stellar winds are \textrm{obtained from the analytic formulae given in} \citet{dal13}. After calculating the hydrodynamic evolution in a computational step, the effects of heating and cooling processes are added explicitly. The energy density is updated with the corresponding heating/cooling rates and the time for the step, which is the minimum of the hydrodynamic time given by Courant-Friedrichs-Lewy condition and the cooling time \citep{art96}.

\subsection{the Radiative Transfer}

In our simulation, the central star is the single source for radiation. The radiative transfer of the ionizing EUV radiation ($h\upsilon \geq13.6~eV$) and the dissociating FUV radiation ($11.26~eV \leq~h\upsilon <13.6~eV$) is solved separately. We use the frequency-integrated EUV and FUV photon luminosities of ionizing stars provided by \citet{dia98}. The spectra of \textrm{the} stars are assumed to be black-body. The solar abundances \textrm{are} assumed for all the models in \textrm{this} work \citep{hol89}. In \textrm{the} ionized region, if the local ionizing \textrm{photon flux} is higher than $10^{10}~cm^{-2}~s^{-1}$, \textrm{ionization equilibrium} is adopted. And if the local photons flux is low, the photoionization and recombination are time dependent. \textrm{The equations} for the time-dependent photoionization and recombination in \textrm{the} ionized region are as follow,

\begin{equation}
\frac{\partial n_{H^+}}{\partial t}=n_{H}\sigma_{EUV}F_{EUV}-\alpha n_{H^+}^2-\frac{\partial n_{H^+}v_r}{\partial r}-\frac{\partial n_{H^+}v_z}{\partial z}-\frac{n_{H^+}v_r}{r}
\end{equation}

\begin{equation}
\frac{\partial n_{H}}{\partial t}=-n_{H}\sigma_{EUV}F_{EUV}+\alpha n_{H^+}^2-\frac{\partial n_{H}v_r}{\partial r}-\frac{\partial n_{H}v_z}{\partial z}-\frac{n_{H}v_r}{r}
\end{equation}

where $n_{H^+}$ and $n_H$ are number densities of ionized hydrogen and atomic hydrogen, respectively. $\sigma_{EUV}$ and $\alpha$ are the absorption and recombination coefficients, and $F_{EUV}$ is the flux of EUV photons. In the H II region, we consider the photoionization heating as the only heating process \citep{spi78}. The "on the spot" approximation is assumed in treating ionizing photons. We use the empirical cooling curve \textrm{for gas with solar abundances} given by \citet{mel02} who consider the cooling due to the collisional excitation of hydrogen and metal atoms and the recombination radiation of hydrogen atoms.

For the PDR, photoelectric heating, heating from photodissociation, reformation of and FUV pumping of $H_2$ molecules, and \textrm{cosmic rays} are considered in the simulations as heating processes of the gas. Cooling processes considered include atomic fine-structure line emission of [O I] $63\mu m$, [O I] $146\mu m$, [C II] $158\mu m$, [Fe II] $26\mu m$, [Fe II] $35\mu m$ and [Si II] $35\mu m$, the rotational and vibrational transitions of $CO$ and $H_2$, dust recombination and gas-grain collisions \citep{hol79,tie85,hol89,bak94,hos06}.
In \textrm{the} neutral region, the photodissociation and the reformation are always time dependent because the time scales are longer than the computational time step. In addition, the photoionizations and recombinations of metals are also time dependent. The equations for the photodissociation and the reformation are shown below,

\begin{equation}
\frac{\partial n_{H}}{\partial t}=2(R_{diss,H_2}-R_{f,H_2})-\frac{\partial n_{H}v_r}{\partial r}-\frac{\partial n_{H}v_z}{\partial z}-\frac{n_{H}v_r}{r}
\end{equation}

\begin{equation}
\frac{\partial n_{H_2}}{\partial t}=-R_{diss,H_2}+R_{f,H_2}-\frac{\partial n_{H_2}v_r}{\partial r}-\frac{\partial n_{H_2}v_z}{\partial z}-\frac{n_{H_2}v_r}{r}
\end{equation}

\begin{equation}
\frac{\partial n_{C^+}}{\partial t}=R_{diss,CO}-R_{f,CO}-\frac{\partial n_{C^+}v_r}{\partial r}-\frac{\partial n_{C^+}v_z}{\partial z}-\frac{n_{C^+}v_r}{r}
\end{equation}

\begin{equation}
\frac{\partial n_{CO}}{\partial t}=-R_{diss,CO}+R_{f,CO}-\frac{\partial n_{CO}v_r}{\partial r}-\frac{\partial n_{CO}v_z}{\partial z}-\frac{n_{CO}v_r}{r}
\end{equation}

where $n_{H_2}$, $n_{C^+}$ and $n_{CO}$ are the number densities of $H_2$, $C^+$ and $CO$, respectively. $R_{diss,H_2}$ and $R_{diss,CO}$ are the dissociation rates of $H_2$ and $CO$. $R_{f,H_2}$ and \textrm{$R_{f,CO}$} are the reformation rates of $H_2$ and $CO$.
When computing the dissociation rate of $H_2$ molecules, we adopt the self-shielding approximation introduced in \citet{dra96}. This approximation is only valid when the $H_2$ column density exceeds $10^{14}~cm^{-2}$ and FUV $H_2$ absorption lines are optically thick \citep{hol99}. The \textrm{column density of a massive star} formation region easily exceeds $5\times10^{20}~cm^{-2}$ \citep{woo89}. Therefore, such an approximation can be applied. The method in \citet{hol99} is used to calculate the reformation rate of hydrogen molecules. The dissociation and the reformation of $CO$ molecules are computed using the methods provided in \citet{lee96} and \citet{nel97}, respectively. In our calculation, the fractional abundances of $H$ and $CO$ are defined as $X_{H^+}=\frac{n_{H^+}}{n_{H^+}+n_H+2n_{H_2}}$, $X_{H_2}=\frac{2n_{H_2}}{n_{H^+}+n_H+2n_{H_2}}$ and $X_{CO}=\frac{n_{CO}}{n_{CO}+n_{C^+}+n_C}$, respectively.



When calculating the cooling due to the fine-structure line emission, the optical depth needs to be considered \citep{tie85,hol89}. In our simulation, the escape probability of fine-structure line photons is calculated using the method provided by \citet{dej80}. However, it is time consuming if the optical depths are calculated at every directions. So an approximation method is used to estimate the optical-depth, which only \textrm{calculates} the depths and escape probabilities along the axial direction. The same escape probability is used in all directions for simplicity. We compared the line flux from this method with values from calculations without the approximation and found the difference is less than $30\%$. \textrm{Such a difference} is small compared with the flux differences among different lines, therefore can be neglected. In order to assess the reliability of our models, two test cases presented in \textrm{the} Appendix are compared with the analytical solution in \citet{spi78} and the numerical results of \citet{hos06} and \citet{mac15}. \textrm{The comparisons} confirm the reliability.

The profiles of [Ne II] $12.81\mu m$ line and $H_2$ $S(2)$ line are computed by considering thermal broadening and convolved with the Gaussian profile at the resolution of the SOFIA/EXES instruments ($R=86,000$ \citep{sof15}). The position-velocity diagrams in this paper are also presented at this resolution. The [Ne II] line shows the information of the velocity field in the ionized region. The $H_2$ line is used as \textrm{a probe of} the PDR. The dense shell \textrm{bounded} by the IF and the shock front (SF) is the only place in the PDR where \textrm{the} gas density and temperature are \textrm{high enough to excite the $H_2$ line}. There is no obvious velocity structure existing in the region outside of the SF. Therefore, we will not discuss dynamics of this "undisturbed" region.

In this paper, it is assumed that the H II region is at the near side of the molecular cloud and observers are viewing the cometary H II region from the tail to the head and looking into the molecular clouds. Hence, a blue-shifted velocity means that the gas mainly moves \textrm{in} a direction from the head to the tail and flows away from the molecular cloud, and a red-shifted velocity suggests the gas mainly moves \textrm{in} the opposite direction. \textrm{An inclination angle} of $0^o$ means that the line of sight is parallel to the \textrm{symmetry axis} of the H II region. The inclination angle increases when the line of sight rotates clockwise from the \textrm{symmetry axis}.

\section{Results and Discussions}
\label{sect:result}
\subsection{the Evolution of the H II region and the PDR}
\label{sect:front}

\subsubsection{Initial set-up}

First, we simulate the time evolution of the cometary H II region \textrm{together} with the surrounding PDR in a \textrm{bow-shock} model and a \textrm{champagne-flow} model. In both models, the effective temperature of the star is assumed to be $35,000~K$. The photon luminosities of ionizing radiation ($h\upsilon\geq13.6~eV$) and dissociating radiation ($11.26~eV\leq h\upsilon<13.6~eV$) of the star are $10^{48.10}~s^{-1}$ and $10^{48.33}s^{-1}$, respectively. The \textrm{mass-loss} rate and the terminal velocity of the stellar wind are $\dot{M}=3.56\times10^{-7}~M_\odot yr^{-1}$ and $v_w=1986.3~km~s^{-1}$, respectively. These parameters are consistent with a $21.9~M_\odot$ star \citep{dia98,dal13}. In the \textrm{bow-shock} model, we assume the density and the temperature of the uniform medium to be $n=8000~cm^{-3}$ and $T=10~K$ at the start. The star moves at a speed of $v_\ast=15~km~s^{-1}$ with respect to the molecular cloud. In the \textrm{champagne-flow} model, the density of the cloud material is assumed to follow an exponential law as $n(z)=n_0exp(z/H)$. Here z is the coordinate pointing from the surface to the center of the molecular cloud. The number density $n_0$ is assumed to be $8000~cm^{-3}$, and the scale height is $H=0.2~pc$.
The stellar velocity is assumed to be zero in the \textrm{champagne-flow} model. These values of parameters are selected so that the resulting H II region has a similar size as that in the \textrm{bow-shock} model.

\subsubsection{Ionization front and dissociation front evolution}

\begin{figure}[!htp]
\centering
\includegraphics[scale=.45]{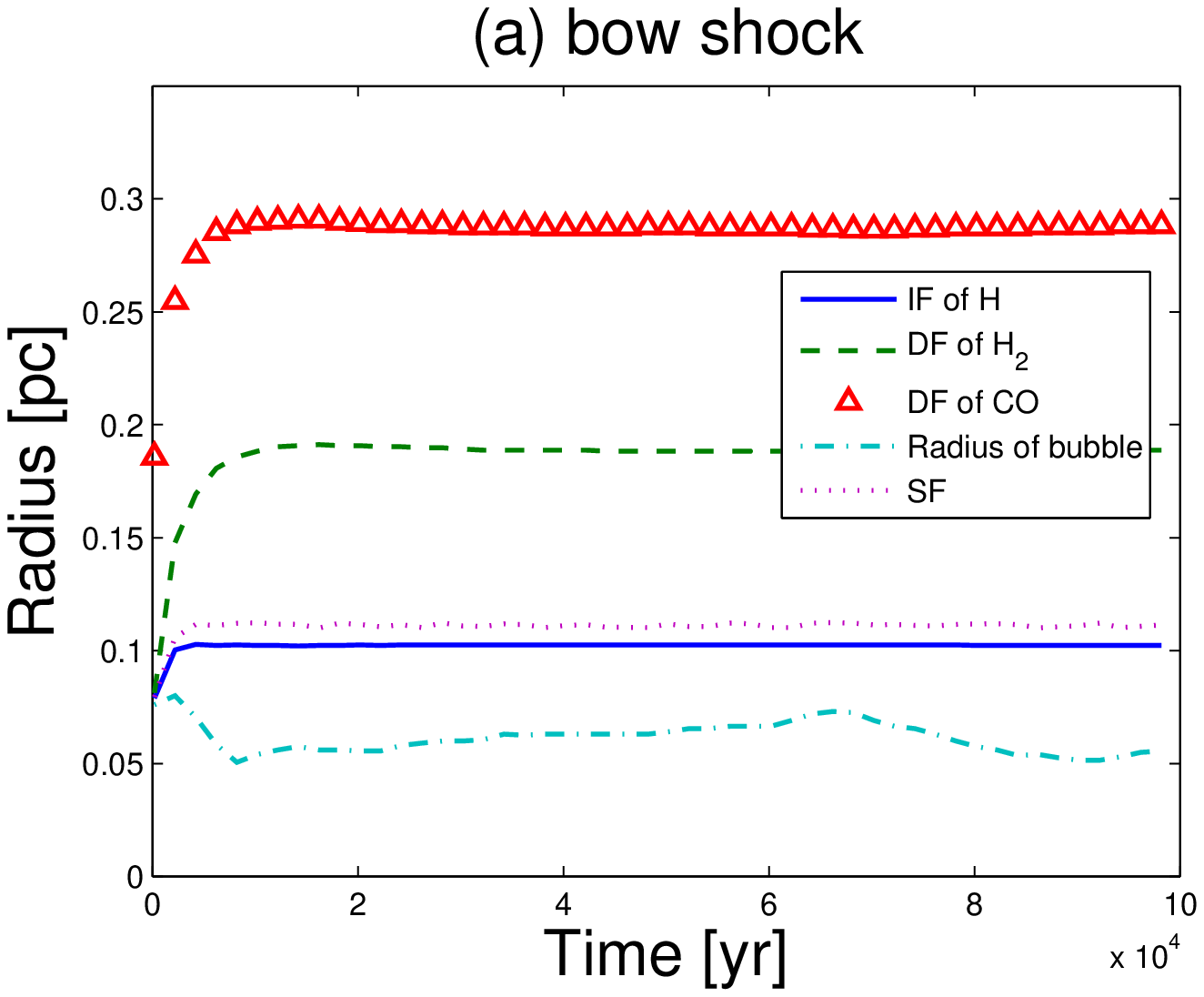}
\includegraphics[scale=.45]{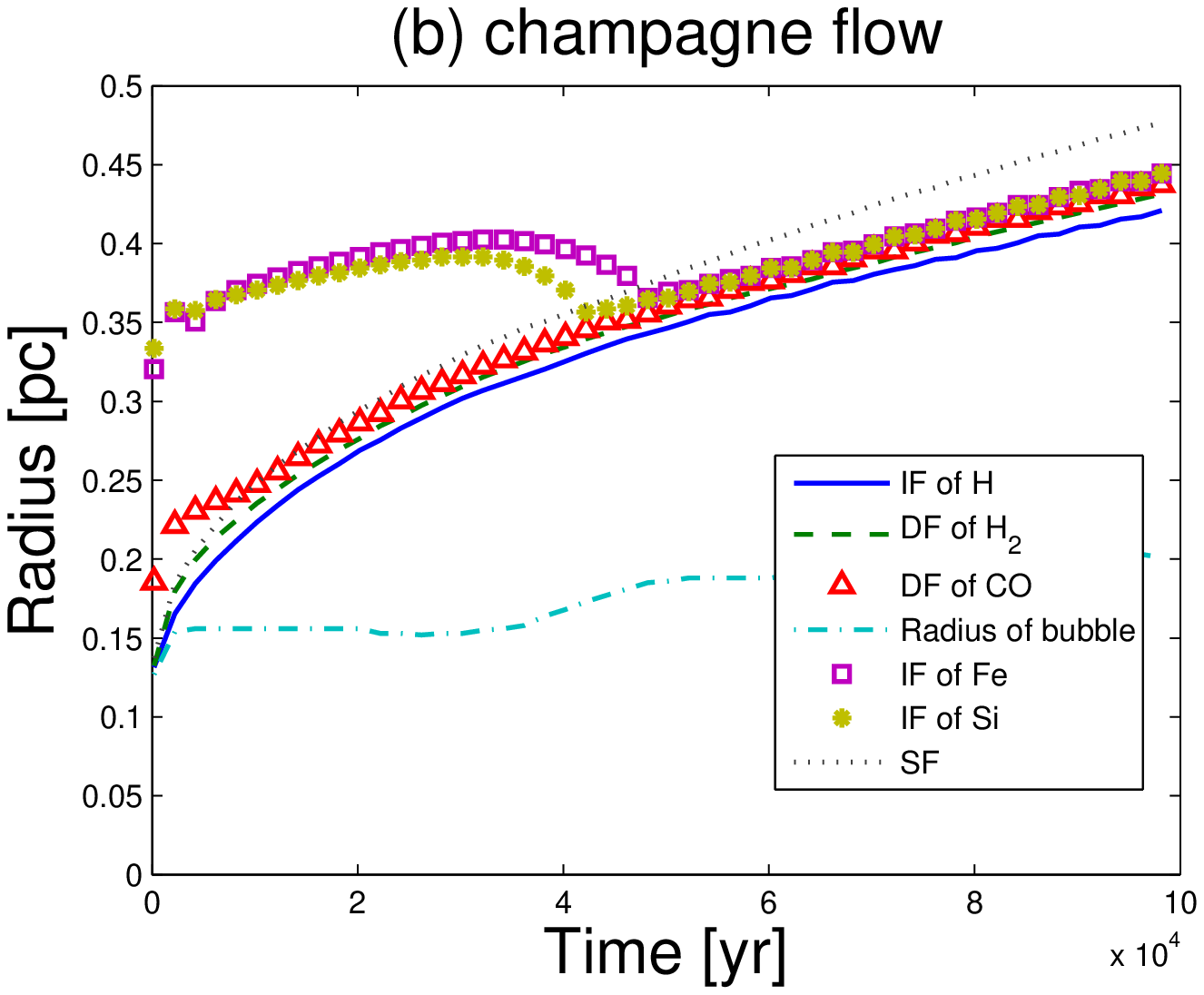}
\caption{The Radii of ionization fronts and dissociation fronts from the star at the direction of $\theta=0^o$ against time for the \textrm{bow-shock} model (left) and the \textrm{champagne-flow} model (right).}
\begin{flushleft}
\end{flushleft}
\label{fig_modfront}
\end{figure}

In Figure \ref{fig_modfront}, the locations of various ionization fronts and dissociation fronts on the \textrm{symmetry axis} are plotted \textrm{as a function of} time for both models. The figure shows that The distances between the star, the ionization front and the dissociation front in the \textrm{bow-shock} model remain roughly constant while these fronts move away from the star in the \textrm{champagne-flow} model. Moreover, in the \textrm{champagne-flow} model, the dissociation fronts (DFs) of $H_2$ and $CO$ and the ionization fronts of $Fe$ and $Si$ are all beyond the shock front at the beginning. \textrm{During the} evolution of the \textrm{champagne-flow} model, these fronts gradually fall into the dense shell and move away from the star. This difference \textrm{between the two models is mainly due to the different column} densities of the compressed shell. In the \textrm{bow-shock} model, the column density of the shell tends to be a stable value. However, the shell column density in the \textrm{champagne-flow} model gradually increases with time. It should be noted that the bow shock caused by a relatively low speed ($\leq10~km~s^{-1}$) star needs more time to reach \textrm{a} quasi-stationary state. Before this state, the H II region also grows \textrm{with} time. So does the column density of the swept-up shell. This reduces the intensity of the \textrm{FUV photons, and so the radii of} the dissociation fronts decrease. Eventually, the dense shell will overtake the dissociation fronts. 


\subsubsection{Cooling rate evolution in the PDR}

\begin{figure}[!htp]
\centering
\includegraphics[scale=.45]{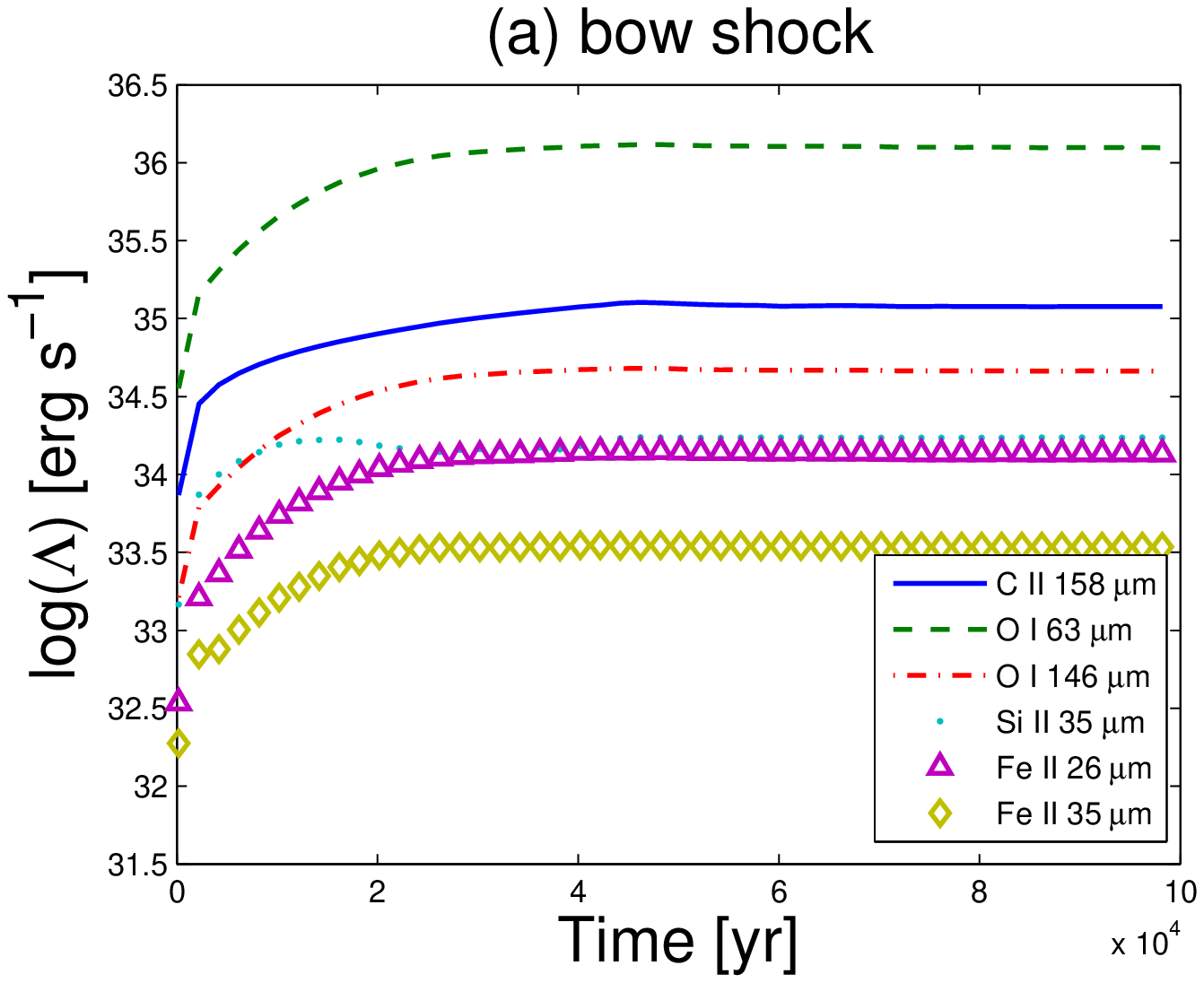}
\includegraphics[scale=.45]{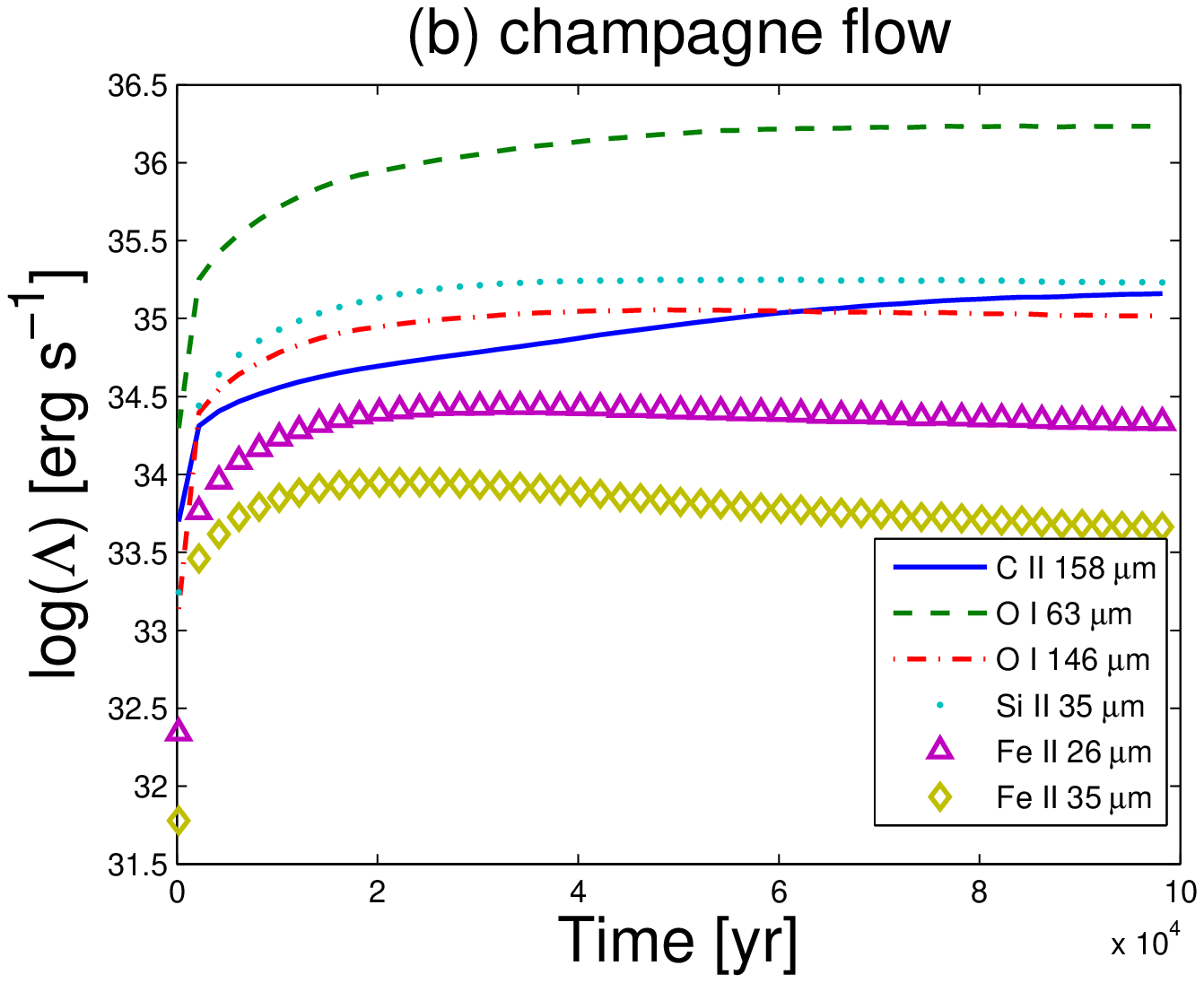}
\caption{The variation of the cooling rate due to the fine-structure lines in PDR for the \textrm{bow-shock} model (left) and the \textrm{champagne-flow} model (right).}
\begin{flushleft}
\end{flushleft}
\label{fig_modcooling}
\end{figure}

The temporal evolutions of the cooling rates of different fine-structure lines \textrm{are plotted} in Figure \ref{fig_modcooling}. The calculation shows that in both models the cooling rates increase rapidly at early times and approach stable values after $40,000~yr$. Among all the lines under consideration, the [O I] $63 \mu m$ line emission is the main cooling process in the PDR. the cooling rate of the [O I] $146 \mu m$ line emission is less than $7\%$ of that of the [O I] $63 \mu m$ line. The cooling due to the [Fe II] $35 \mu m$ line emission is less than $1\%$ of that of the [O I] $63 \mu m$ \textrm{line and is the lowest among all the lines considered}. In the \textrm{bow-shock} model, the cooling due to [C II] $158 \mu m$ line emission is the second most important. But, in the \textrm{champagne-flow} model, the cooling due to [Si II] $35 \mu m$ line emission is higher than that of the [C II] line. For most of these lines, the cooling rates in the \textrm{champagne-flow} model are generally higher than in the \textrm{bow-shock} model due to higher densities in the PDRs of the \textrm{champagne-flow} models.




\subsection{Bow-shock models}
\label{sect:bow shock}

We simulate two \textrm{bow-shock} models in the uniform medium. The parameters assumed in model A are the same as those in the \textrm{bow-shock} model mentioned in \ref{sect:front}. In model B, the $21.9~M_\odot$ star is replaced with a $40.9~M_\odot$ star. For the new star, the ionizing and dissociating photon luminosities are $10^{48.78}~s^{-1}$ and $10^{48.76}~s^{-1}$, respectively. The effective temperature is $40,000~K$. The stellar wind mass-loss rate and the terminal velocity are adopted to be $\dot{M}=9.93\times10^{-7}~M_\odot yr^{-1}$ and $v_w=2720.1~km~s^{-1}$, respectively. \textrm{In both models}, the stellar velocity is $v_\ast=15~km~s^{-1}$. Again, these parameters are taken from \citet{dia98} and \citet{dal13}. We carry out the simulation in the rest frame of the star in order to use the same procedures \textrm{as in the champagne-flow cases. The calculated} velocities are \textrm{then} converted to the values in the rest frame of the molecular cloud. In all models, the z-axis is parallel to the \textrm{symmetry axis} and the positive direction is from the tail to the head of the cometary H II region. The position of the star is at $(r,z)=(0,0)$.

\begin{figure}[!htp]
\centering
\includegraphics[scale=.5]{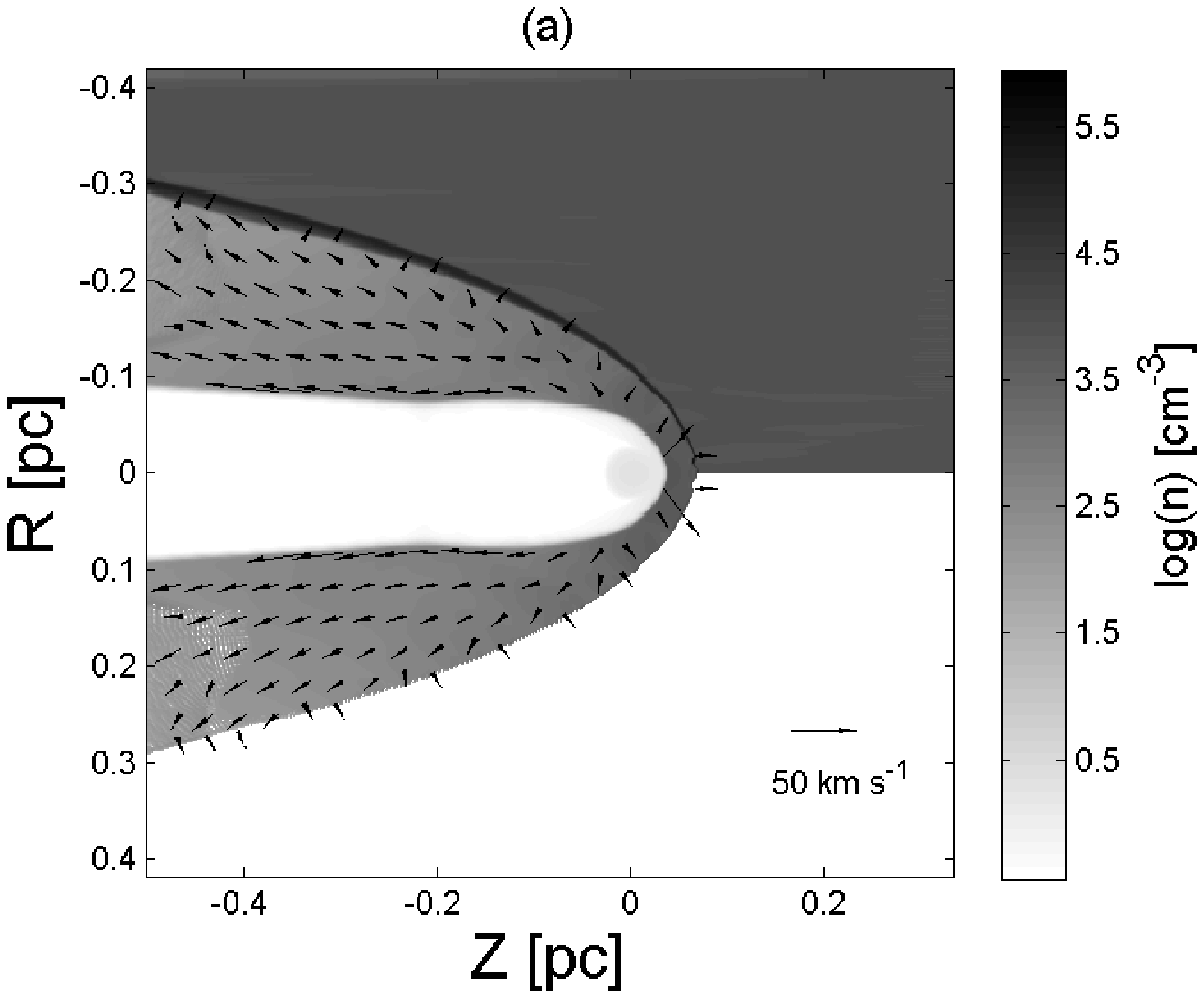}
\includegraphics[scale=.5]{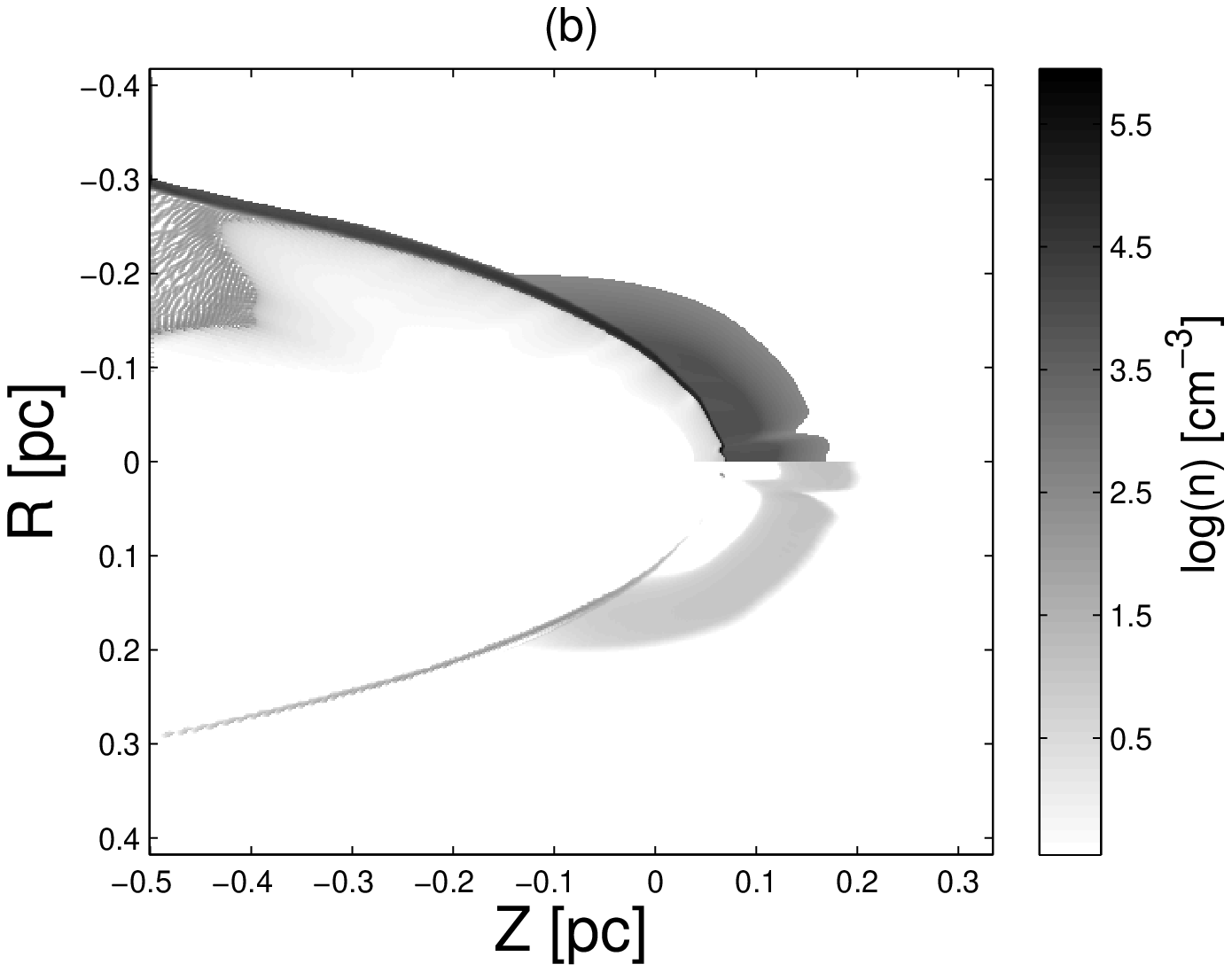}
\includegraphics[scale=.5]{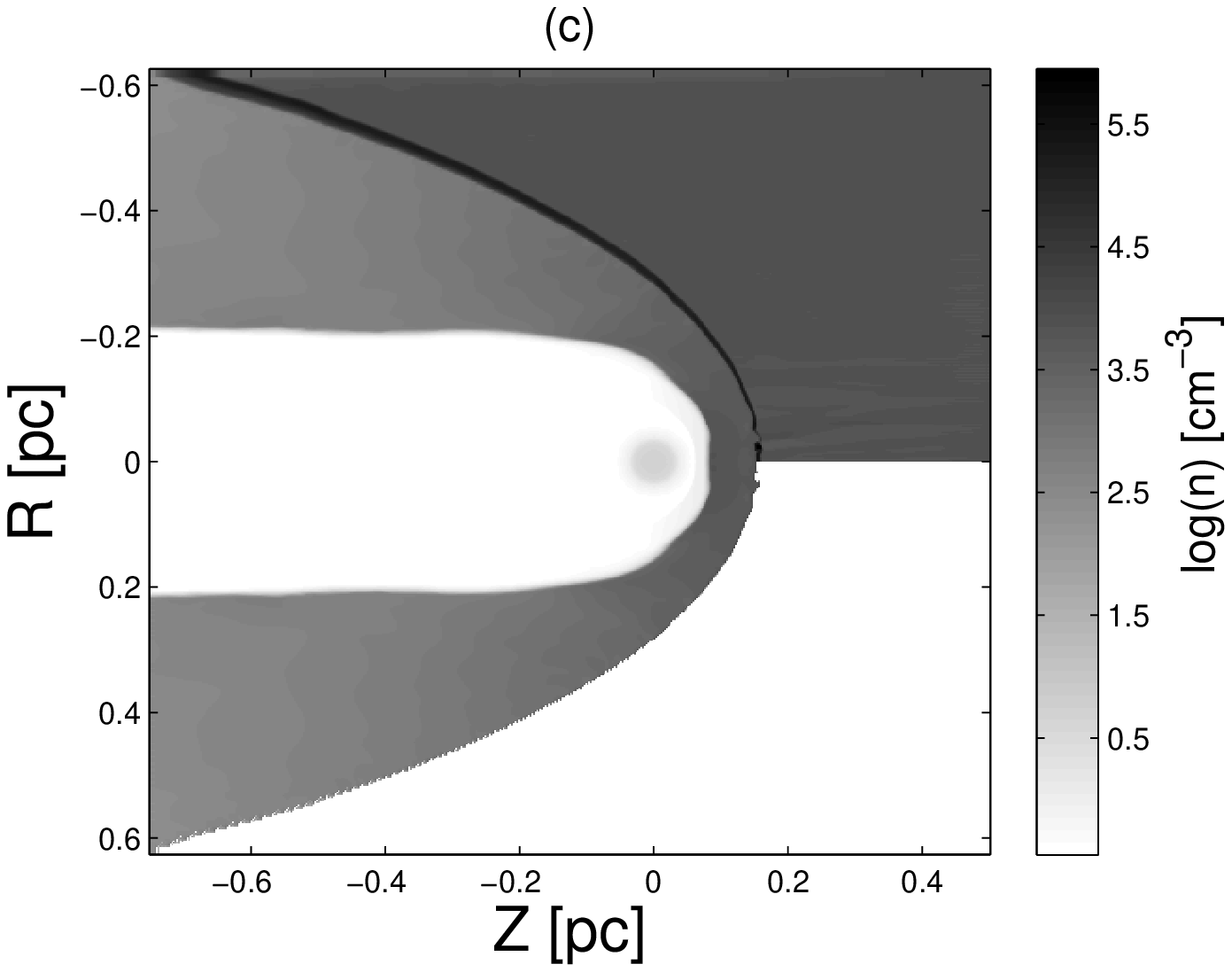}
\includegraphics[scale=.5]{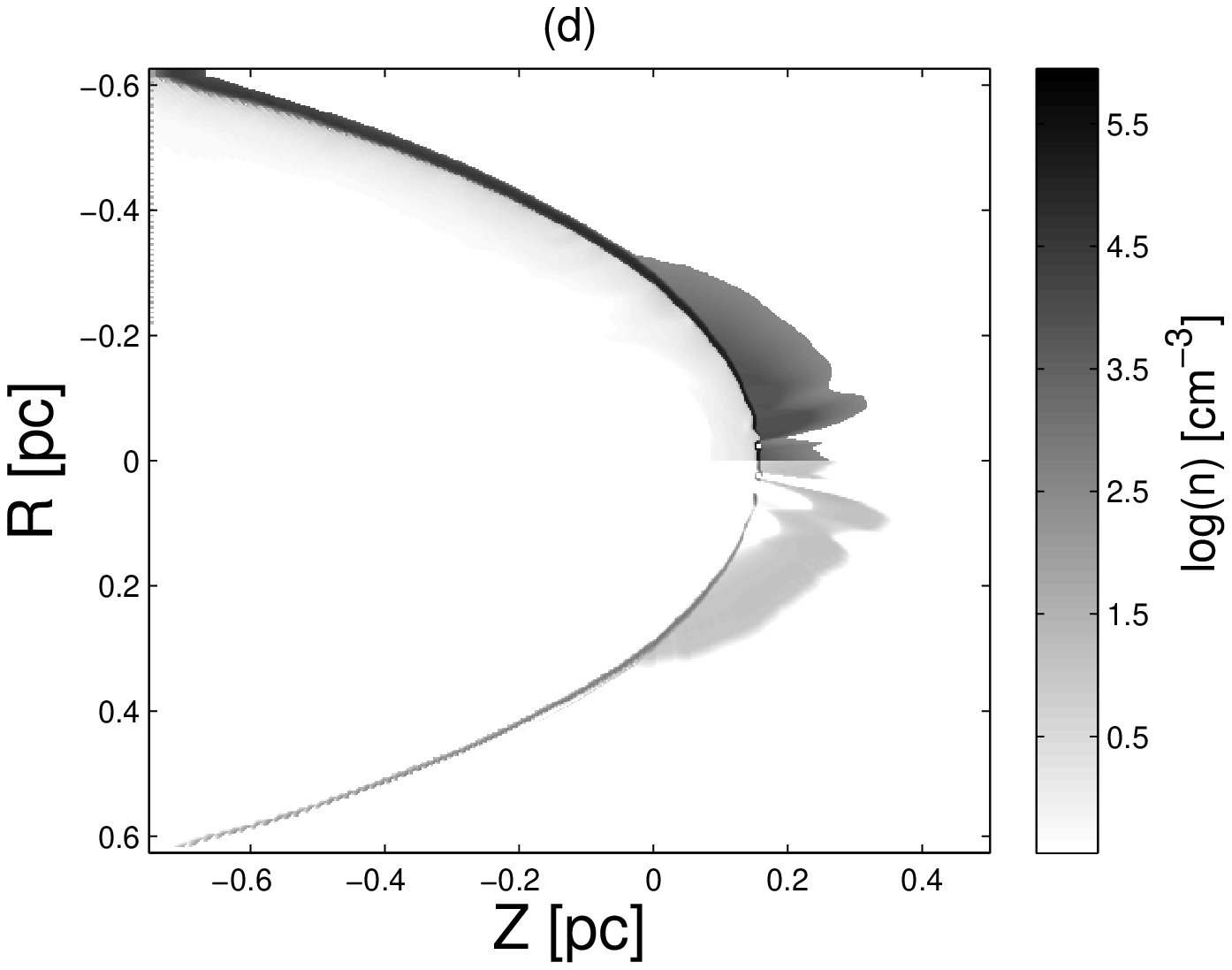}
\caption{The number density distributions of all materials, ionized hydrogen, hydrogen atoms and hydrogen molecules at the rotational level of $J=4$ in \textrm{models} A and B at the age of $100,000~yr$. The top panels and the bottom panels are for \textrm{models} A and B, respectively. The left panels show the densities of all materials (top half) and the $H^+$ ions (bottom half). The right panels show those of $H$ atoms (top half) and $H_2$ molecules (bottom half). The arrows show the velocity field, and only velocities higher than $1~km~s^{-1}$ are presented, but the velocity field in the \textrm{stellar-wind} bubble is not shown.}
\begin{flushleft}
\end{flushleft}
\label{fig_modab}
\end{figure}

We cease the simulations at $100,000~yr$. The density distributions of all materials, the $H$ atoms, the $H^+$ ions and the warm $H_2$ molecules in \textrm{models} A and B at \textrm{the} age of $100,000~yr$ are presented in \textrm{Figure \ref{fig_modab}, together with the velocity fields}. We compared the distributions at $100,000~yr$ with those at age of $80,000~yr$, which are not presented in the current work, and found that the appearance of the cometary H II region does not changed significantly for more than \textrm{twenty thousand years. So we conclude that the bow shock} has entered a quasi-stationary stage. In model A, a low-density and hot \textrm{stellar-wind} bubble ($T>10^6~K$ and $n<5~cm^{-3}$) forms around the moving star. The photoionized region enveloped by a dense shell surrounds the hot bubble. In this region, the densities are $n\sim200-20000~cm^{-3}$, and the temperatures are close to $T=10000~K$. A shell outside of the photoionized H II region is made of the compressed neutral gas with density and temperature $n\sim50000~cm^{-3}$ and $T\sim400~K$. The temperature in the neutral region outside of the shell decreases gradually from $\sim500~K$ near the IF to $\sim50~K$ in the depth of the molecular cloud. The top panels of Figure \ref{fig_modab} show that a part of ionized gas in the tail is recombining. This is because the ionizing photon flux decreases \textrm{with distance} from the star. The $H_2$ dissociation front keeps a distance from the dense shell ahead of the apex of H II region, but it falls into the dense shell at the sides because the flux of dissociation photons reduces along the shell due to the increasing distance and the higher column density of the shell at the sides. In model B, the structure of the H II region is similar to that in model A. However, because of the presence of a more massive star in model B, the sizes of the photoionized region and \textrm{stellar-wind} bubble are larger. And there is no indication of recombining gas in the tail because the ionizing photon flux is still high enough to keep all material ionized there. The figure also shows that the PDR in model B is not significantly bigger than that in model A. This is due to the longer distance between the star and the shell and the higher density of the shell in model B. 

\subsubsection{Profiles and diagrams of the [Ne II] $12.81$ line}

\begin{figure}[!htp]
\centering
\includegraphics[scale=.33]{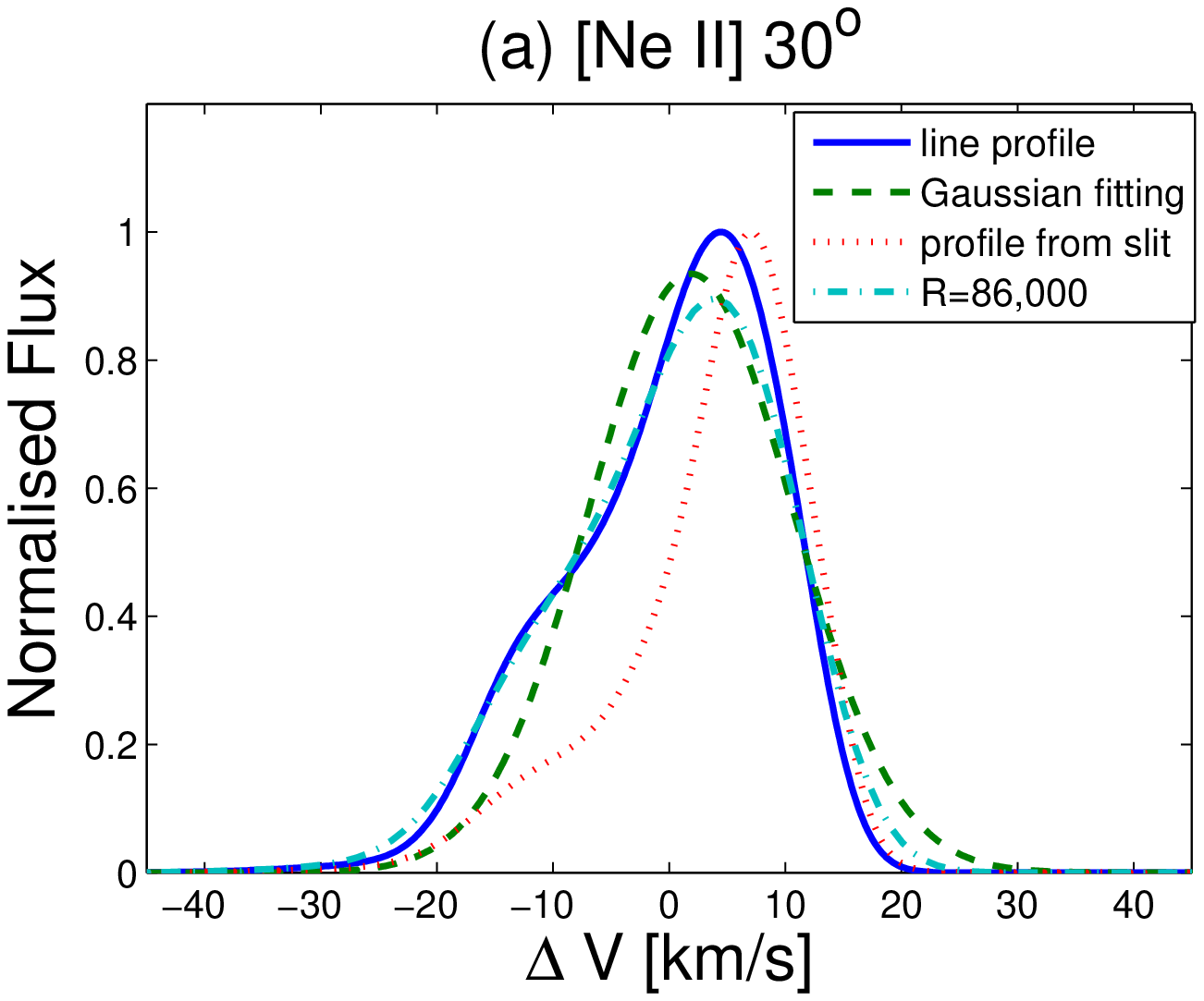}
\includegraphics[scale=.33]{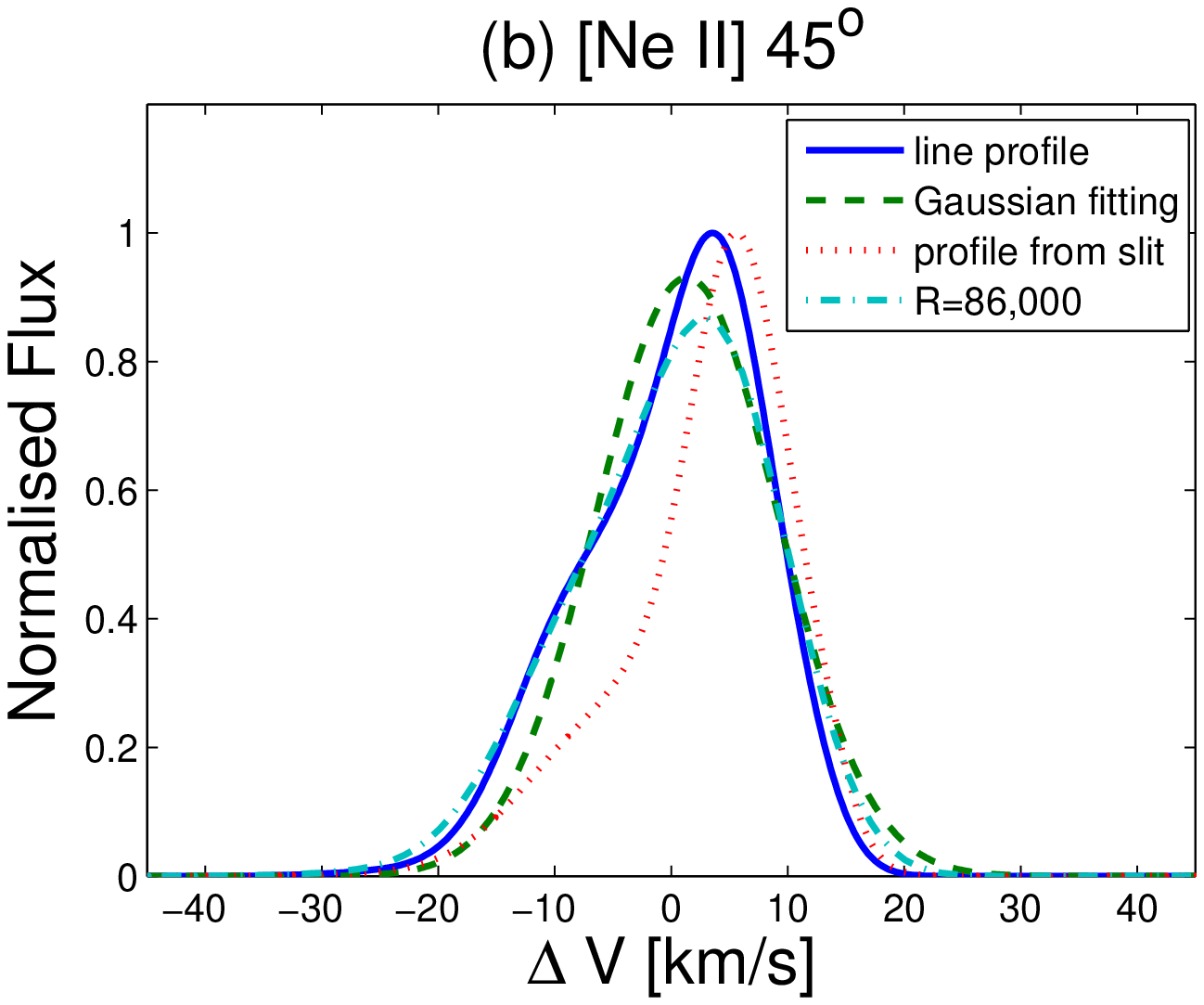}
\includegraphics[scale=.33]{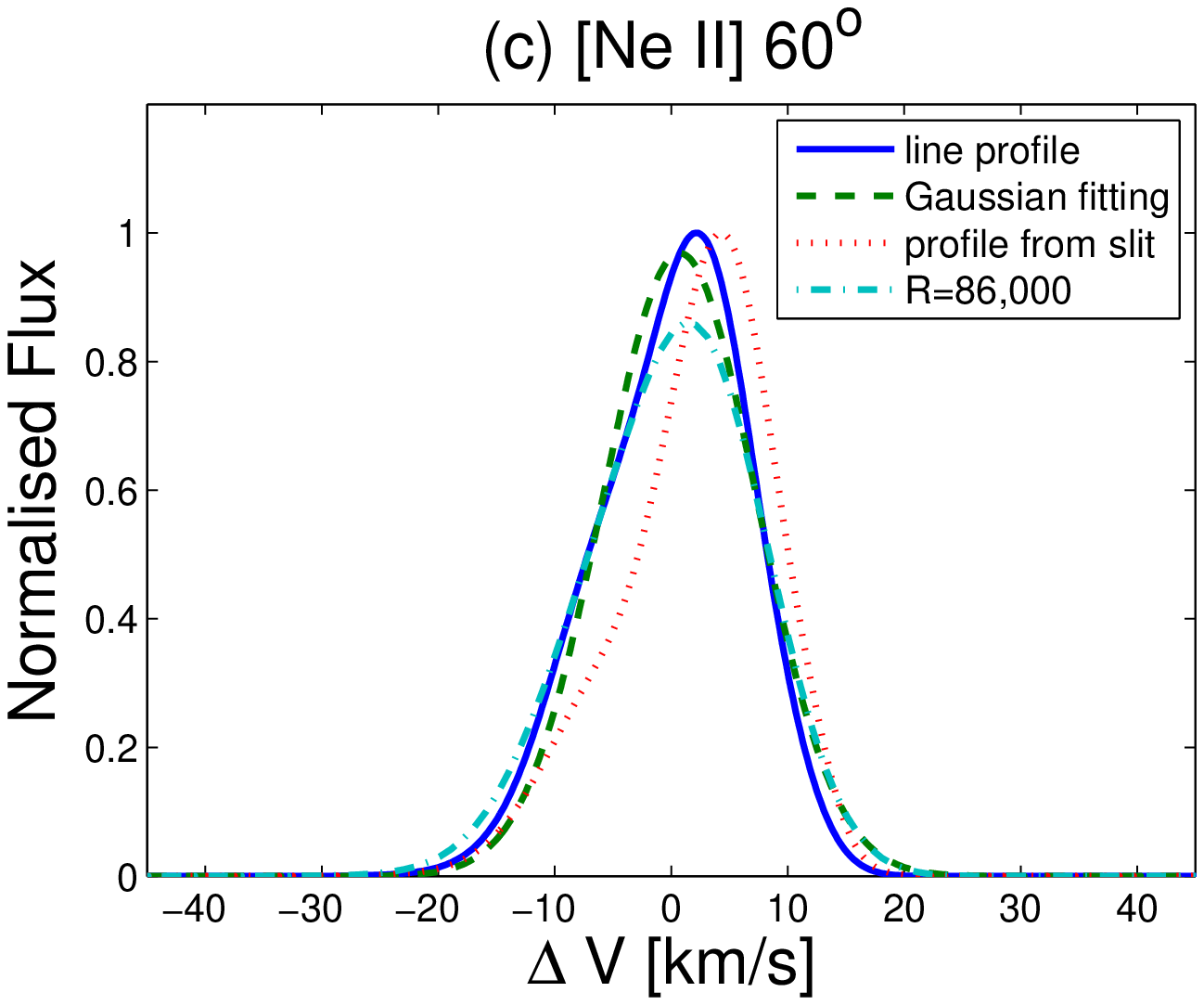}
\includegraphics[scale=.33]{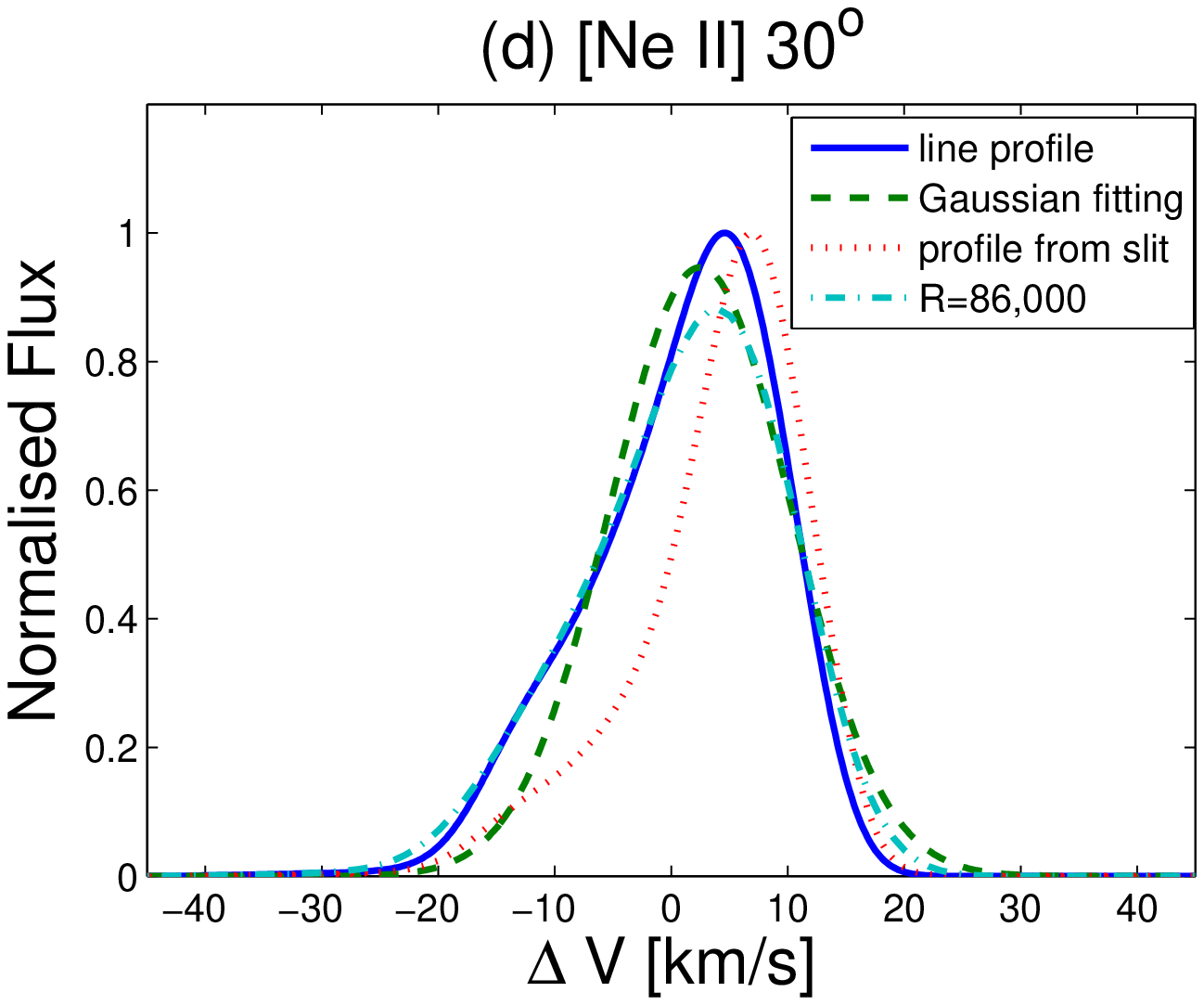}
\includegraphics[scale=.33]{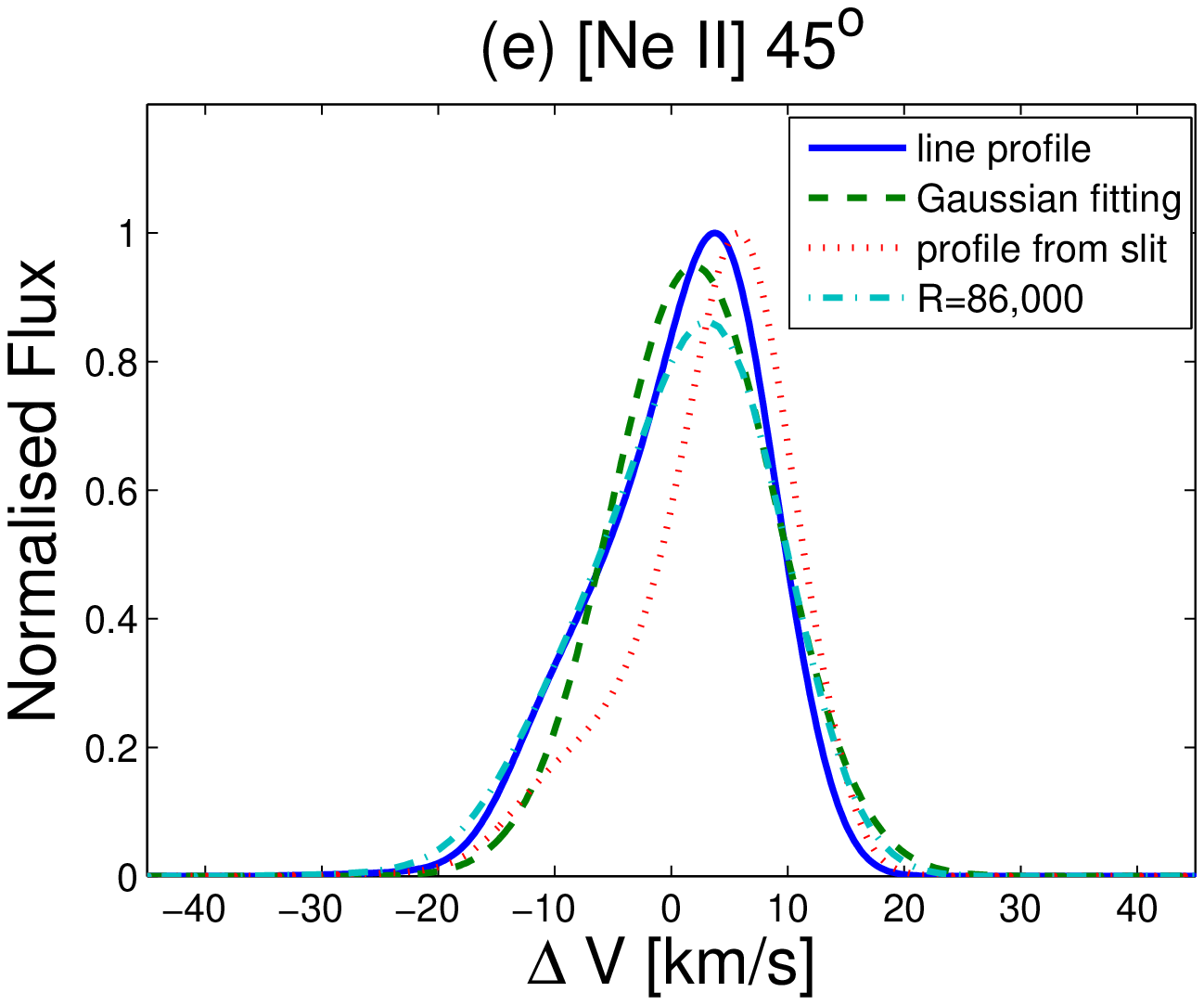}
\includegraphics[scale=.33]{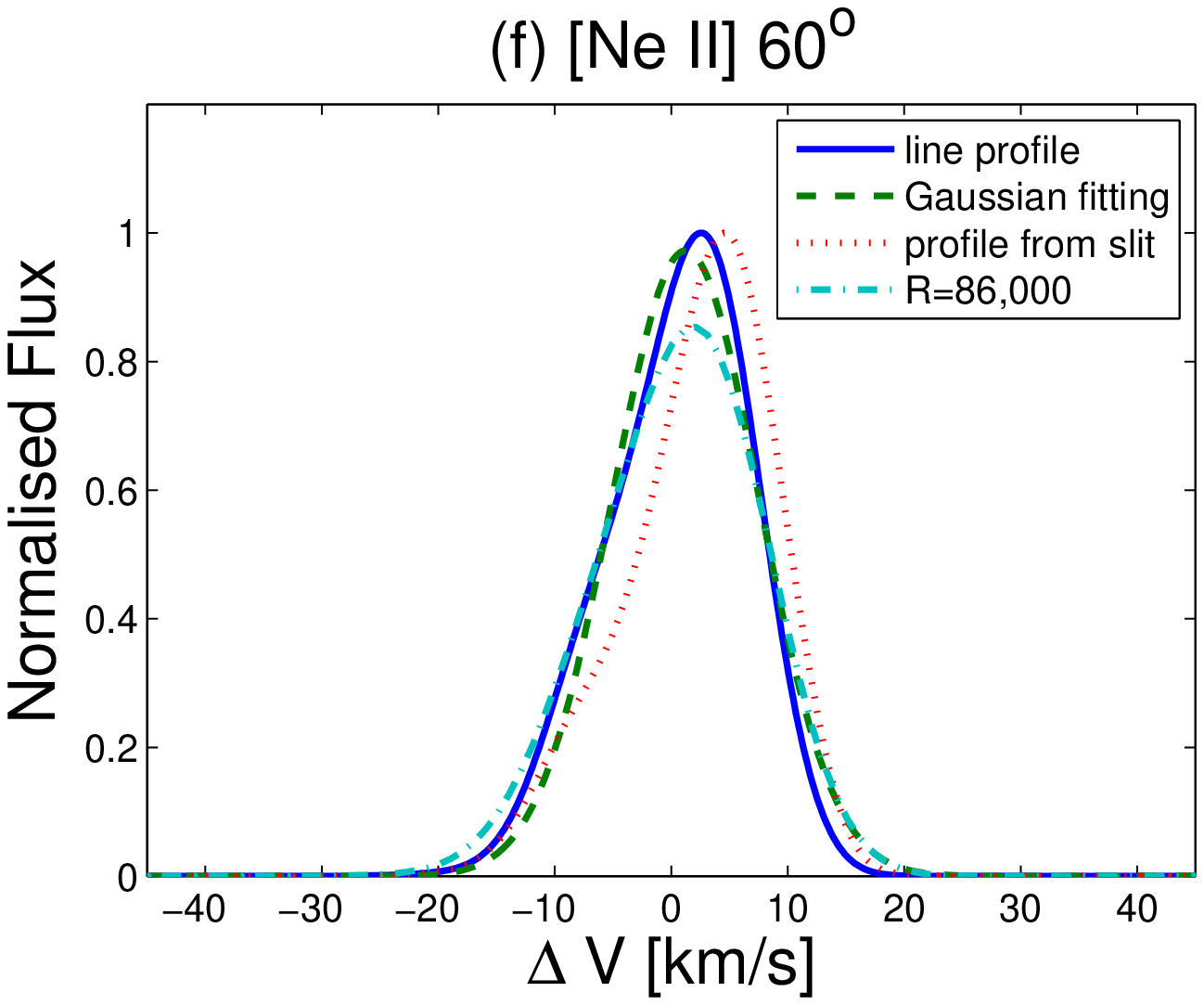}
\caption{Profiles of the [Ne II] line (solid lines), the Gaussian fitting to the profiles (dashed lines) and the profiles at the resolution of $R=86,000$ (dot-dashed lines) at three inclination angles ($\theta=30^o$, $45^o$ and $60^o$). The [Ne II] line profiles from the slit along the \textrm{symmetry axis} of the projected 2D image are also plotted (dotted lines). The top panels are plotted for model A, and the bottom panels are for model B}
\begin{flushleft}
\end{flushleft}
\label{fig_modabfit}
\end{figure}

The profiles of the [Ne II] $12.81$ line 
for the inclinations of $\theta=30^o,~45^o$, and $60^o$ are presented in Figure \ref{fig_modabfit}. \textrm{The differences between the profiles from model A and model B are not large}. We also show the properties of the [Ne II] line in Table \ref{tab_modab}. The computed profiles of the [Ne II] line are all negative-skewed. The skewnesses of the line profiles in model A are $-0.59,~-0.50,~and~-0.33$ for the inclinations of $\theta=30^o,~45^o,~and~60^o$, respectively. And those in model B are $-0.61,~-0.51,~and~-0.34$ for the three inclination angles. It seems that the skewness of the [Ne II] line profiles in \textrm{bow-shock} models correlates with the inclination but not the stellar mass. The flux weighted central velocities (FWCVs) of the [Ne II] line profiles are $0.03,~0.02,~0.02~km~s^{-1}$ in model A and $0.91,~0.74,~0.52~km~s^{-1}$ in model B for $\theta=30^o,~45^o,~\textrm{and}~60^o$, respectively. The larger EUV flux and the stronger stellar wind in model B result in a larger proportion of the red-shifted ionized gas, therefore slightly more red-shifted FWCVs of the line profiles. Because Gaussian fitting of the line profiles are the routine procedure in observations, we provide the fitting results in the current work as well. In \textrm{models} A and B, the centers of the fitting curves are all red-shifted ($1.87,~1.35$, and $0.73~km~s^{-1}$ for $\theta=30^o,~45^o$, and $60^o$ in model A, $2.51,~1.96$, and $1.25~km~s^{-1}$ in model B). Both the centers of the fitting curves and the FWCVs in the two \textrm{bow-shock} models are much lower than the projected stellar velocities along the line of sight. This is because the pressure gradient caused by the density gradient formed in the H II region leads to an acceleration generally along the direction toward the tail. This acceleration causes the ionized gas to move slower than the star along the symmetric axis. Although a component of acceleration also exists along the direction perpendicular to the axis. Apparently, this perpendicular acceleration does not contribute significantly to the projected velocity component along the line of sight. The details of this phenomenon has been discussed in \citet{zhu15}. The full width half maximum (FWHM) of the fitting \textrm{curves} in \textrm{models} A and B is $20.66,~18.13,~15.51~km~s^{-1}$ and $18.33,~16.66,~14.82~km~s^{-1}$, respectively. It is obvious that the FWHM decreases \textrm{with increasing} angle. In Figure \ref{fig_modabfit}, the [Ne II] line profile becomes more Gaussian with the increasing inclination angle. The [Ne II] line profiles from the slit (20 cells wide) along the \textrm{symmetry axis} of the projected 2D image for the three inclinations in \textrm{models} A and B are also presented in Figure \ref{fig_modabfit}. The profiles from the slit are obviously more red-shifted than the corresponding profiles from the whole H II region. This is because the influence of the dense and red-shifted gases in the head of H II region is enforced, and the contribution to the profiles from the low-density blue-shifted region is weakened when the line profiles are only computed from the slit \citep{zhu15}. In addition, the profiles at the resolution of SOFIA/EXES instrument are shown in Figure \ref{fig_modabfit}. Because the thermal broadening of $Ne^+$ ions is comparable to the broadening at the slit resolution, these 'realistic' profiles are not very different from the 'ideal' [Ne II] line profiles.

\begin{figure}[!htp]
\centering
\includegraphics[scale=.33]{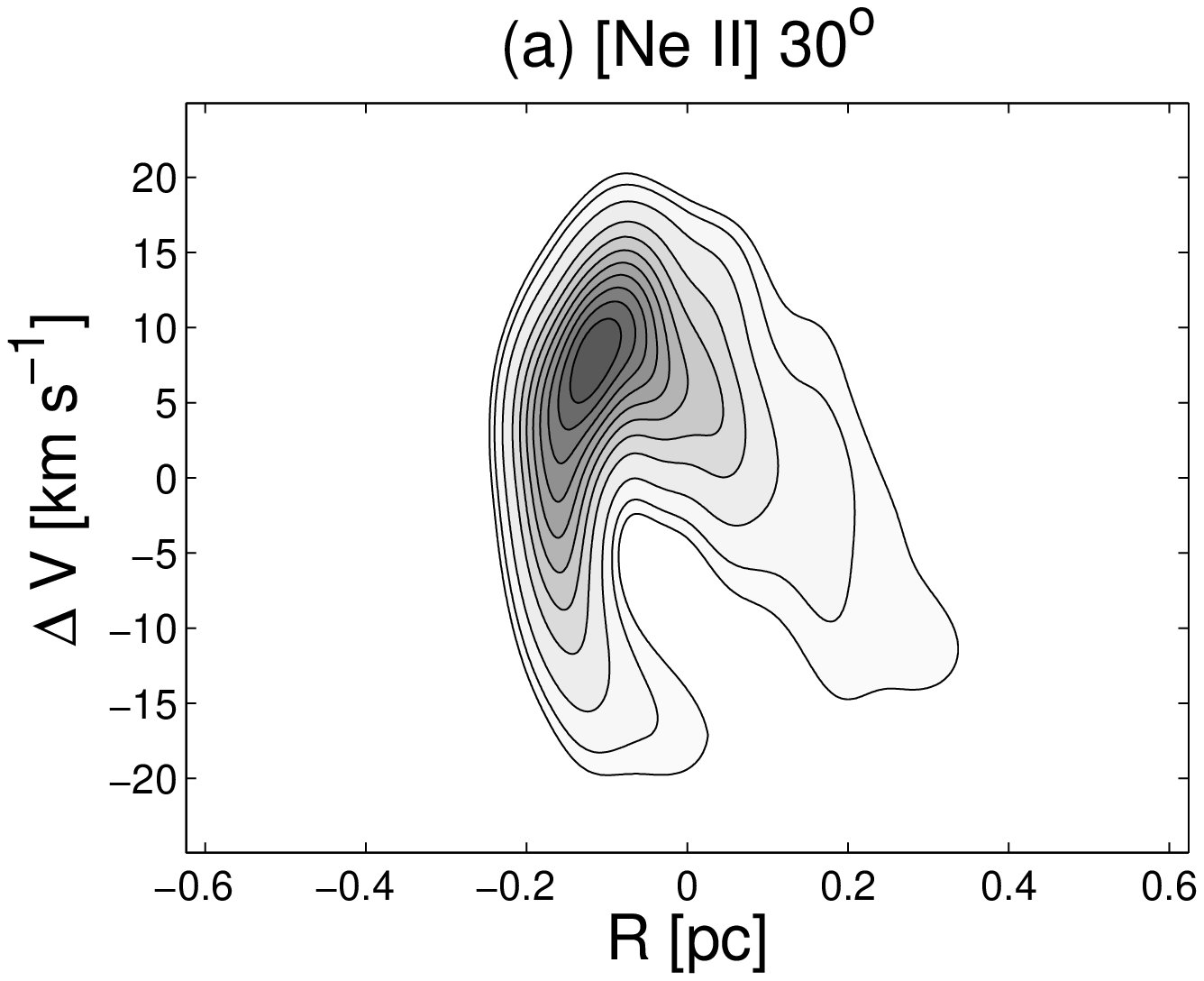}
\includegraphics[scale=.33]{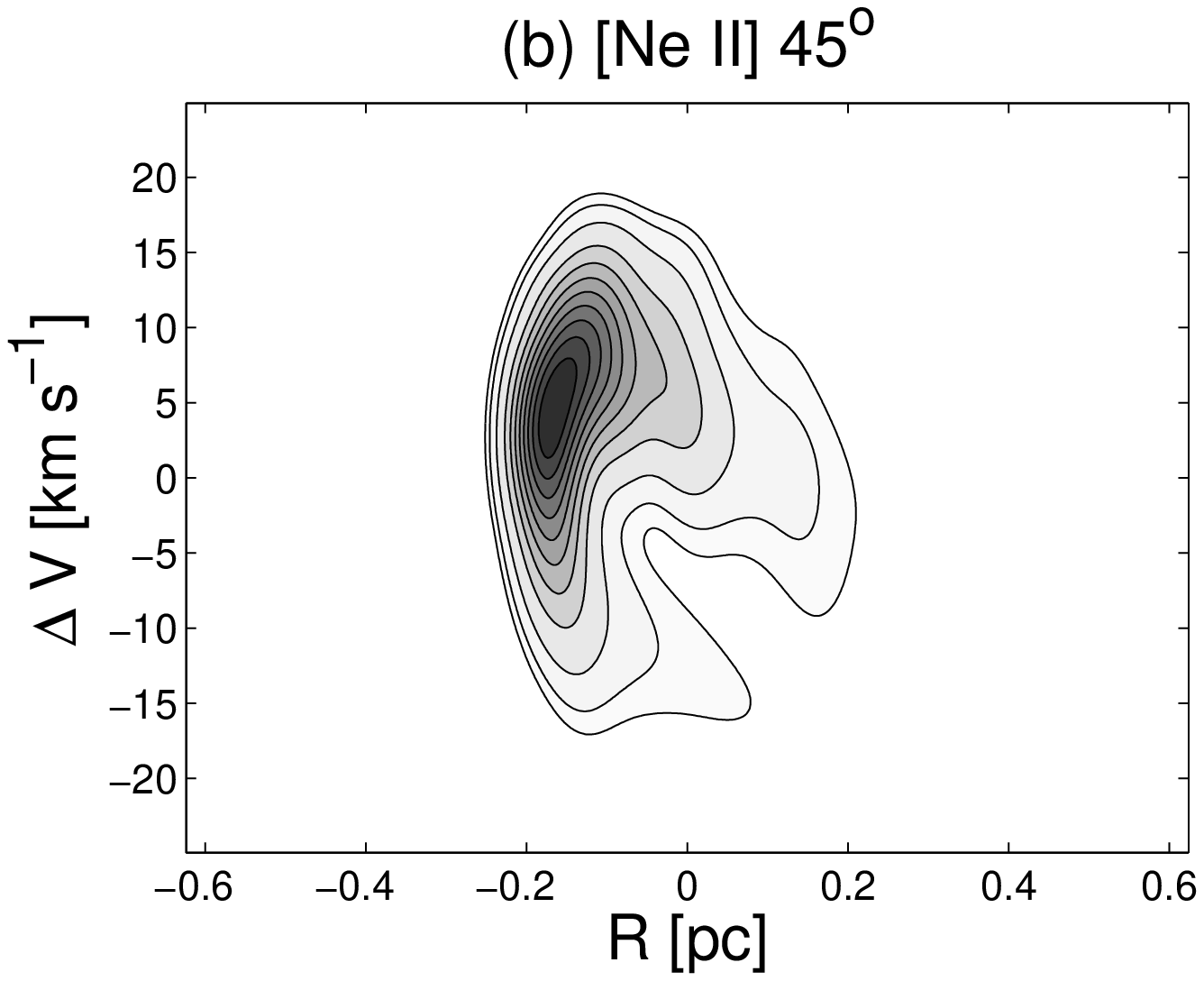}
\includegraphics[scale=.33]{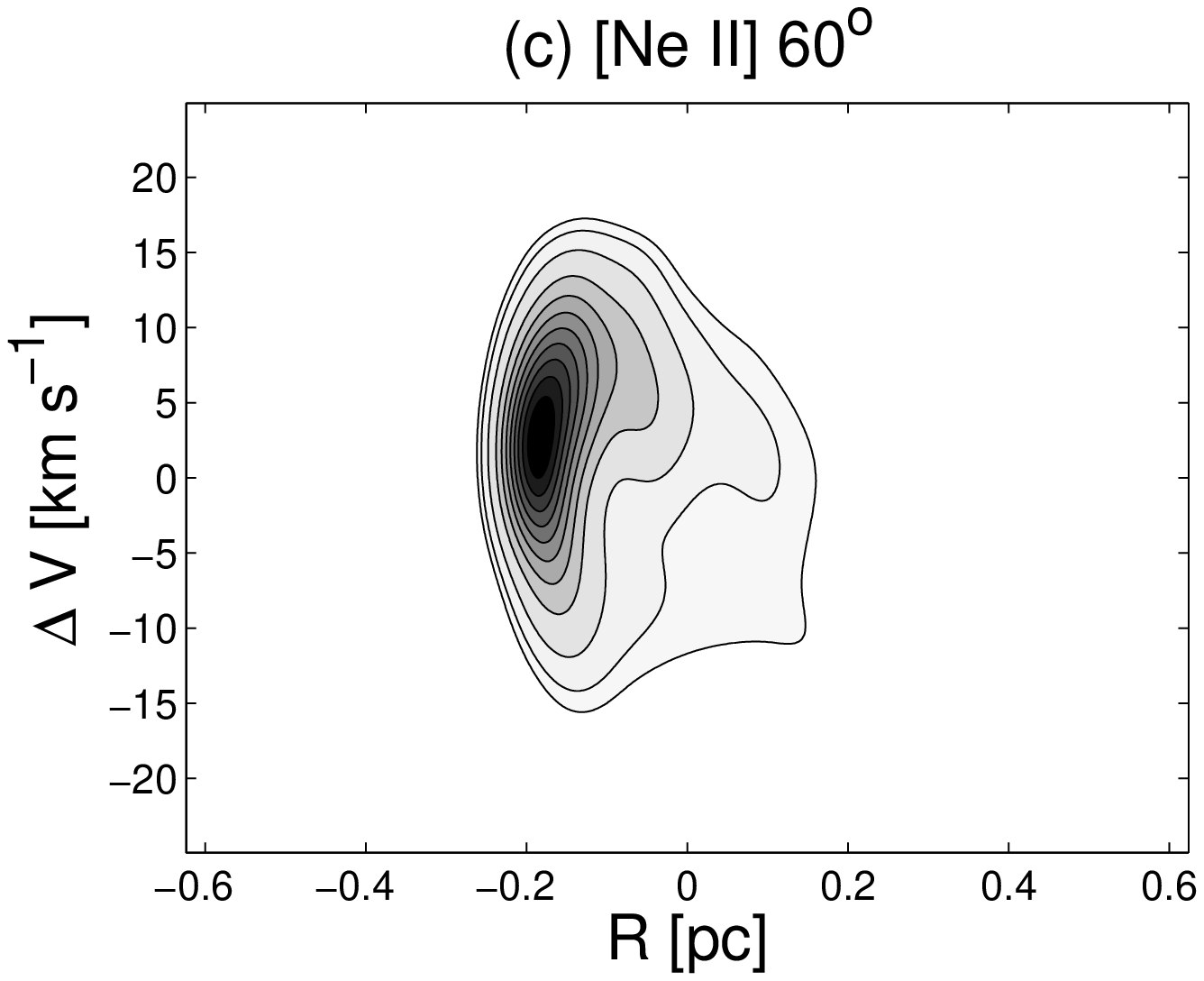}
\includegraphics[scale=.33]{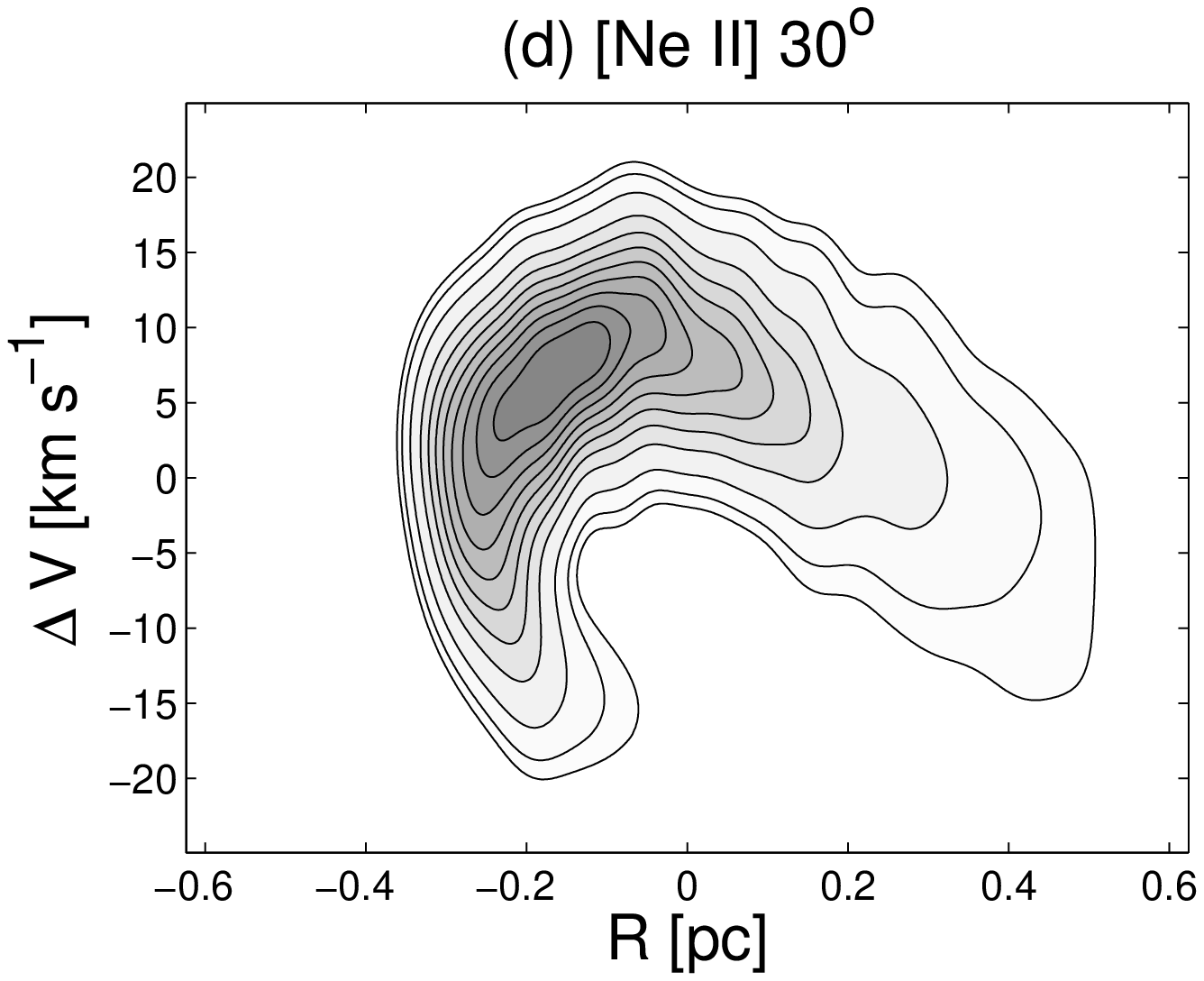}
\includegraphics[scale=.33]{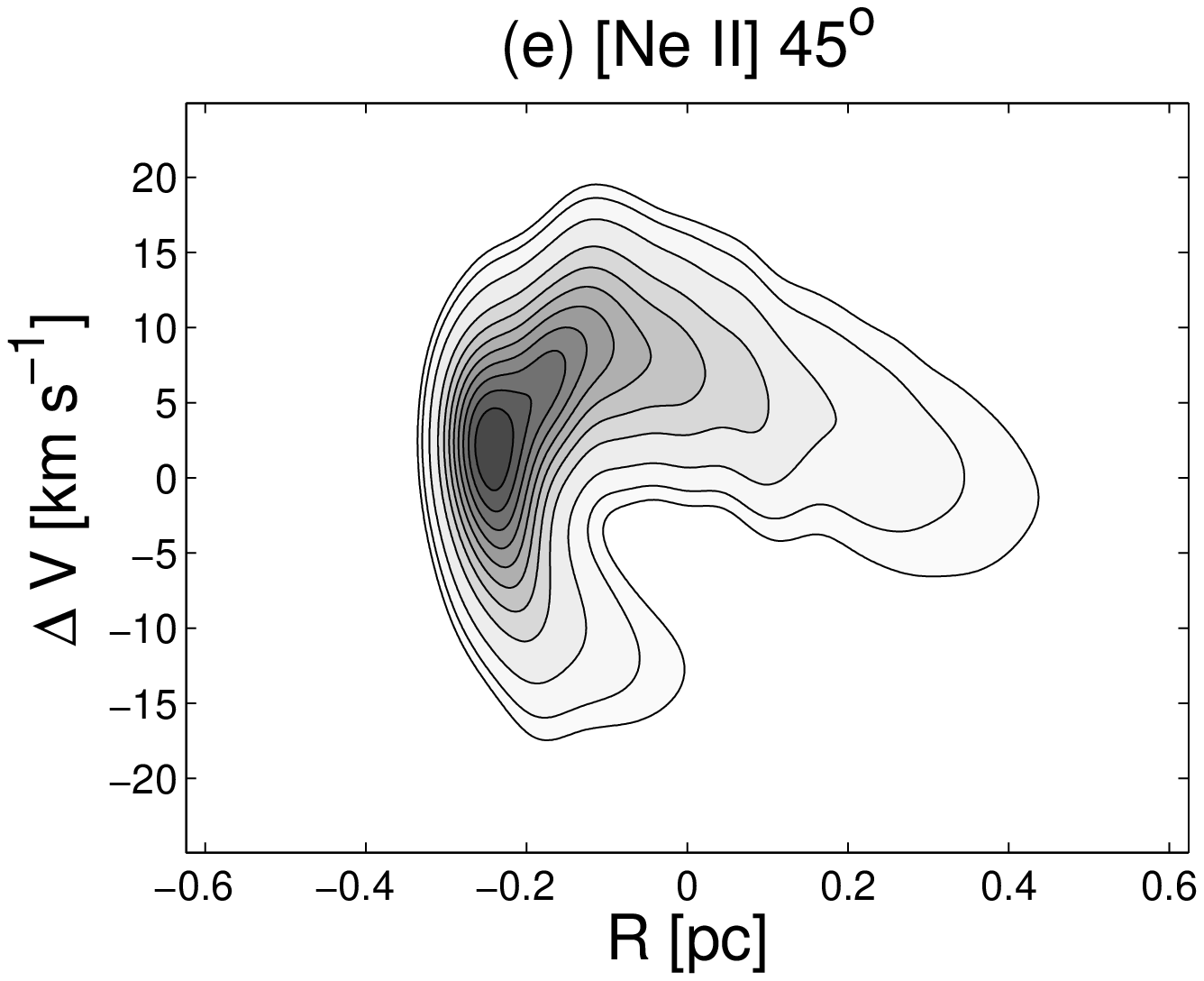}
\includegraphics[scale=.33]{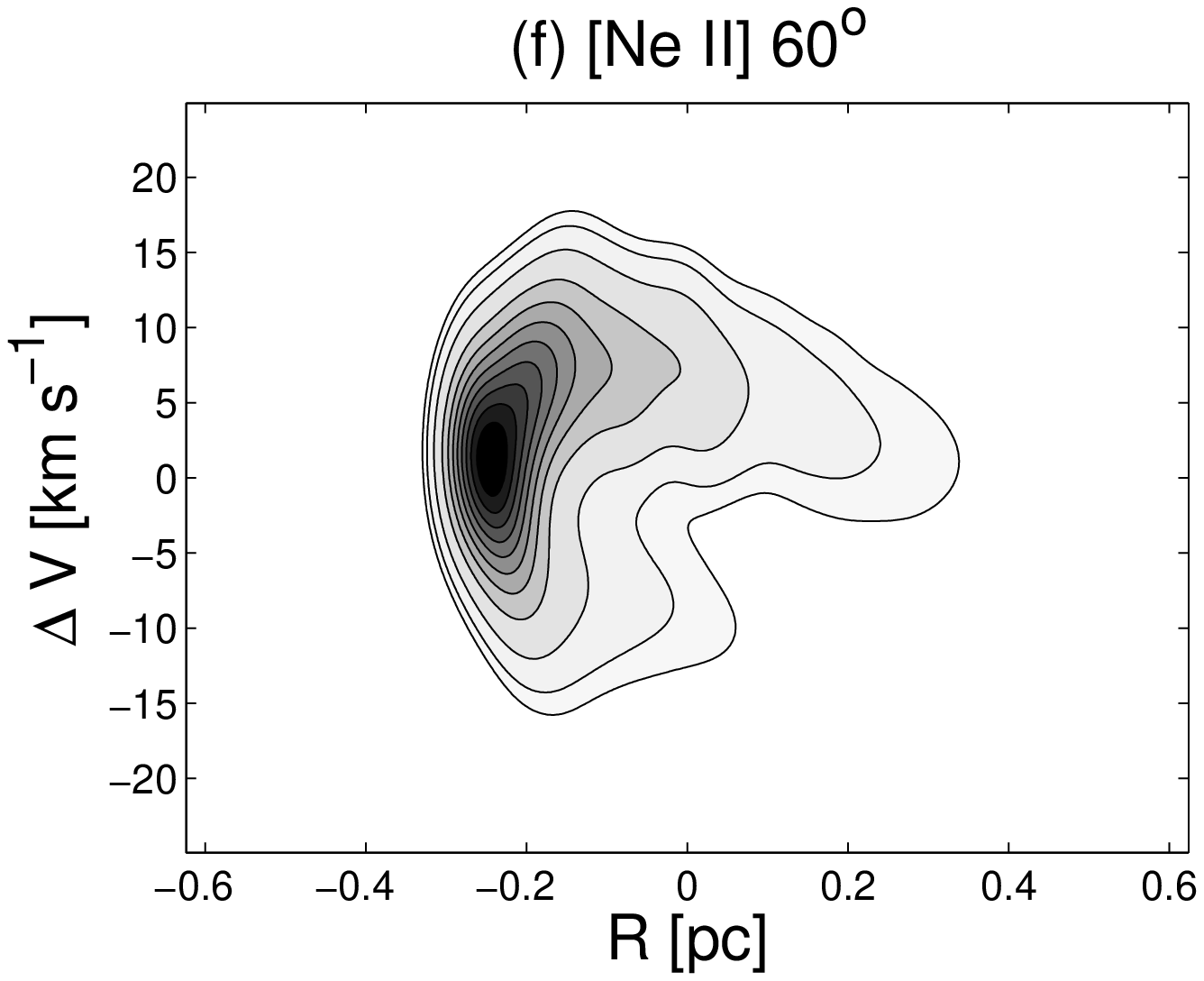}
\caption{The position-velocity diagrams of the [Ne II] $12.81~\mu m$ line from the slit along the \textrm{symmetry axis} of the projected 2D image for three inclination angles ($\theta=30^o$, $45^o$ and $60^o$). The top panels are presented for model A, and the bottom panels are for model B. The contour levels are at $3,~5,~10,~20,~30,~40,~50,~60,~70,~80,~90\%$ of the emission peaks in each panel.}
\begin{flushleft}
\end{flushleft}
\label{fig_modabpvNe}
\end{figure}

In Figure \ref{fig_modabpvNe}, the position-velocity diagrams of the [Ne II] line from the slit along the \textrm{symmetry axis} of the projected 2D image in \textrm{models} A and B are shown for three \textrm{inclination angles} $\theta=30^o,~45^o$, and $60^o$. It can be seen that the line emission distribution is made of two branches. These two branches are contributed by the near and far sides of the shell like H II region. Because the density in the hot \textrm{stellar-wind} bubble is extremely low compared with the density of the photoionized H II region, the contribution from the bubble to the [Ne II] line luminosity is negligible. This causes \textrm{the empty area} between the two branches. In the p-v diagrams of the [Ne II] line, the peaks are all red-shifted and gradually move towards $0~km~s^{-1}$ with \textrm{increasing angles. The range} of the velocity distribution of the line \textrm{becomes narrower with} the increasing angle so that the emission peak becomes more concentrated.  
In model A, the density of the $Ne^+$ ions is the highest at the apex ($n\sim10000~cm^{-3}$ and $X(Ne^+)=0.84$) and decreases to lower values in the tail ($n\sim300~cm^{-3}$ and $X(Ne^+)=0.80$). The emission from the apex is also higher than from the rest part of the ionized region. As the inclination angle increases, the projection of the head region on the slit becomes narrower. Hence, the bright areas are more concentrated in the p-v diagram for the inclinations of $\theta=60^o$ than in the diagrams \textrm{for lower inclination angles}. In model B, \textrm{the large size} of the H II region leads to large projected images. The separations between two branches in p-v diagrams of model B are larger than those of model A. This is due to the larger cavity and opening angle of the H II region in model B.

\subsubsection{Profiles and diagrams of the $H_2~S(2)$ line}

\begin{figure}[!htp]
\centering
\includegraphics[scale=.33]{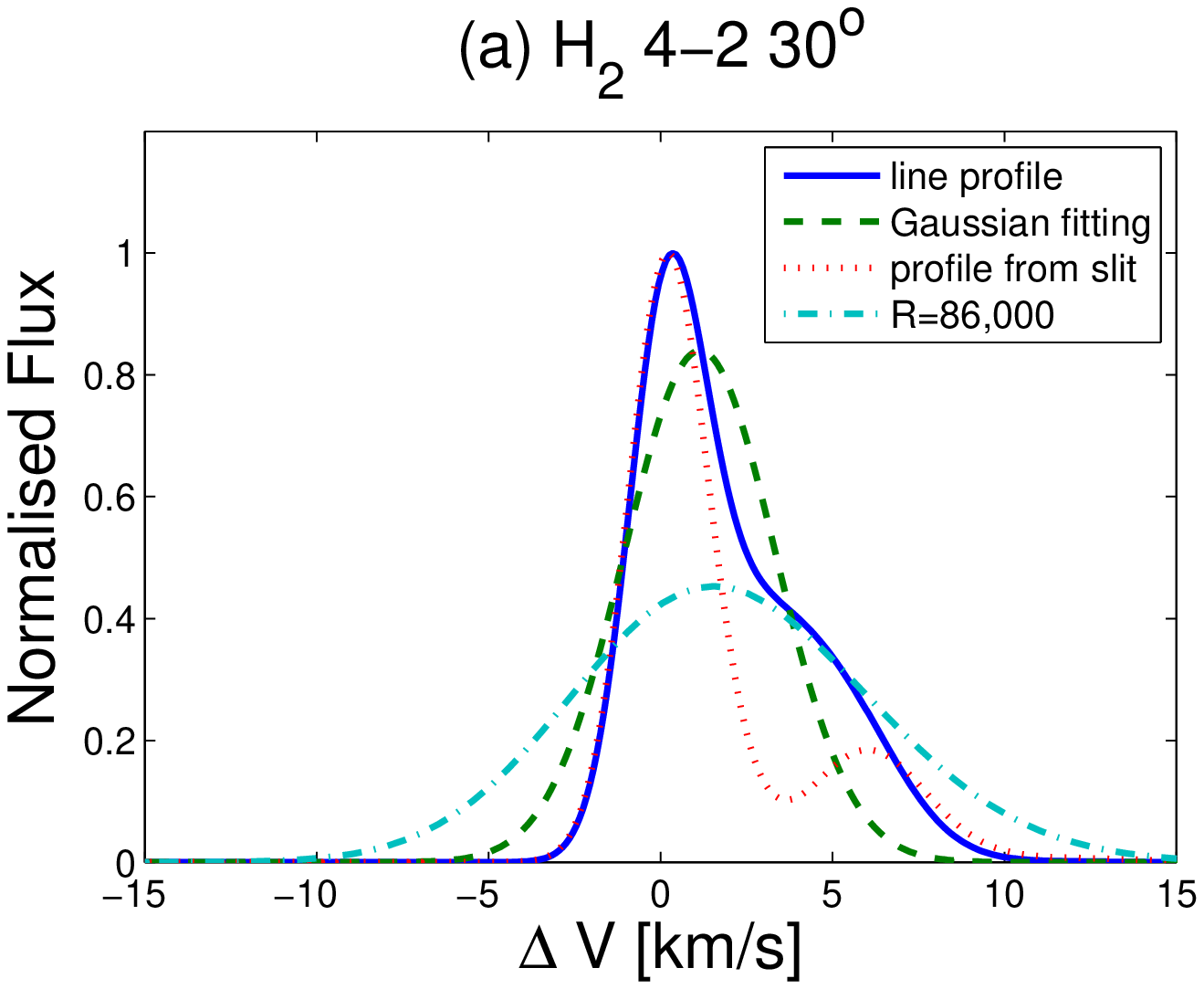}
\includegraphics[scale=.33]{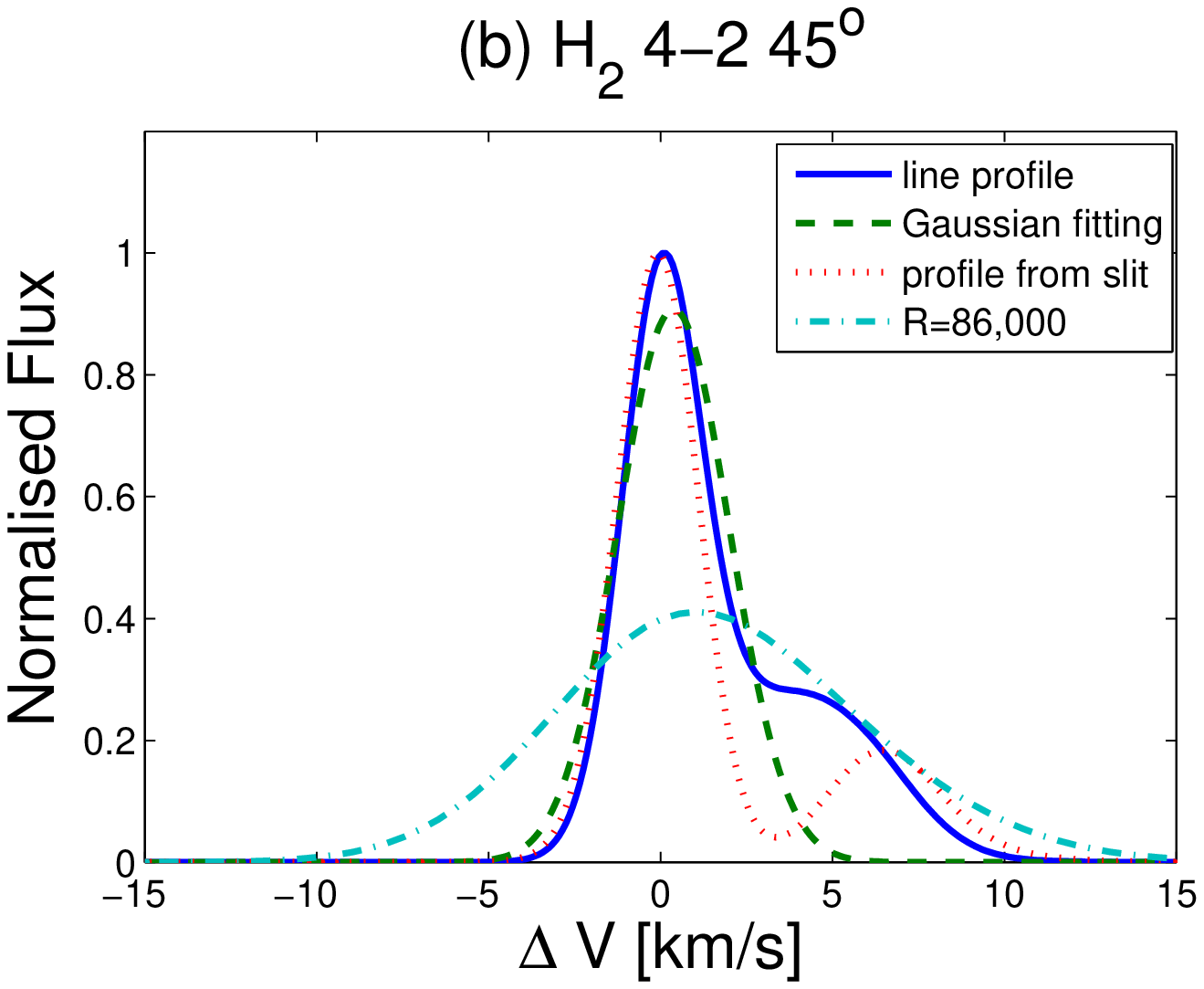}
\includegraphics[scale=.33]{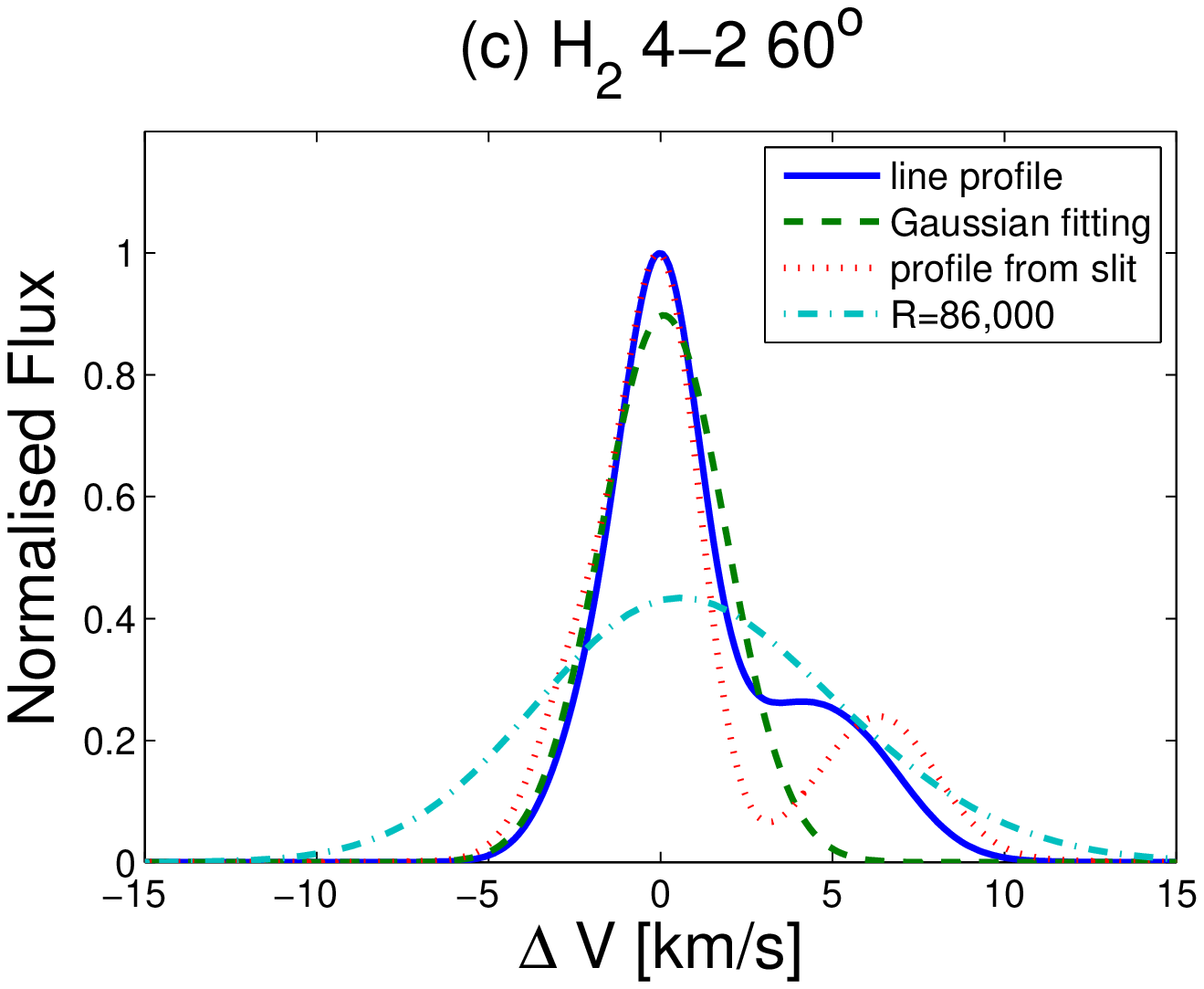}
\includegraphics[scale=.33]{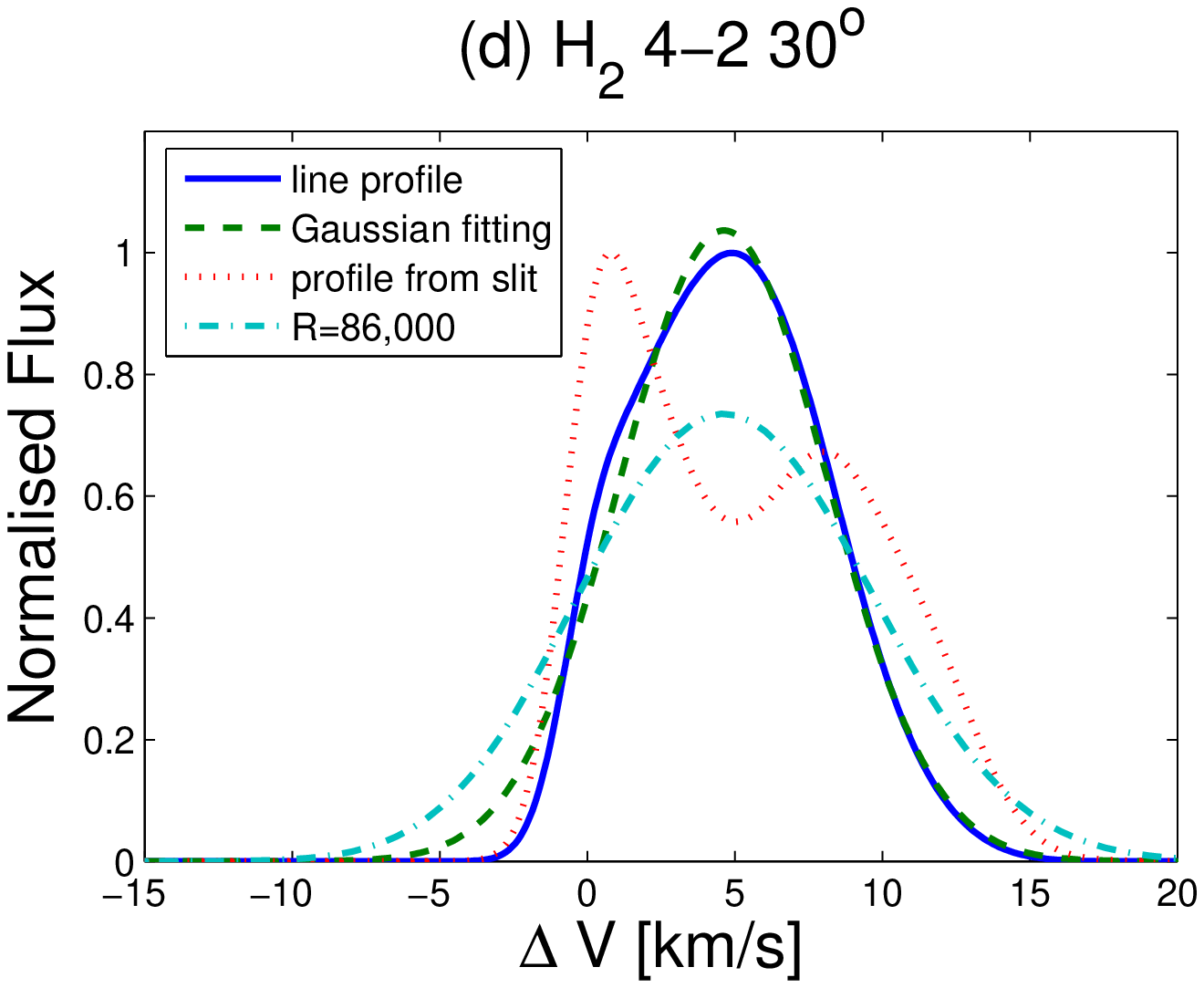}
\includegraphics[scale=.33]{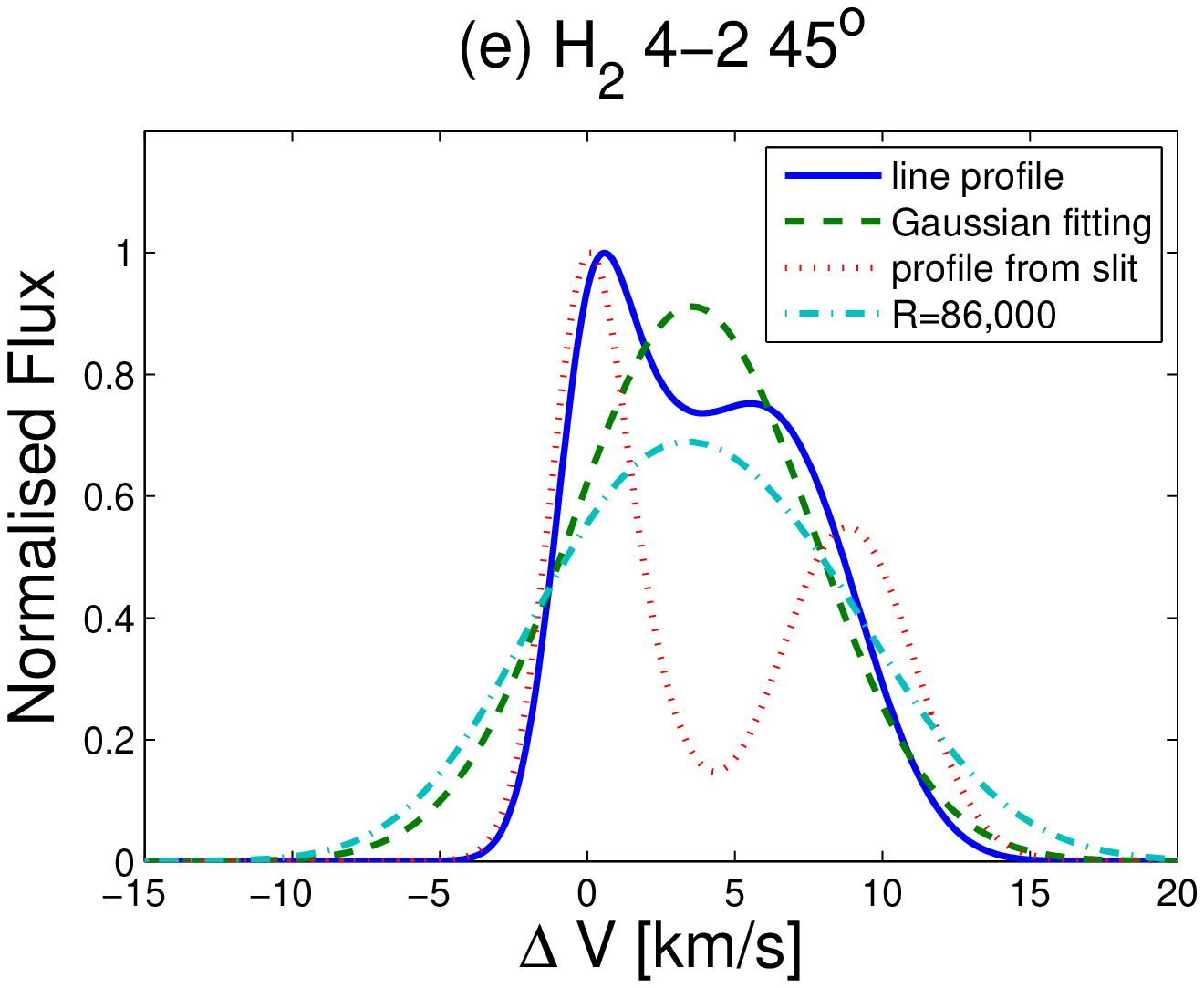}
\includegraphics[scale=.33]{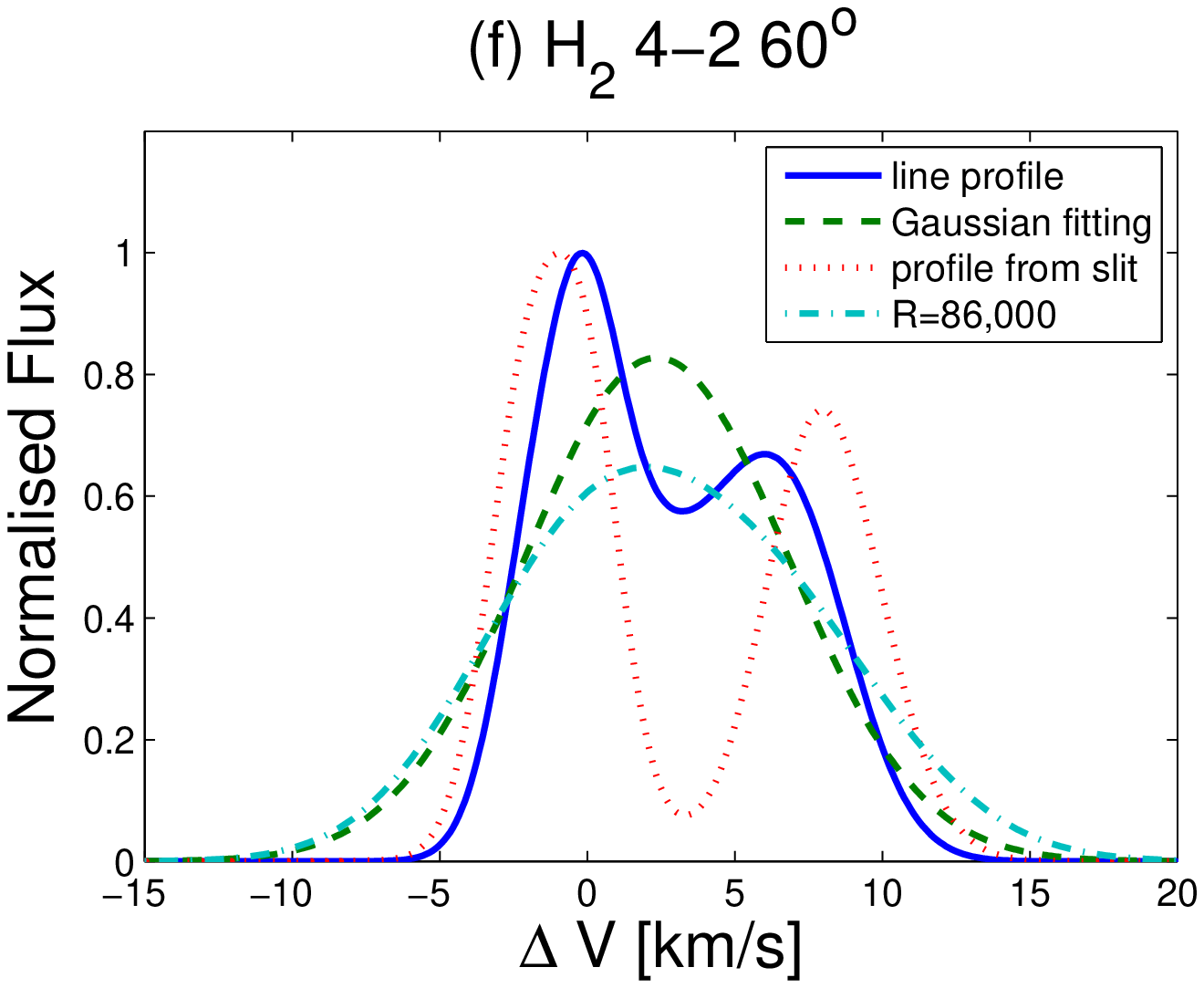}
\caption{Profiles of the $H_2~S(2)$ line (solid lines), the Gaussian fitting to the profiles (dashed lines) and the profiles at the resolution of $R=86,000$ (dot-dashed lines) at three inclination angles ($\theta=30^o$, $45^o$ and $60^o$). The $H_2~S(2)$ line profiles from the slit along the \textrm{symmetry axis} of the projected 2D image are also presented (dotted lines). The top panels are plotted for model A, and the bottom panels are shown for model B.}
\begin{flushleft}
\end{flushleft}
\label{fig_modabfitH}
\end{figure}

The profiles of the $H_2$ $S(2)$ line for three inclinations of $\theta=30^o,~45^o$ and $60^o$ are shown in Figure \ref{fig_modabfitH}. The properties of this line are shown in Table \ref{tab_modab}. It is different from the [Ne II] line in that the profiles of the $H_2$ line in the two \textrm{bow-shock} models are positive-skewed. The skewnesses of the $H_2$ $S(2)$ line profiles in model A are $0.72,~0.86,~0.75$ for $\theta=30^o,~45^o,~60^o$, respectively. And those in model B are $0.22,~0.29$, and $0.26$ for the three inclination angles. It is obvious that no general trend of skewness as in [Ne II] line case exists in $H_2~S(2)$ line. At the same time, profiles of the $H_2~S(2)$ line are generally far from a Gaussian distribution. \textrm{In particular}, the line profiles for the inclination of $45^o$ and $60^o$ in model B are double-peaked. Therefore, the skewness of the $H_2~S(2)$ line profile is not a good indicator of the inclination angle. The FWCVs of the $H_2$ $S(2)$ line profiles are $1.94,~1.58$, and $1.12~km~s^{-1}$ for the inclination angles $\theta=30^o,~45^o$, and $60^o$ in model A, respectively. Those in model B are $4.81,~3.93$, and $2.78~km~s^{-1}$ for the corresponding inclinations. And the centers of the Gaussian fitting for the three inclinations are $1.14,~0.38,~0.09~km~s^{-1}$ in model A and $4.63,~3.55,~2.36~km~s^{-1}$ in model B. All these values of the $H_2$ line are much less than the stellar velocities. However, a decreasing trend with \textrm{increasing inclination} is still observed.
The FWHMs of the fitting curves in \textrm{models} A and B are $5.16,~3.82,~4.25~km~s^{-1}$ and $8.27,~9.54,~10.45~km~s^{-1}$ for the inclinations of $30^o,~45^o$ and $60^o$, separately. The FWCVs and the centers of the fitting curves in the $H_2~S(2)$ line in model B are higher than those in model A, and the FWHMs in model B are also larger than in model A. This is because the proportion of the warm molecular gas compressed in the shell is higher in model B. It is worth noting that the FWHM of the $H_2$ $S(2)$ line does not decrease \textrm{with increasing angle}. This is \textrm{different from the} [Ne II] line case. The FWHM in model B even increases with the inclination angles. The line profiles of the $H_2$ line from the slit along the \textrm{symmetry axis} of the projected 2D image are shown in Figure \ref{fig_modabfitH} as well. We find that the double peak is more apparent in the profiles from the slit than in those from the whole neutral region. This is because the contribution from the gas above and below the plane of the slit is reduced. This part of gas has velocities distributed in between two \textrm{the} extreme velocities represented by the two peaks. The computed $H_2$ line profiles are convolved with the thermal and instrumental broadening kernel to predict the \textrm{line profiles of simulated observations}. The results show that the broadening \textrm{does not significantly degrade our ability to extract the necessary kinematical information, since the line center} can still be estimated reliably from observations.

\begin{figure}[!htp]
\centering
\includegraphics[scale=.33]{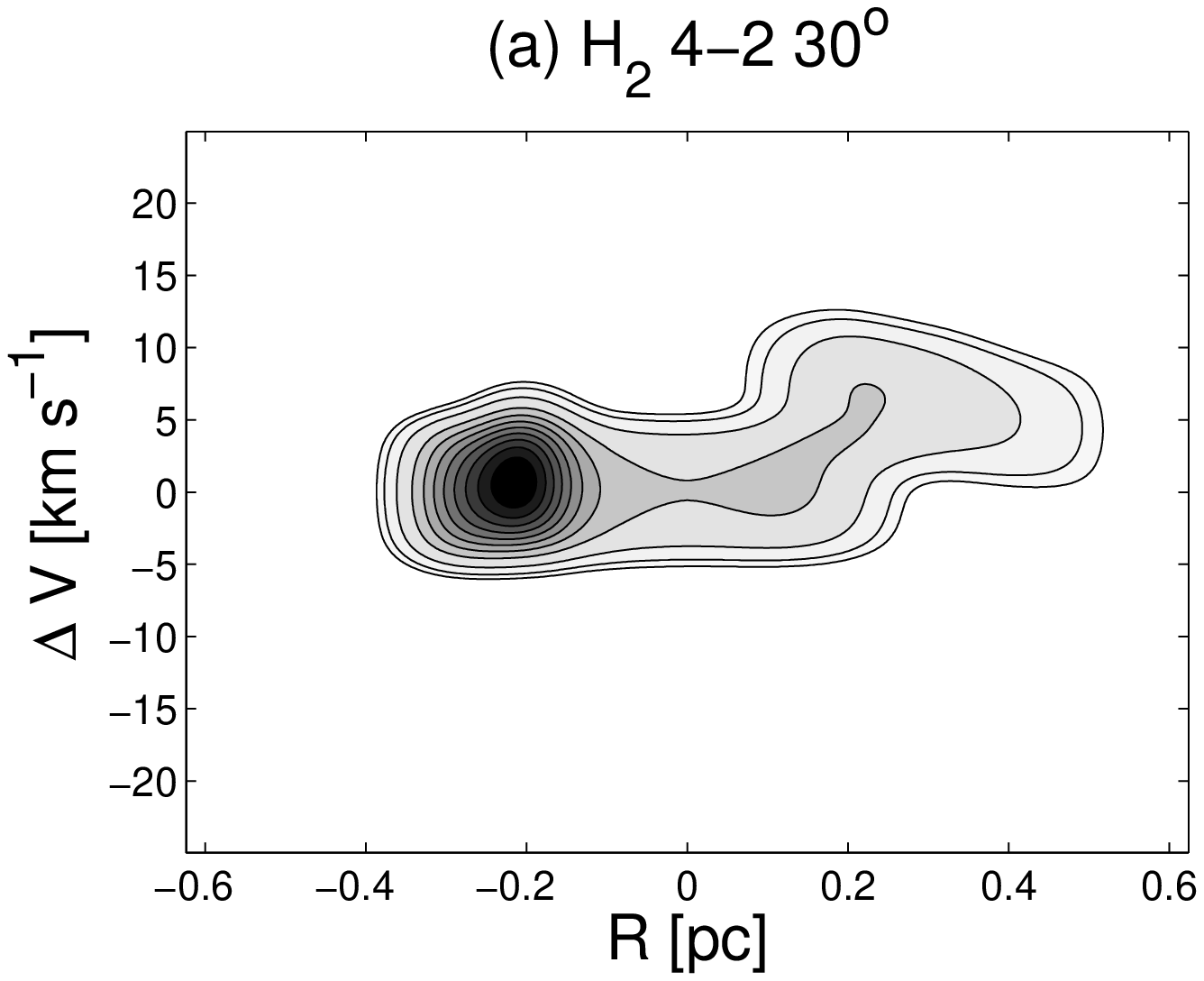}
\includegraphics[scale=.33]{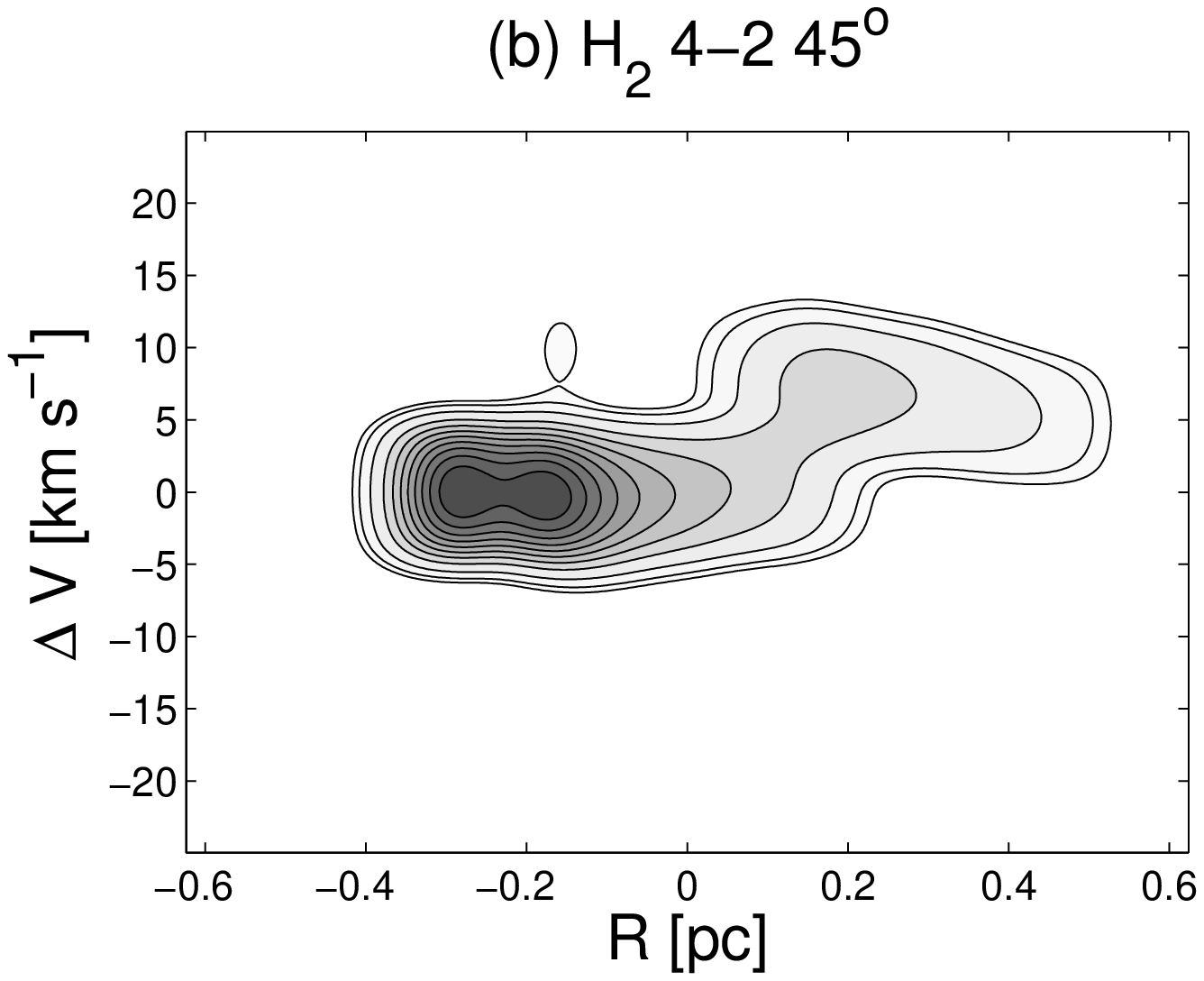}
\includegraphics[scale=.33]{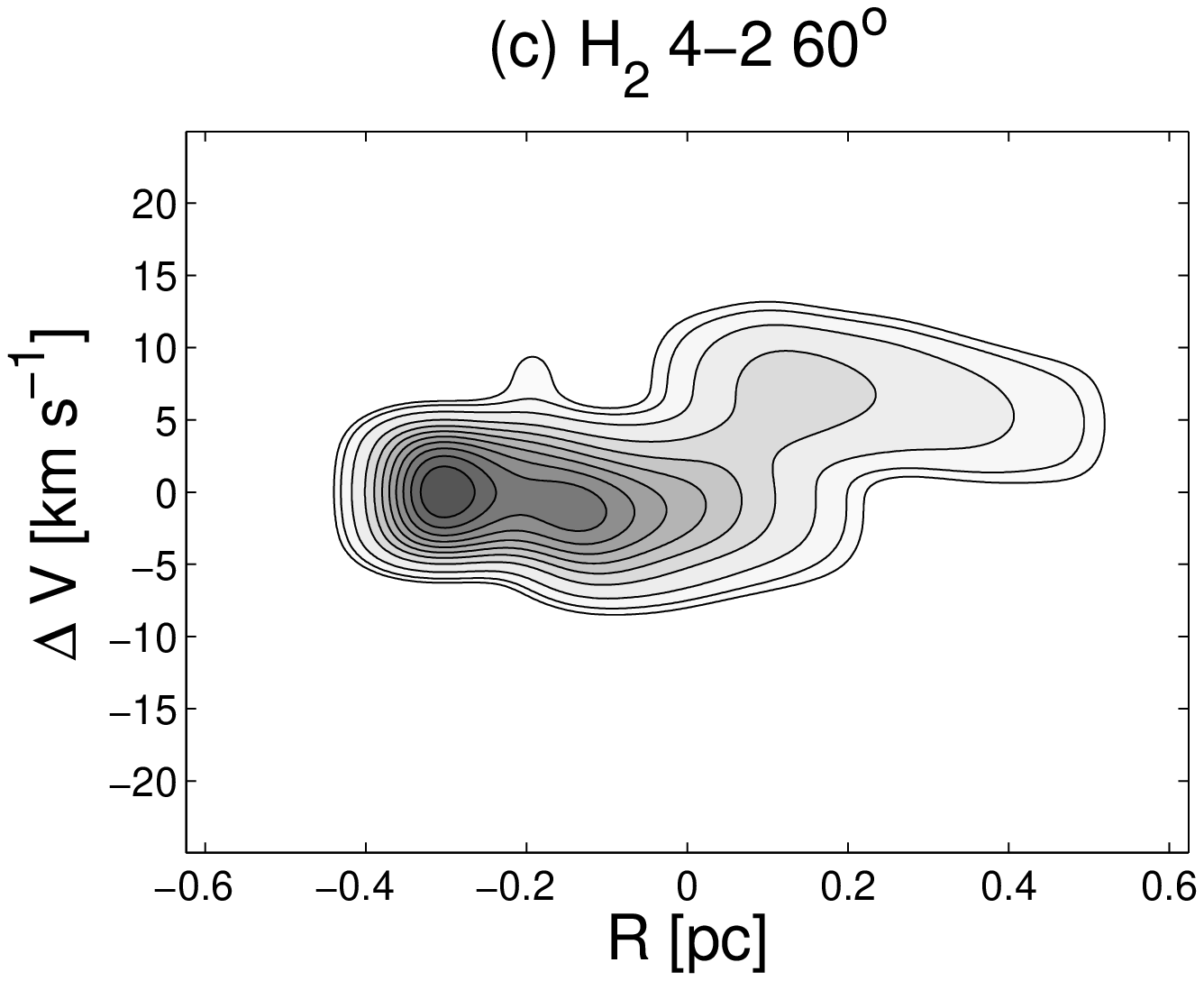}
\includegraphics[scale=.33]{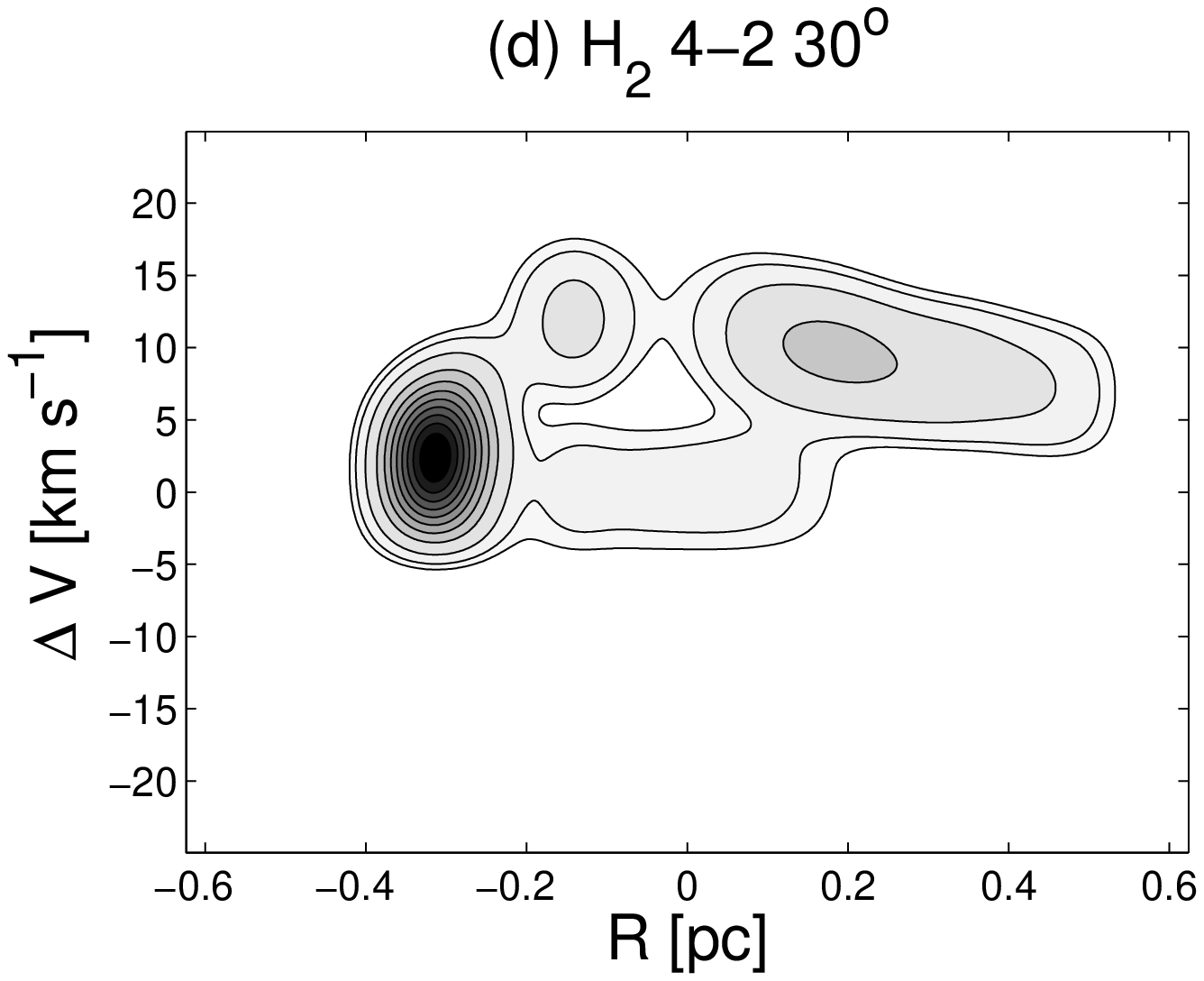}
\includegraphics[scale=.33]{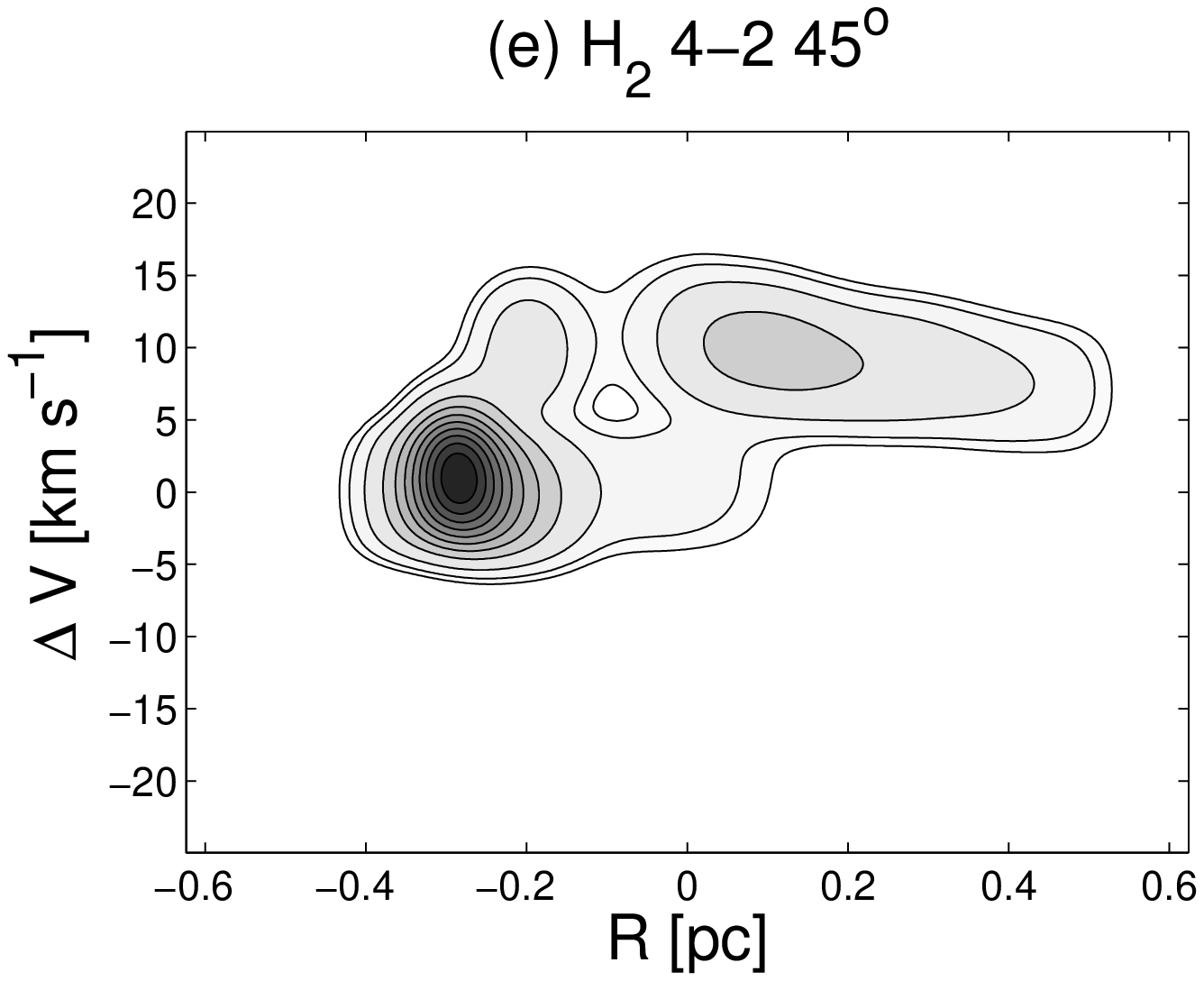}
\includegraphics[scale=.33]{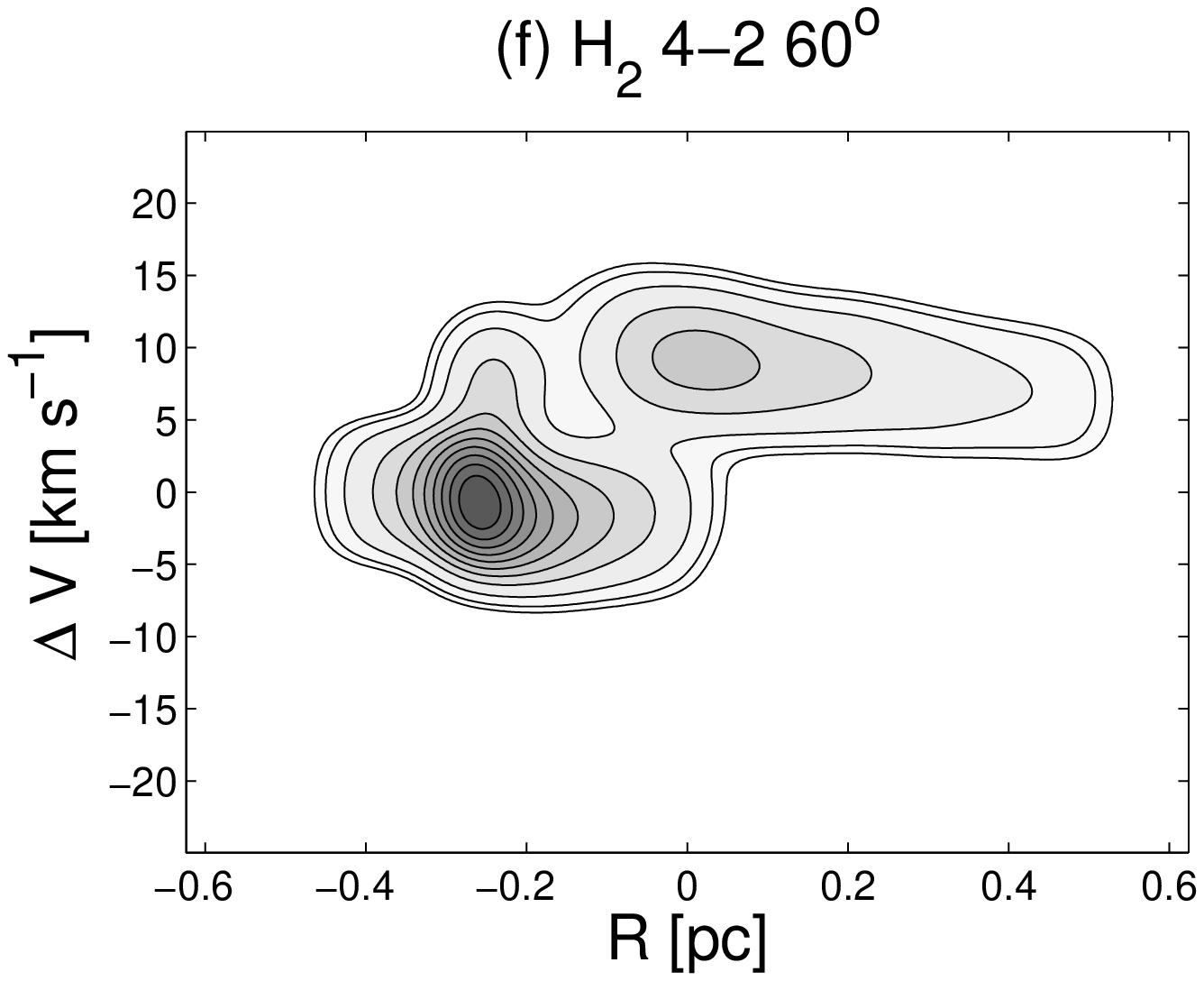}
\caption{The position-velocity diagrams of the $H_2$ $4-2$ rotational line from the slit along the \textrm{symmetry axis} of the projected 2D image for three inclination angles ($\theta=30^o$, $45^o$ and $60^o$). The top panels and the bottom panels are shown for model A and model B, respectively. The contour levels are at $3,~5,~10,~20,~30,~40,~50,~60,~70,~80,~90\%$ of the emission peaks in each panel.}
\begin{flushleft}
\end{flushleft}
\label{fig_modabpvHmol}
\end{figure}

In Figure \ref{fig_modabpvHmol}, the position-velocity diagrams of the $H_2$ $S(2)$ line from the slit along the \textrm{symmetry axis} of the projected 2D image in \textrm{models} A and B are presented for the inclinations of $\theta=30^o,~45^o,~\textrm{and}~60^o$. \textrm{As already demonstrated in previous figures, a large part} of $H_2$ line emitting warm molecular gas does not exist close to the apex of the shell \textrm{but exists in} the shell at the sides and the undisturbed region beyond the shock front. So the $H_2$ rotational line emission peaks are not obviously red-shifted. These diagrams also show the difference in the projected velocities of the warm molecular gas between the two sides of the cometary H II region. The projected velocity of the molecular gas in the near side of the shell and the region outside of the shell is close to zero, and that from the gas \textrm{at the farther side} of the shell is higher than $5~km~s^{-1}$. In addition, the molecular gas in the apex of the dense shell in model B has a high velocity $\Delta v\geq10~km~s^{-1}$, \textrm{similar to} the stellar velocity. But in model A, the apex of the dense shell is almost fully dissociated. There is \textrm{no such high velocity} component shown in the p-v diagrams for model A. 

\begin{table}[!htp]\footnotesize
\centering
\begin{tabular}{c|ccccc}
\hline
Model A Bow shock $21.9~M_\odot$  \\
\hline
Line & Inclination & FWCV & Skewness & Center of line & FWHM \\
 & & ($km~s^{-1}$) & & ($km~s^{-1}$) & ($km~s^{-1}$) \\
\hline
[Ne II] $12.81 \mu m$ & $30^o$ & 0.03 & -0.59 & 1.87 & 20.69 \\
 & $45^o$ & 0.02 & -0.50 & 1.35 & 18.34 \\
 & $60^o$ & 0.02 & -0.33 & 0.73 & 15.58 \\
\hline
$H_2~4-2$ & $30^o$ & 1.94 & 0.72 & 1.14 & 5.16 \\
 & $45^o$ & 1.58 & 0.86 & 0.38 & 3.82 \\
 & $60^o$ & 1.12 & 0.75 & 0.09 & 4.25 \\
\hline
Model B Bow shock $40.9~M_\odot$  \\
\hline
Line & Inclination & FWCV & Skewness & Center of line & FWHM \\
 & & ($km~s^{-1}$) & & ($km~s^{-1}$) & ($km~s^{-1}$) \\
\hline
[Ne II] $12.81 \mu m$ & $30^o$ & 0.91 & -0.61 & 2.51 & 18.33 \\
 & $45^o$ & 0.74 & -0.51 & 1.96 & 16.66 \\
 & $60^o$ & 0.52 & -0.34 & 1.25 & 14.82 \\
\hline
$H_2~4-2$ & $30^o$ & 4.81 & 0.22 & 4.63 & 8.27 \\
 & $45^o$ & 3.93 & 0.29 & 3.55 & 9.54 \\
 & $60^o$ & 2.78 & 0.26 & 2.36 & 10.45 \\
\hline
\end{tabular}
\caption{Flux weighted central velocities (FWCV), skewness, and the center and FWHMs of Gaussian fitting of the lines for different angles in \textrm{models} A and B\label{tab_modab}}
\end{table}

\subsection{Champagne-Flow Models}

In \textrm{models} C and D, the evolutions of a champagne flow including a stellar wind are simulated. The parameters in model C are the same as those in the \textrm{champagne-flow} model mentioned in \ref{sect:front}. A $40.9~M_\odot$ massive star is assumed in model D, and other parameters in model D are the same as in model C. The simulations of \textrm{models} C and D are also stopped at $100,000~yr$. At that time, the champagne flows have cleared the low-density material from the grid.

\begin{figure}[!htp]
\centering
\includegraphics[scale=.5]{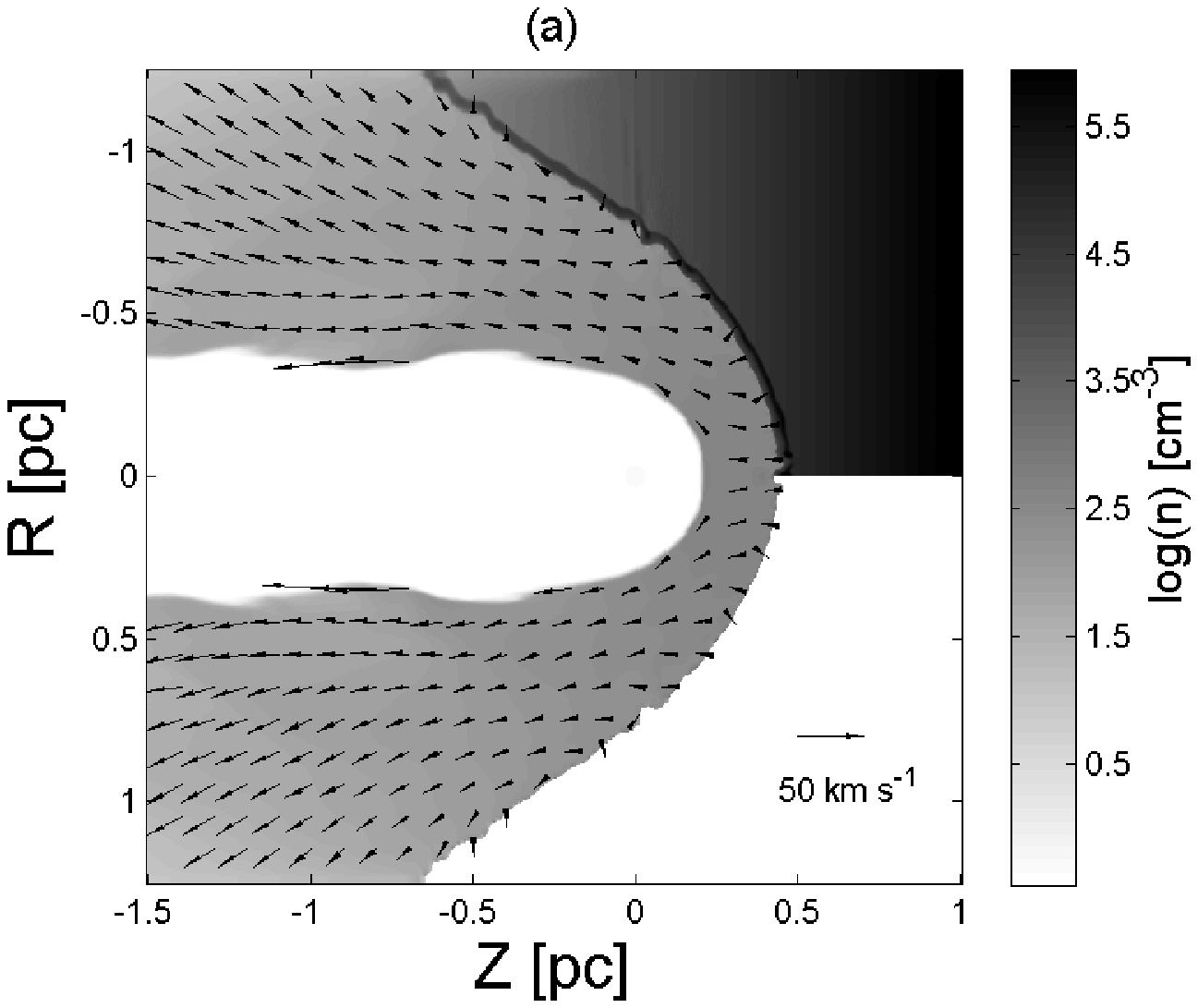}
\includegraphics[scale=.5]{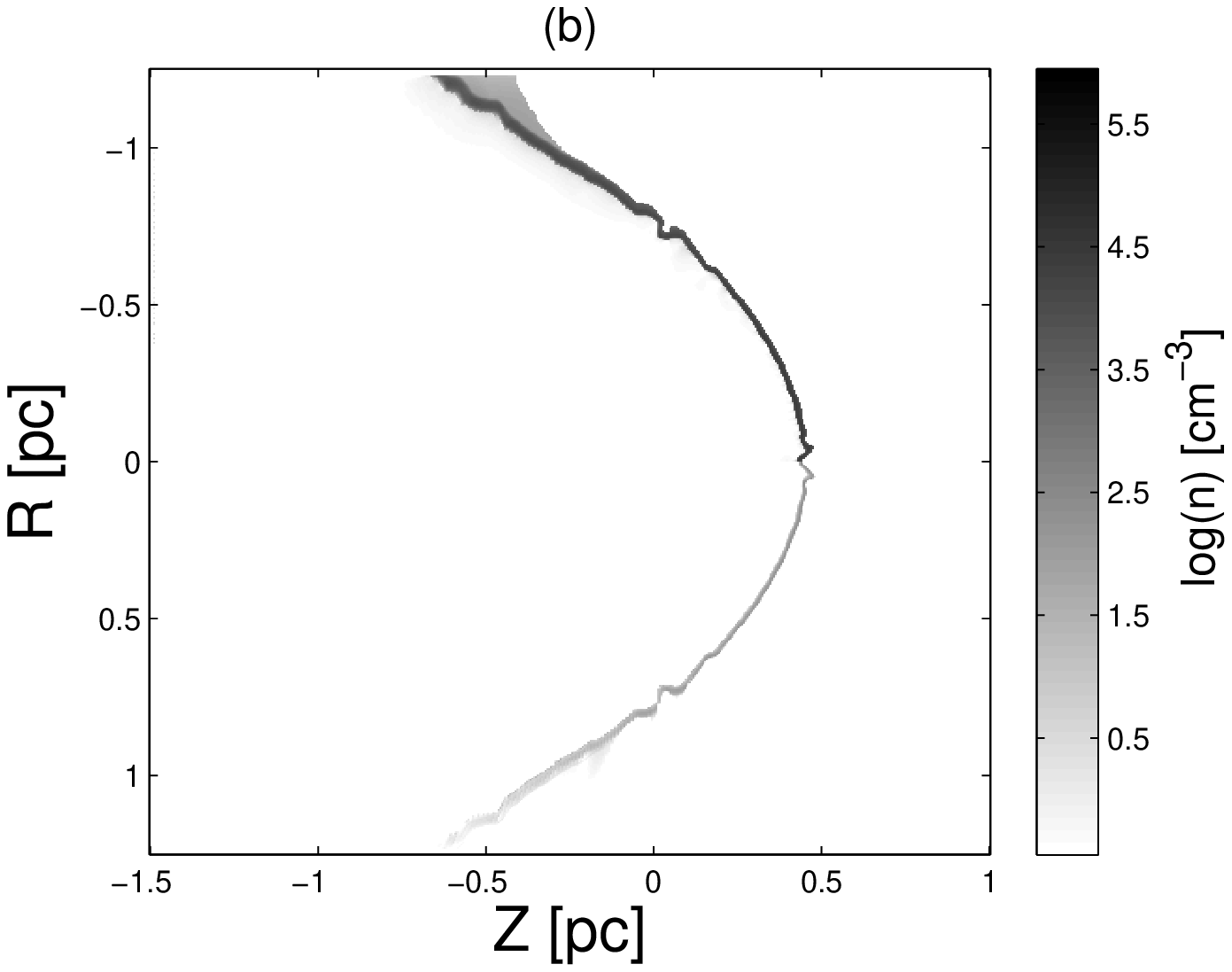}
\includegraphics[scale=.5]{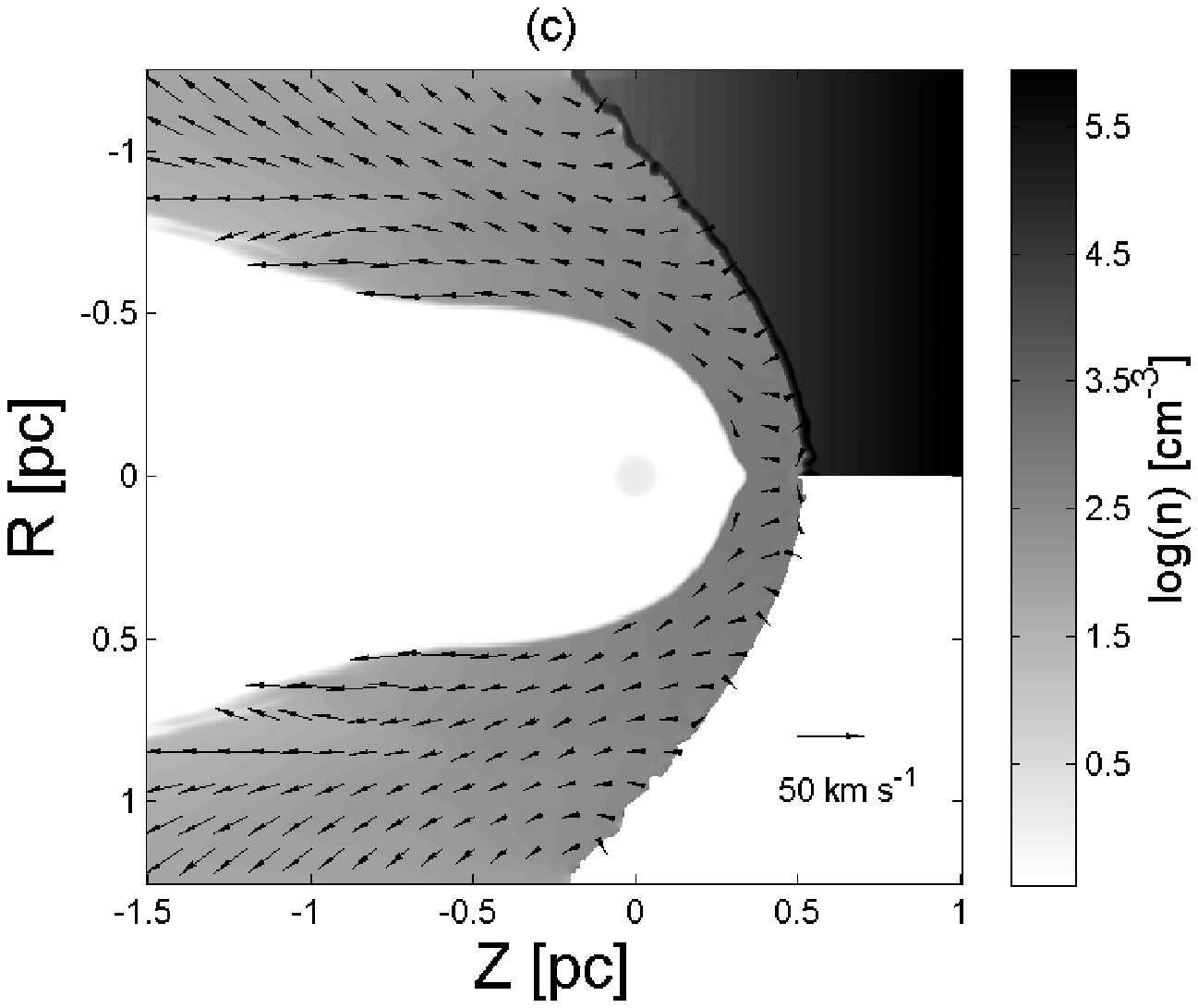}
\includegraphics[scale=.5]{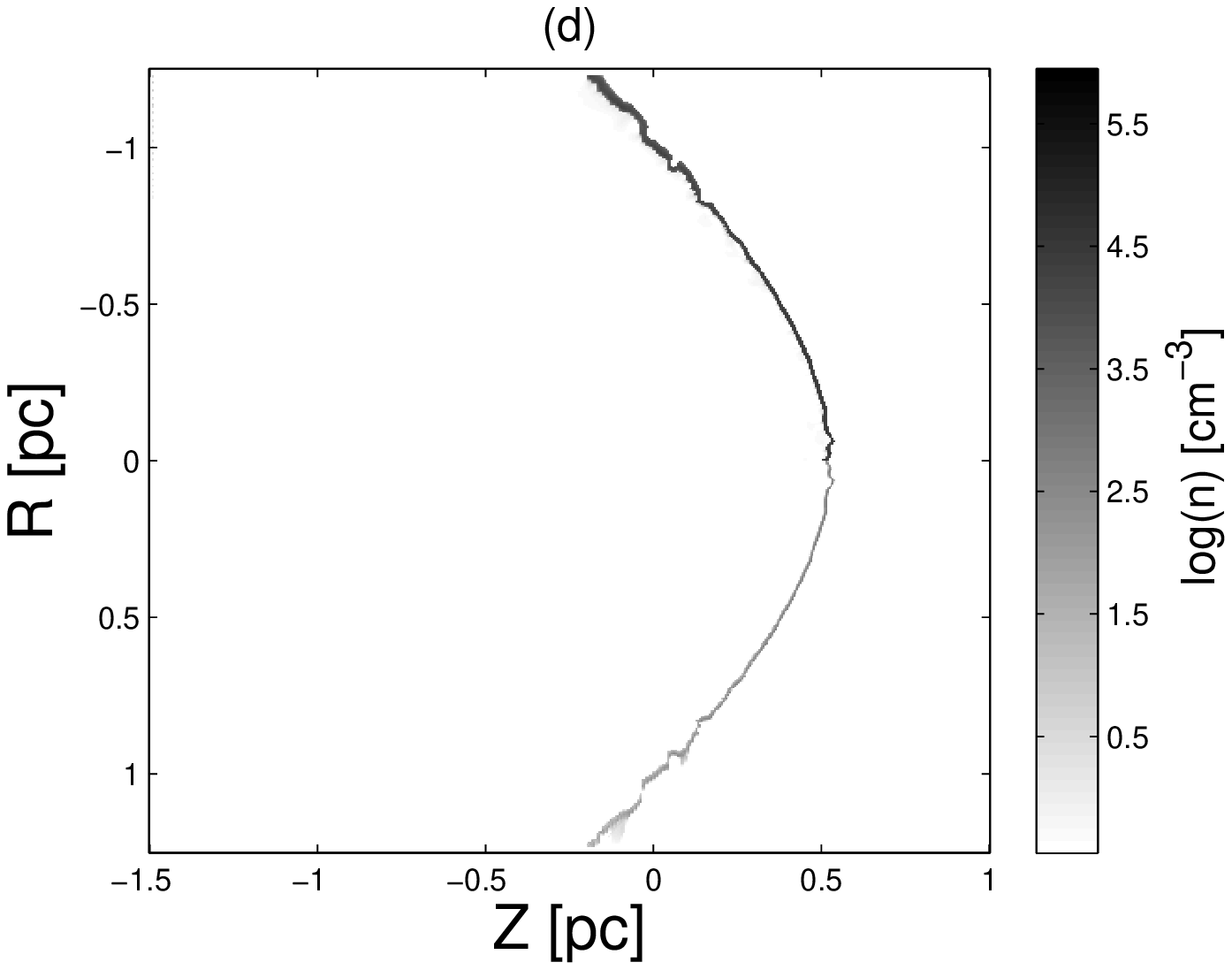}
\caption{The density distributions of all materials, ionized hydrogen, hydrogen atoms and hydrogen molecules at the rotational level of $J=4$ in \textrm{models} C and D at the age of $100,000~yr$. The top panels and the bottom panels are presented for \textrm{models} C and D, respectively. The left panels show the densities of all materials (top half) and the $H^+$ ions (bottom half). The right panels show those of $H$ atoms (top half) and $H_2$ molecules (bottom half). The arrows in the left panels represent velocities of $v\geq1~km~s^{-1}$, but the velocity field in \textrm{stellar-wind} bubbles are not shown.}
\begin{flushleft}
\end{flushleft}
\label{fig_modcd}
\end{figure}

The density distributions in \textrm{models} C and D are presented in Figure \ref{fig_modcd}. The size of the H II region in \textrm{models} C and D is much larger than that in \textrm{models} A and B. Due to the high density \textrm{in the molecular} cloud and the large distance from the shell to the star, the ionization front and the dissociation fronts of $H_2$ and $CO$ are close \textrm{to each other} and all fall into the shell at a certain age. 

\subsubsection{Profiles and diagrams of the [Ne II] $12.81$ line}

\begin{figure}[!htp]
\centering
\includegraphics[scale=.33]{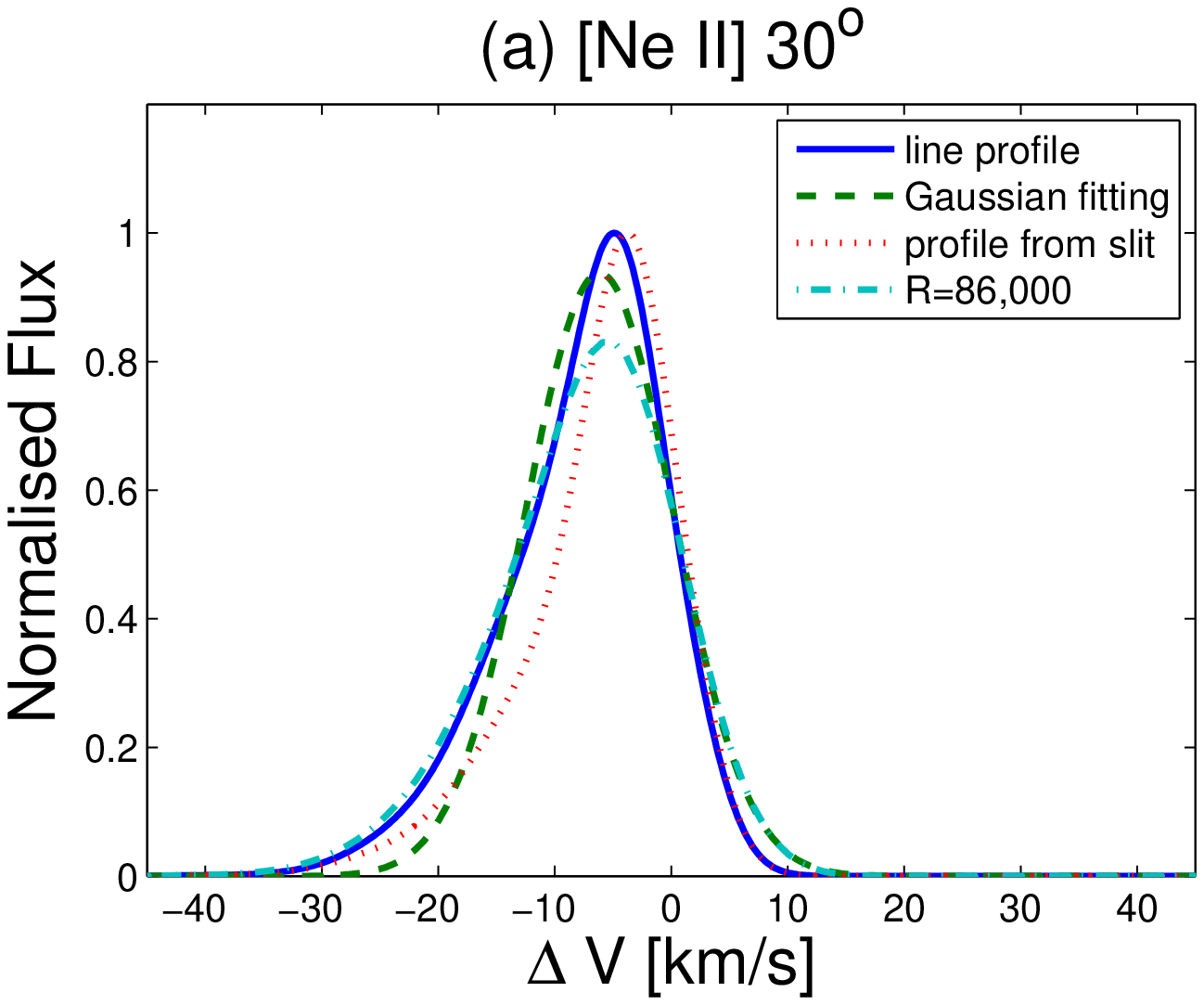}
\includegraphics[scale=.33]{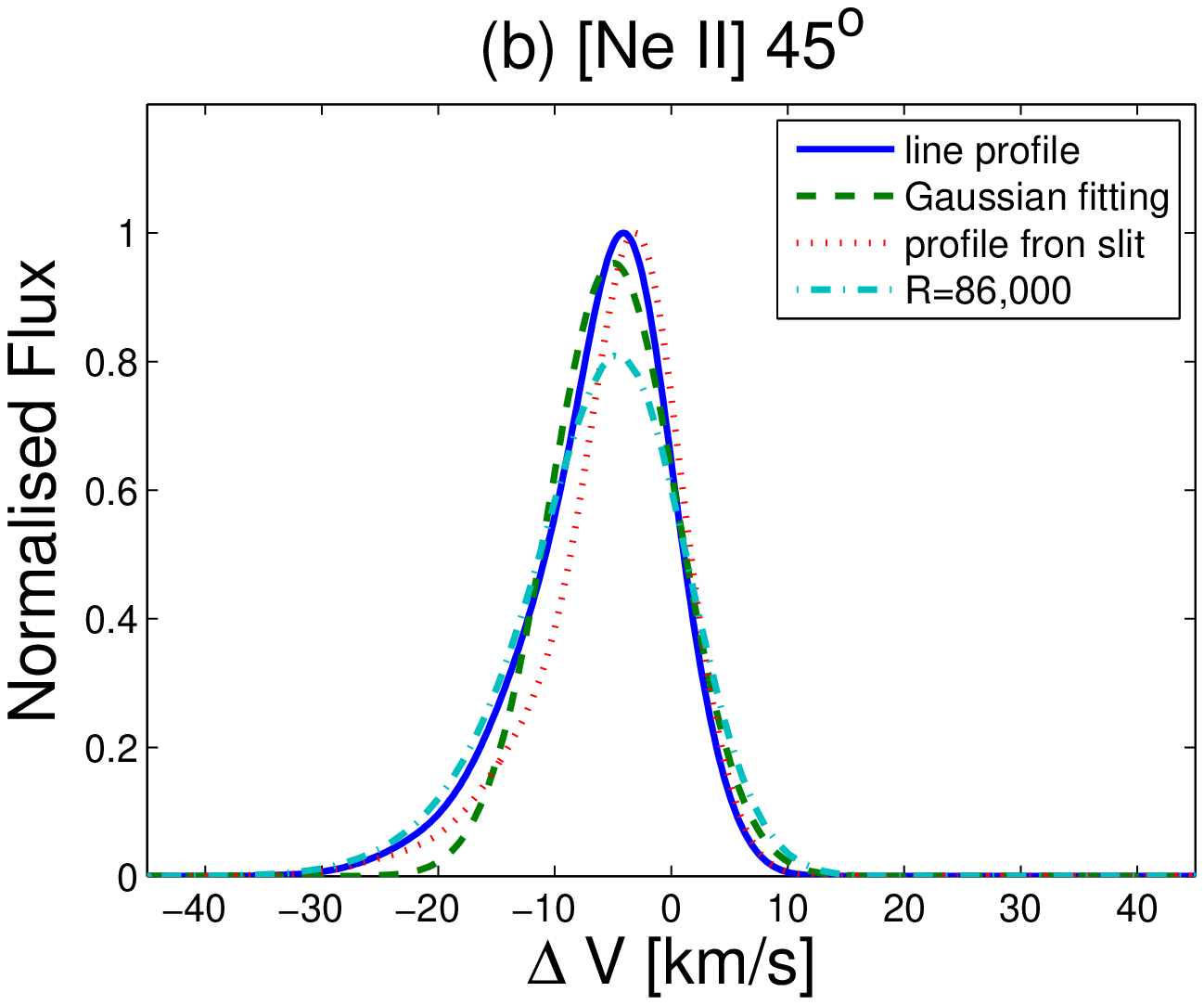}
\includegraphics[scale=.33]{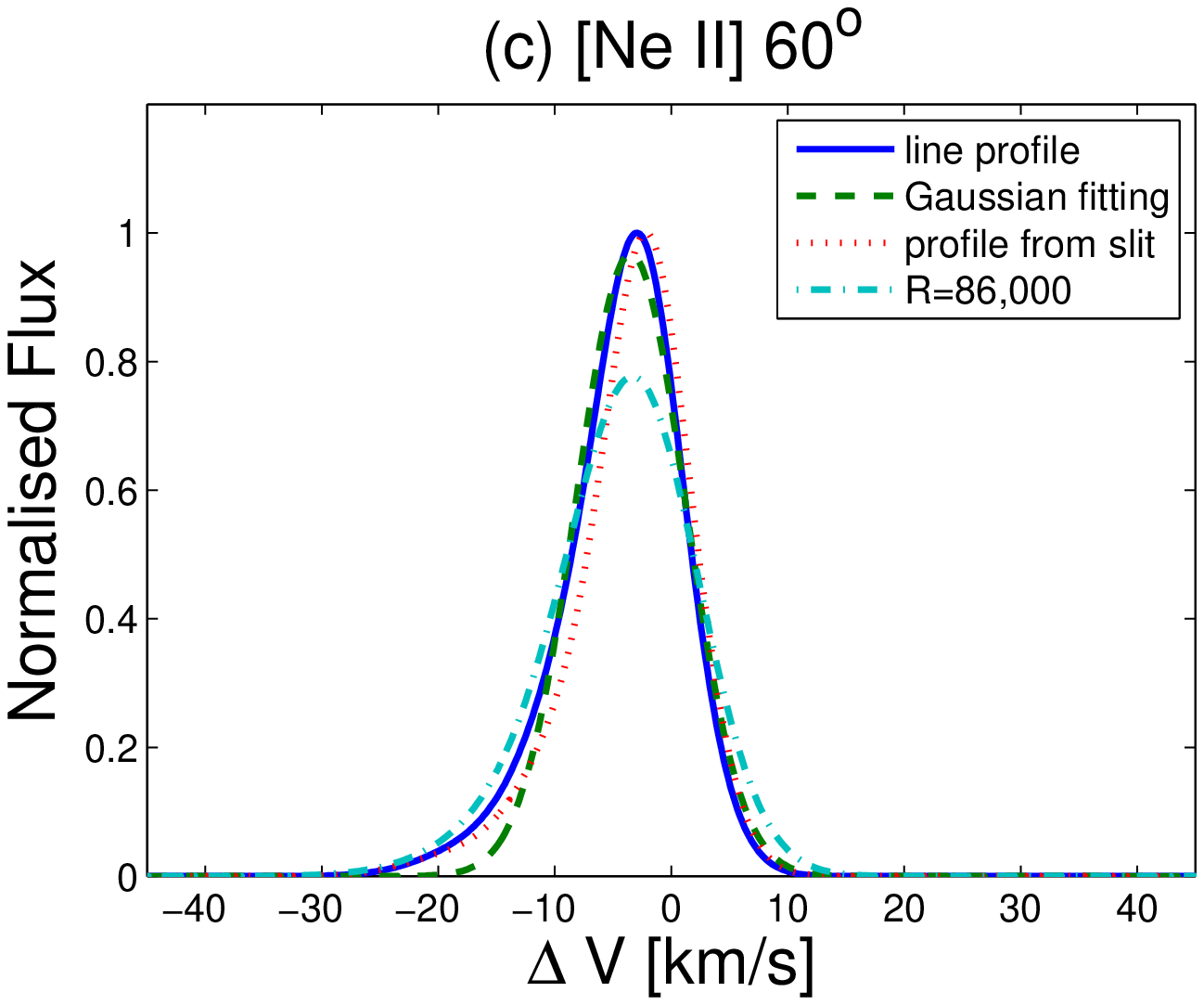}
\includegraphics[scale=.33]{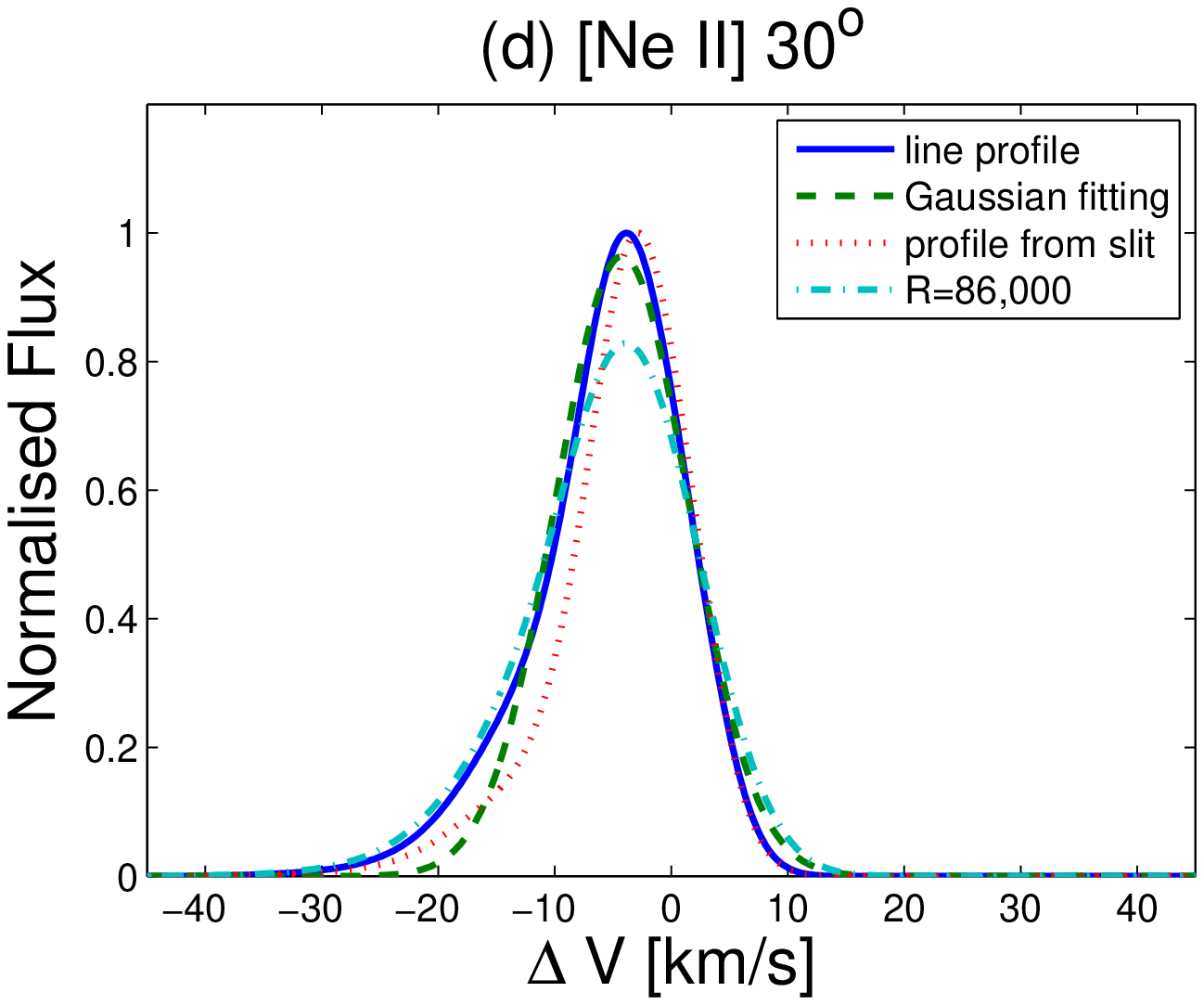}
\includegraphics[scale=.33]{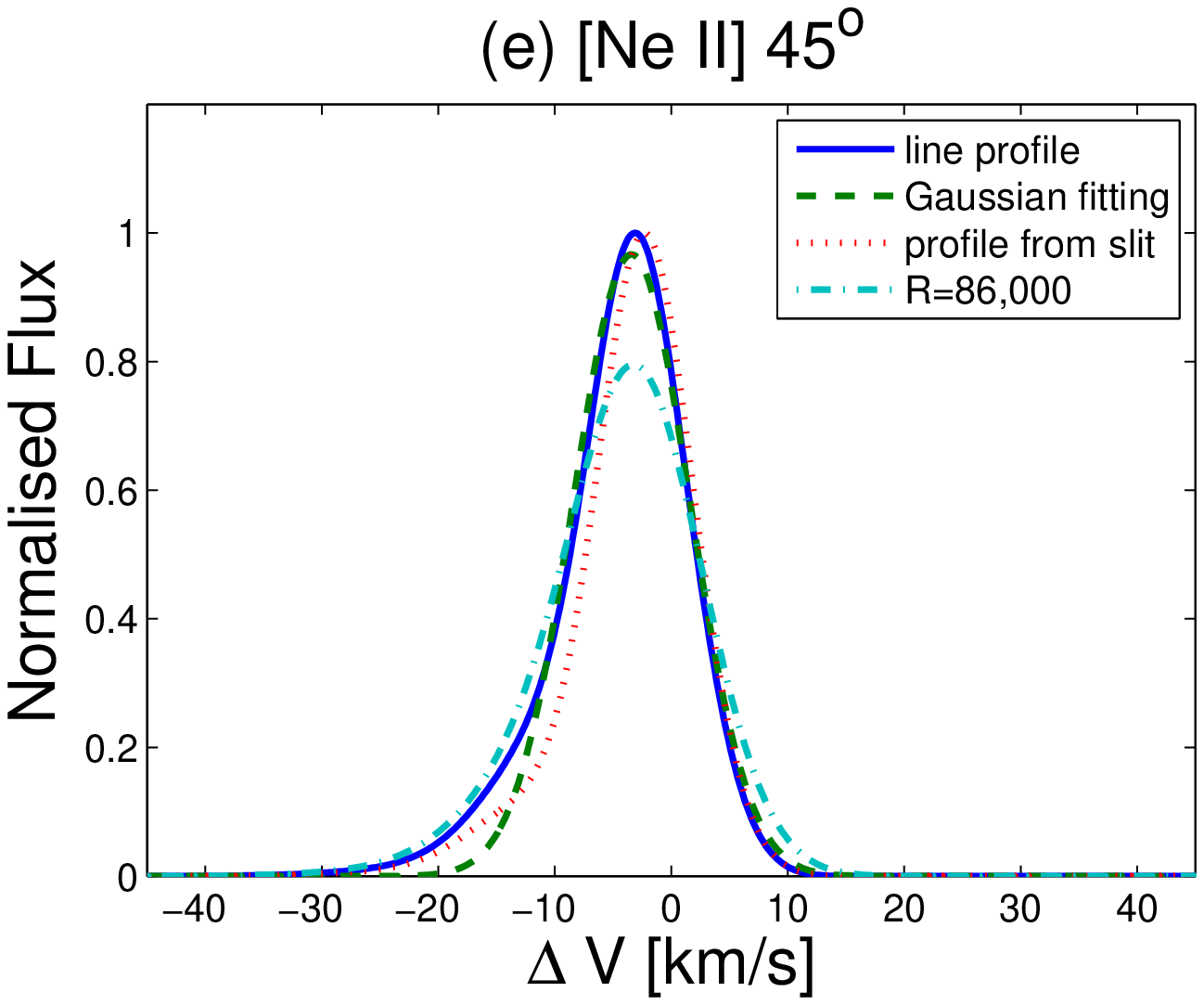}
\includegraphics[scale=.33]{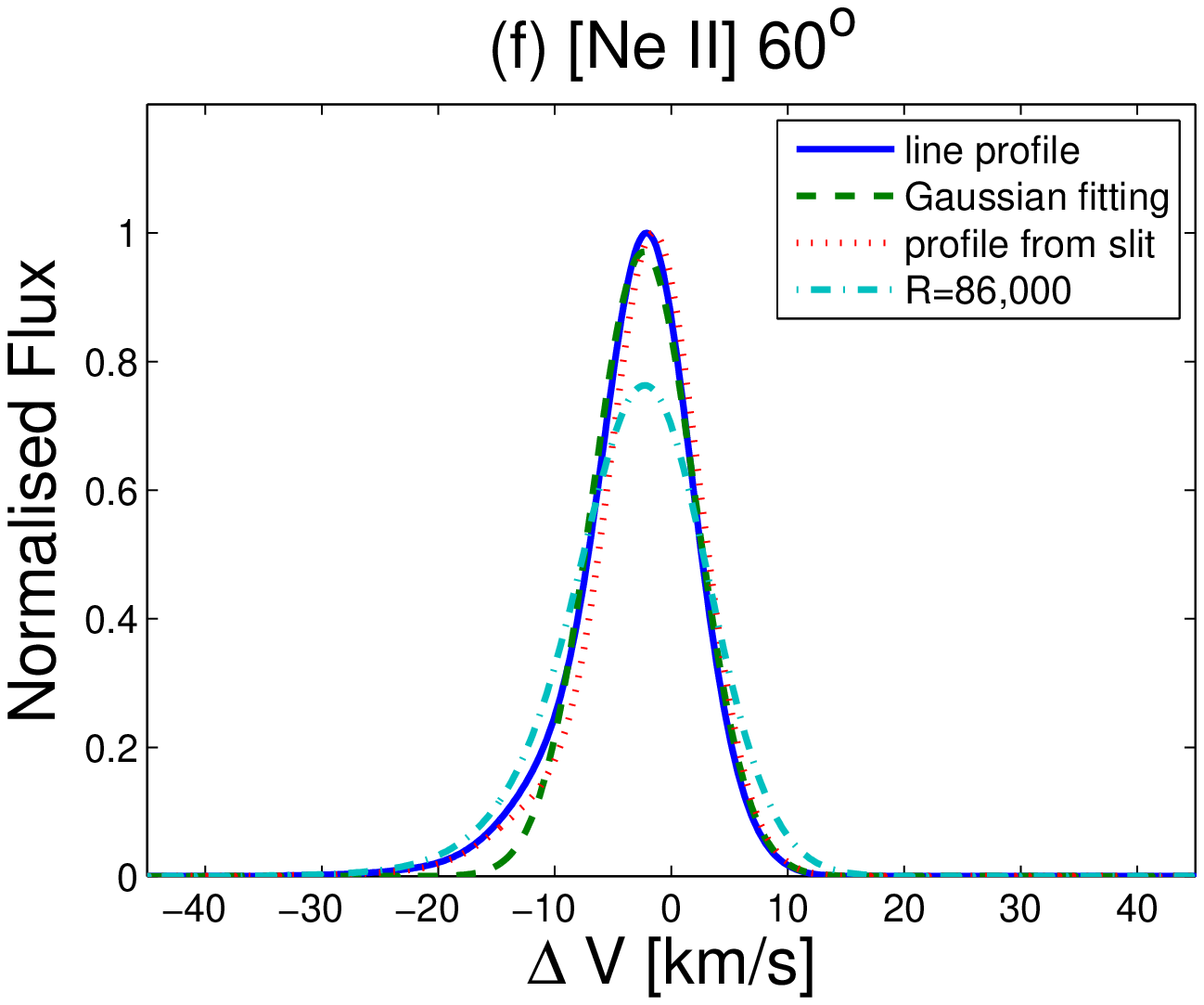}
\caption{Profiles of the [Ne II] line (solid lines), the Gaussian fitting to the profiles (dashed lines) and the profiles at the resolution of $R=86,000$ (dot-dashed lines) at three inclination angles ($\theta=30^o$, $45^o$ and $60^o$). The [Ne II] line profiles from the slit along the \textrm{symmetry axis} of the projected 2D image are also plotted (dotted lines). The top panels are plotted for model C, and the bottom panels are for model D.}
\begin{flushleft}
\end{flushleft}
\label{fig_modcdfit}
\end{figure}

A part of the ionized region in model C and model D is outside of the grid and not shown in Figure \ref{fig_modcd}. However, since more than 80 percent of the [Ne II] line emission in model C is emitted by the region with $z\geq-0.5~pc$, the line emission from the region outside of the grid in \textrm{models} C and D should be small. The profiles of the [Ne II] line and the Gaussian fitting to the profiles for three inclinations in \textrm{models} C and D are presented in Figure \ref{fig_modcdfit}. The properties of the [Ne II] line profiles are listed in Table \ref{tab_modcd}. In model C, the FWCVs of the [Ne II] line profiles are $-7.54,~-6.16,~\textrm{and}~-4.36~km~s^{-1}$ for three inclinations of $30^o,~45^o$, and $60^o$ respectively, and the centers of the Gaussians are $-6.16,~-4.99,~\textrm{and}~-3.49~km~s^{-1}$. In model D, the FWCVs of the [Ne II] line profiles are $-5.42,~-4.42,~\textrm{and}~-3.13~km~s^{-1}$ for the three inclinations, and the centers of the fitting curves are $-4.23,~-3.41,~\textrm{and}~-2.36~km~s^{-1}$. These values are all obviously blue-shifted. The FWHMs of the Gaussian fitting are $14.91,~12.91,~\textrm{and}~11.00~km~s^{-1}$ in model C and $13.43,~11.75,~10.40~km~s^{-1}$ in model D. These FWHMs are apparently lower than those in the \textrm{bow-shock} models. The velocity of the ionized gas in both the \textrm{bow-shock} models and the \textrm{champagne-flow} models approaches $v_z\sim-20~km~s^{-1}$. In the head of the photoionized region, the axial velocities of the ionized gas in the \textrm{bow-shock} models can reach the stellar velocity. But, it is at most $1~km~s^{-1}$ in the \textrm{champagne-flow} models. As the result, the FWHMs of the [Ne II] line in the \textrm{bow-shock} models are larger than those in the \textrm{champagne-flow} models. The [Ne II] line profiles in \textrm{models} C and D are also \textrm{negative skewed like those} in \textrm{models} A and B. The skewnesses in \textrm{models} C and D are $-0.68,~-0.70,~-0.67$ and $-0.76,~-0.81,~-0.79$ for the inclinations of $\theta=30^o,~45^o,~60^o$, respectively. The skewnesses for the inclinations of $\theta=70^o,~80^o$ are also calculated but not listed in Table \ref{tab_modcd}. These skewnesses are $-0.58,~-0.35$ in model C and $-0.57,~-0.34$ in model D for $\theta=70^o,~80^o$, respectively. The trend that the skewness of the [Ne II] line profile increases from more negative values toward zero also exists in the \textrm{champagne-flow} models, but it is not obvious when the inclination angle is smaller than $45^o$. This may be due to large opening angles of the cometary H II region in the \textrm{champagne-flow} models. 
The [Ne II] line profiles from the slit along the \textrm{symmetry axis} of the projected 2D image in \textrm{models} C and D are also shown in Figure \ref{fig_modcdfit}. The profiles are more red-shifted than those from the whole region as in \textrm{models} A and B. \textrm{However, In the \textrm{champagne-flow} models, the profiles from the slit are more similar to those of the entire region}.

\begin{figure}[!htp]
\centering
\includegraphics[scale=.33]{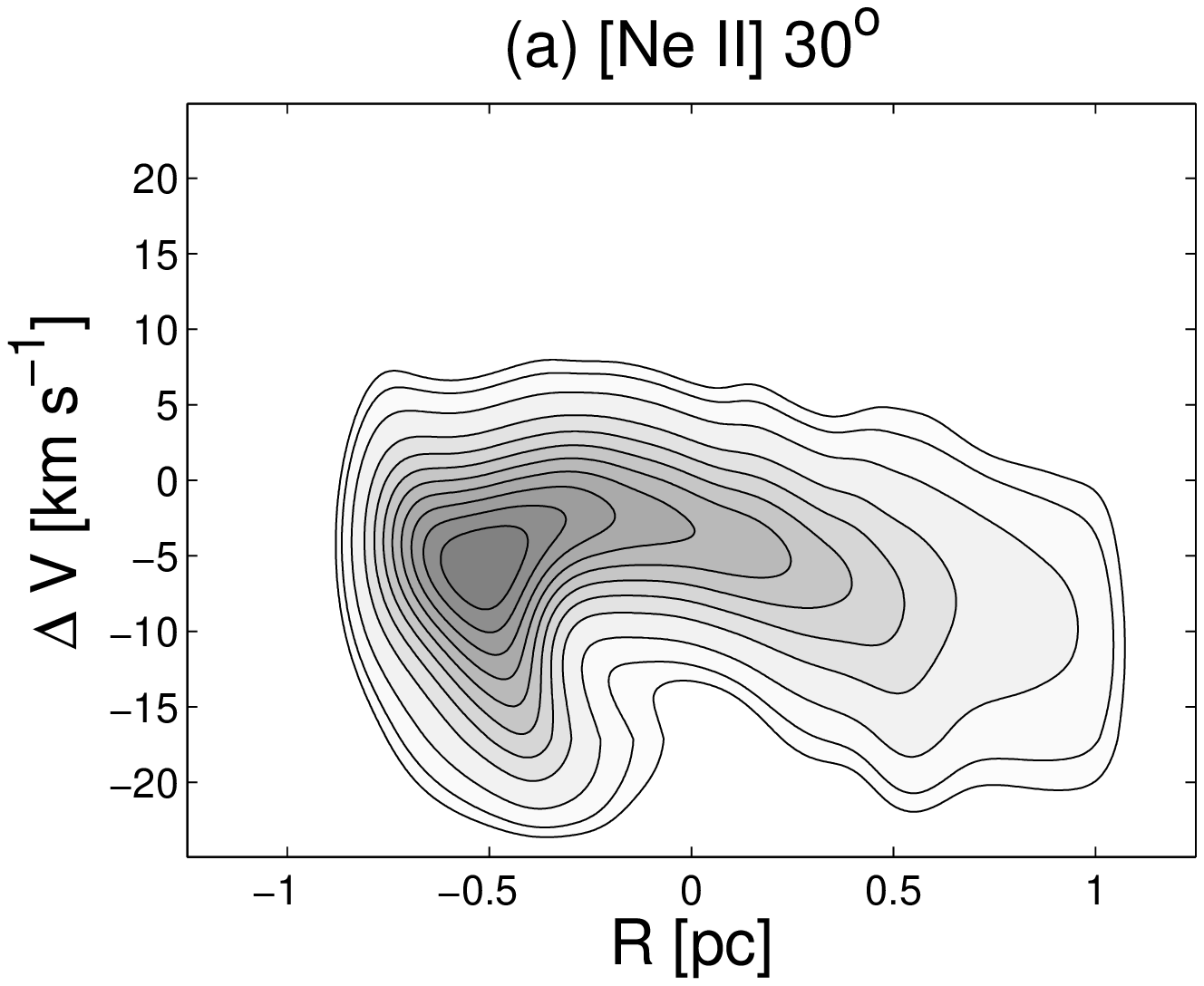}
\includegraphics[scale=.33]{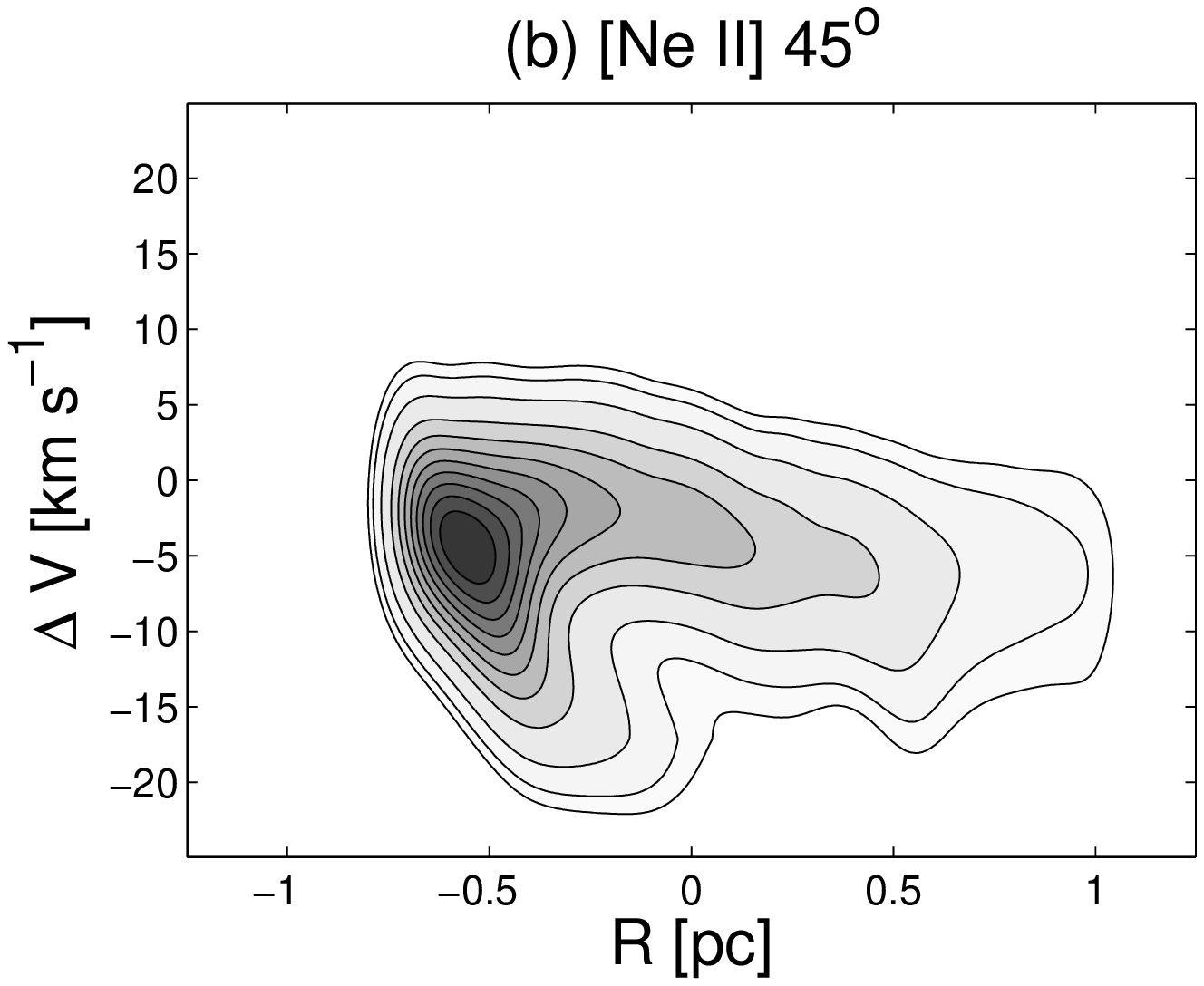}
\includegraphics[scale=.33]{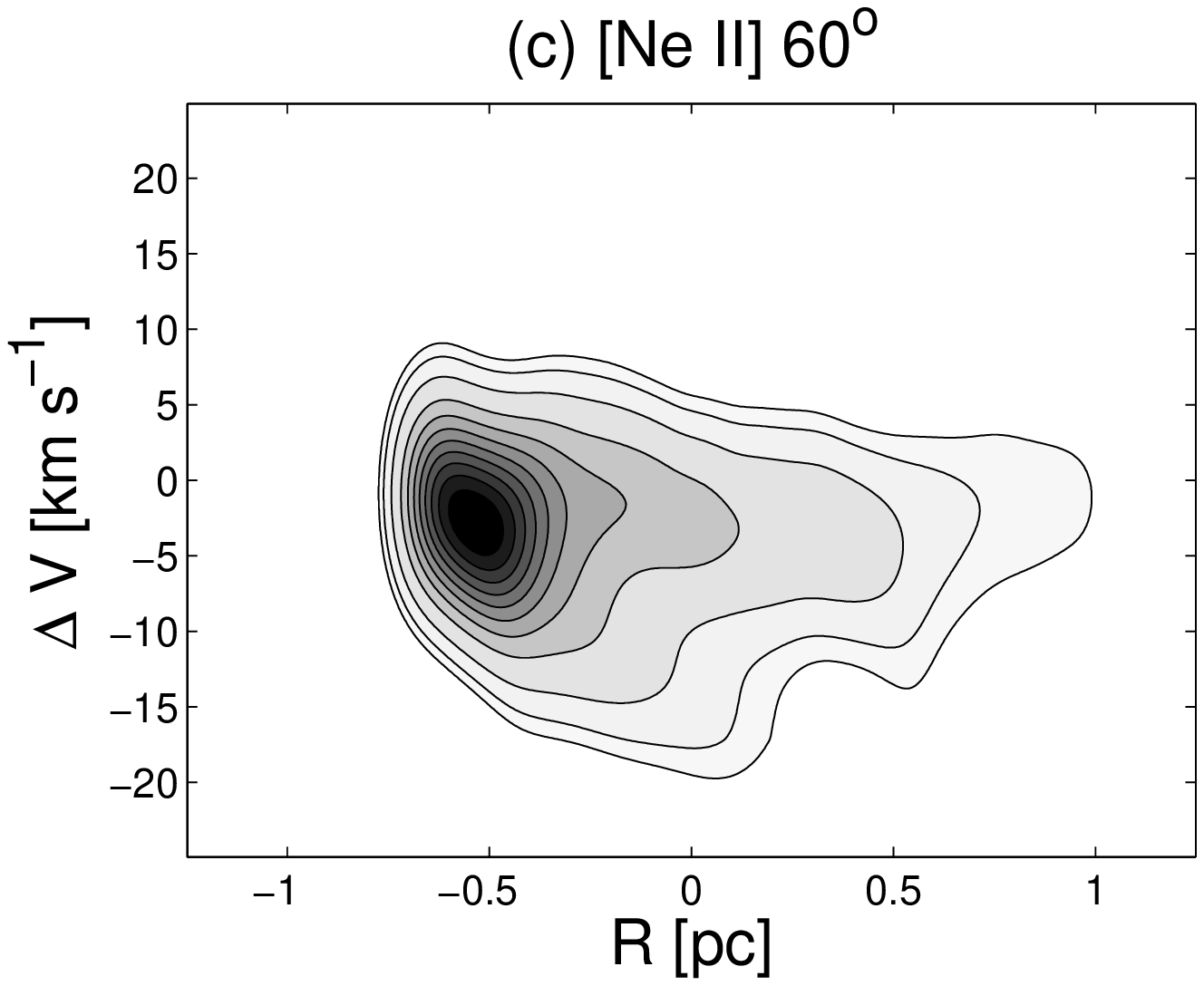}
\includegraphics[scale=.33]{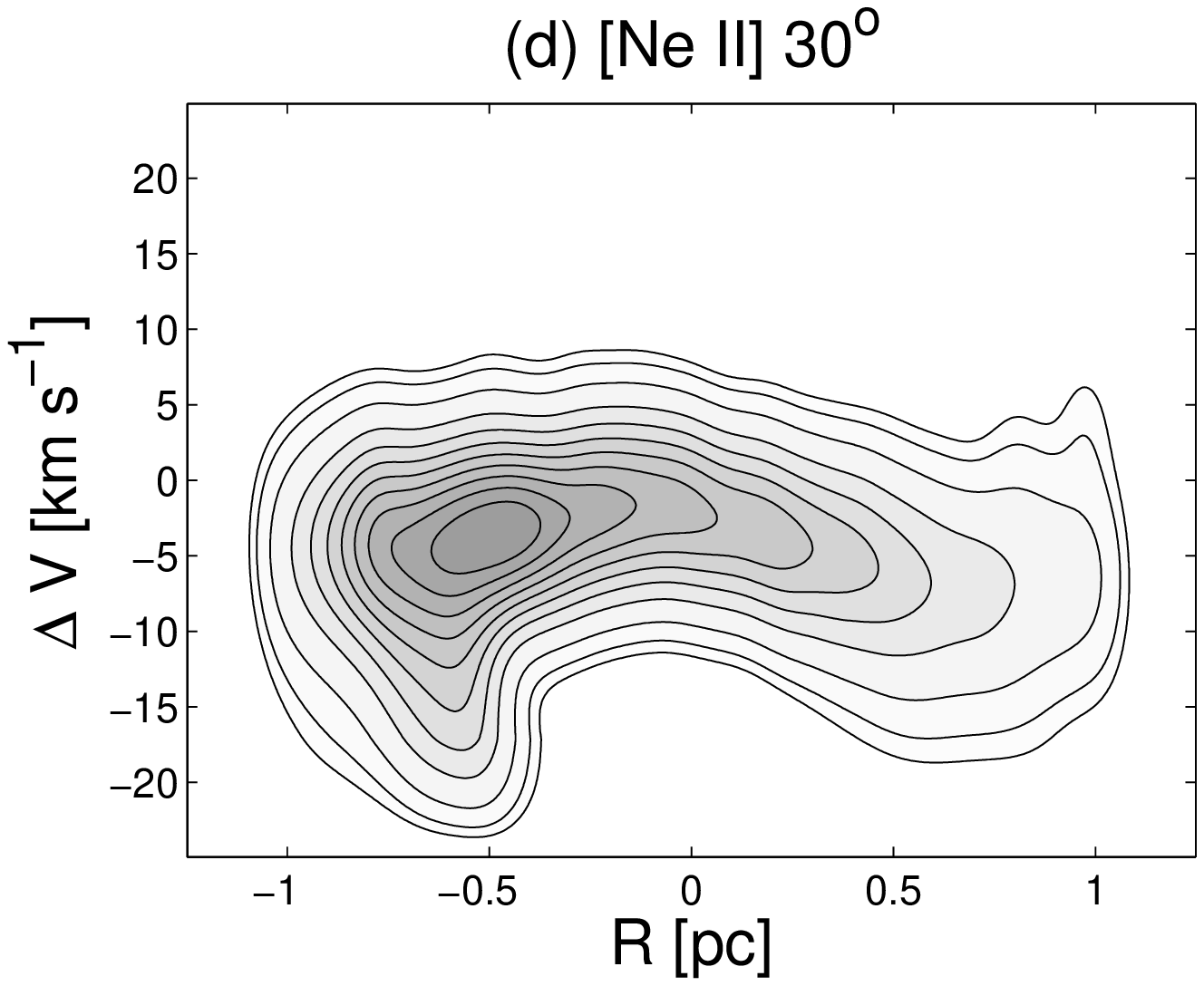}
\includegraphics[scale=.33]{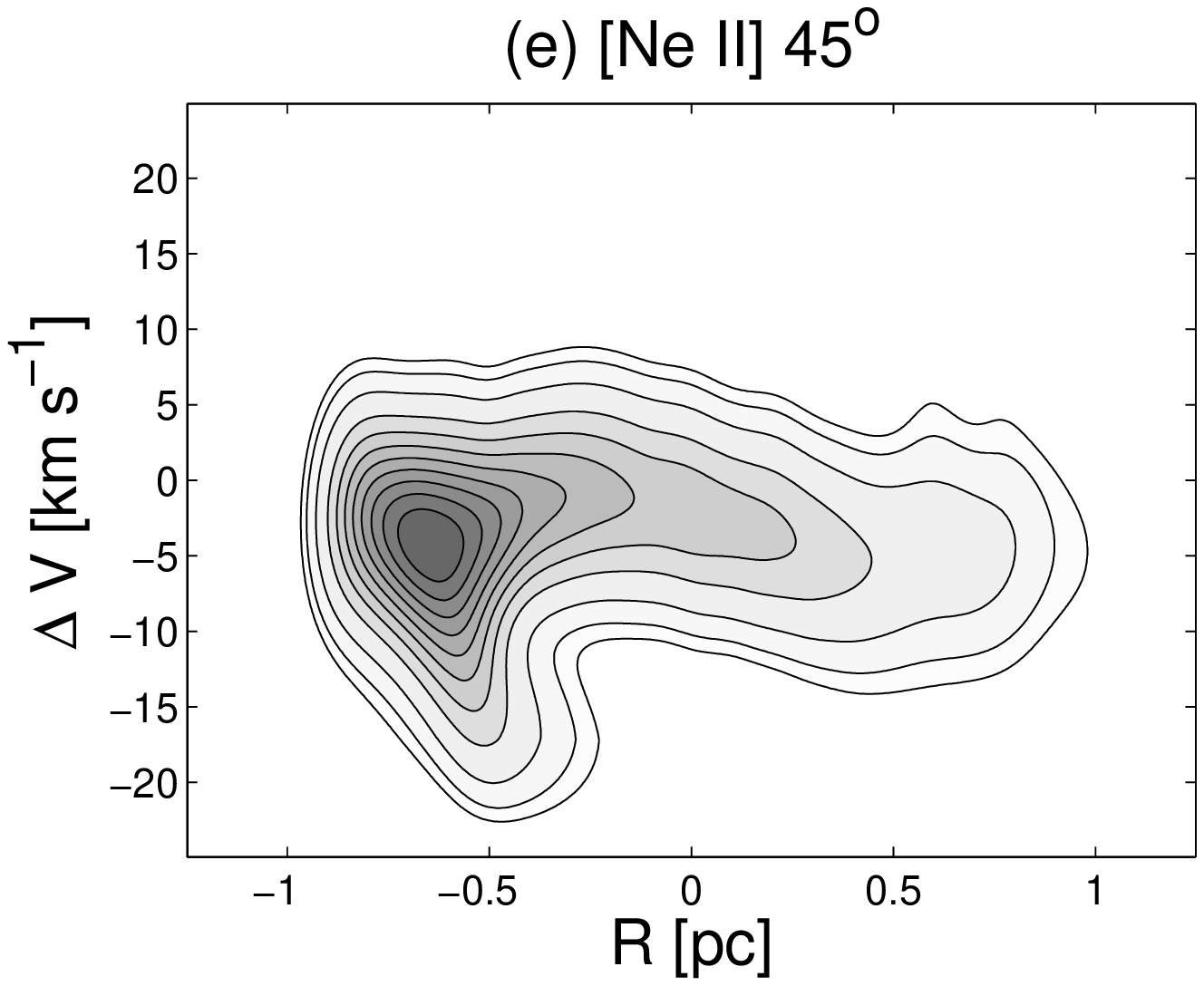}
\includegraphics[scale=.33]{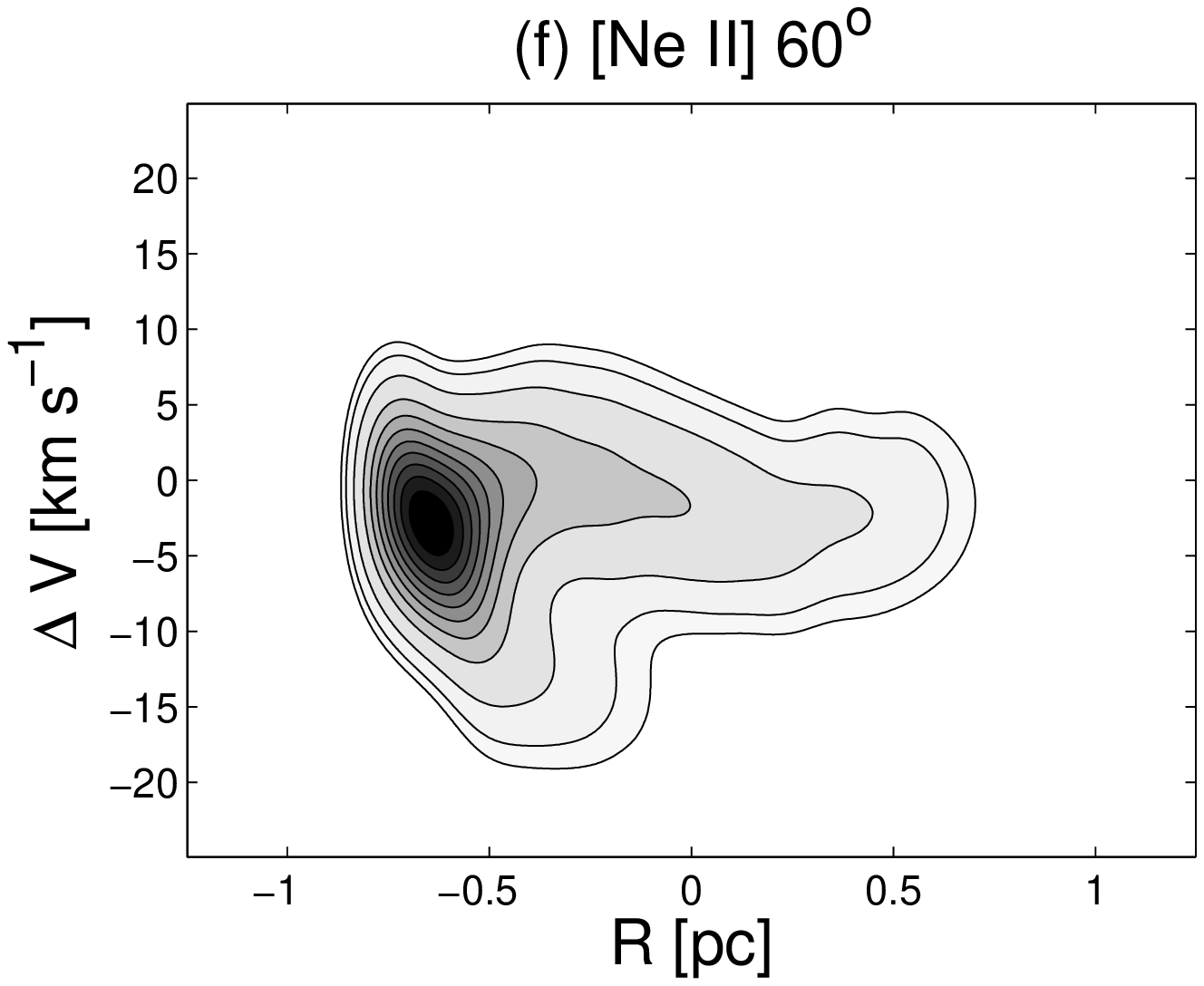}
\caption{The position-velocity diagrams of the [Ne II] $12.81~\mu m$ line from the slit along the \textrm{symmetry axis} of the projected 2D image for three inclination angles ($\theta=30^o$, $45^o$ and $60^o$). The top panels are presented for model C, and the bottom panels are for model D. The contour levels are at $3,~5,~10,~20,~30,~40,~50,~60,~70,~80,~90\%$ of the emission peaks in each panel.}
\begin{flushleft}
\end{flushleft}
\label{fig_modcdpvNe}
\end{figure}

In Figure \ref{fig_modcdpvNe}, the position-velocity diagrams of the [Ne II] line from the slit along the \textrm{symmetry axis} of the projected 2D image for three inclination angles in the two \textrm{champagne-flow} models are presented. In the diagrams, the main part of the flux is located at the position of $\Delta v <0~km~s^{-1}$. In \textrm{models} C and D, only the ionized gas close to the ionization front may have a red-shifted velocity component, and the ionized materials in the rest part of the H II region flow towards the tail. As in the \textrm{bow-shock} models, the emission becomes more concentrated with increasing inclination in \textrm{models} C and D. There are also two branches in every diagram in Figure \ref{fig_modcdpvNe}. In these diagrams, the velocity range of the left branch ($R<0$) is wider than that of the other branch ($R>0$). This is because the gas contributing emission to the left branch is from the near side of the shell. The angle between this side of the shell and the line of sight is smaller. The velocity projection onto the line of sight is more efficient than that of the gas moving approximately perpendicular to the line of sight, which is the case for the gas in the farther side of the shell. 


\subsubsection{Profiles and diagrams of the $H_2~S(2)$ line}

\begin{figure}[!htp]
\centering
\includegraphics[scale=.33]{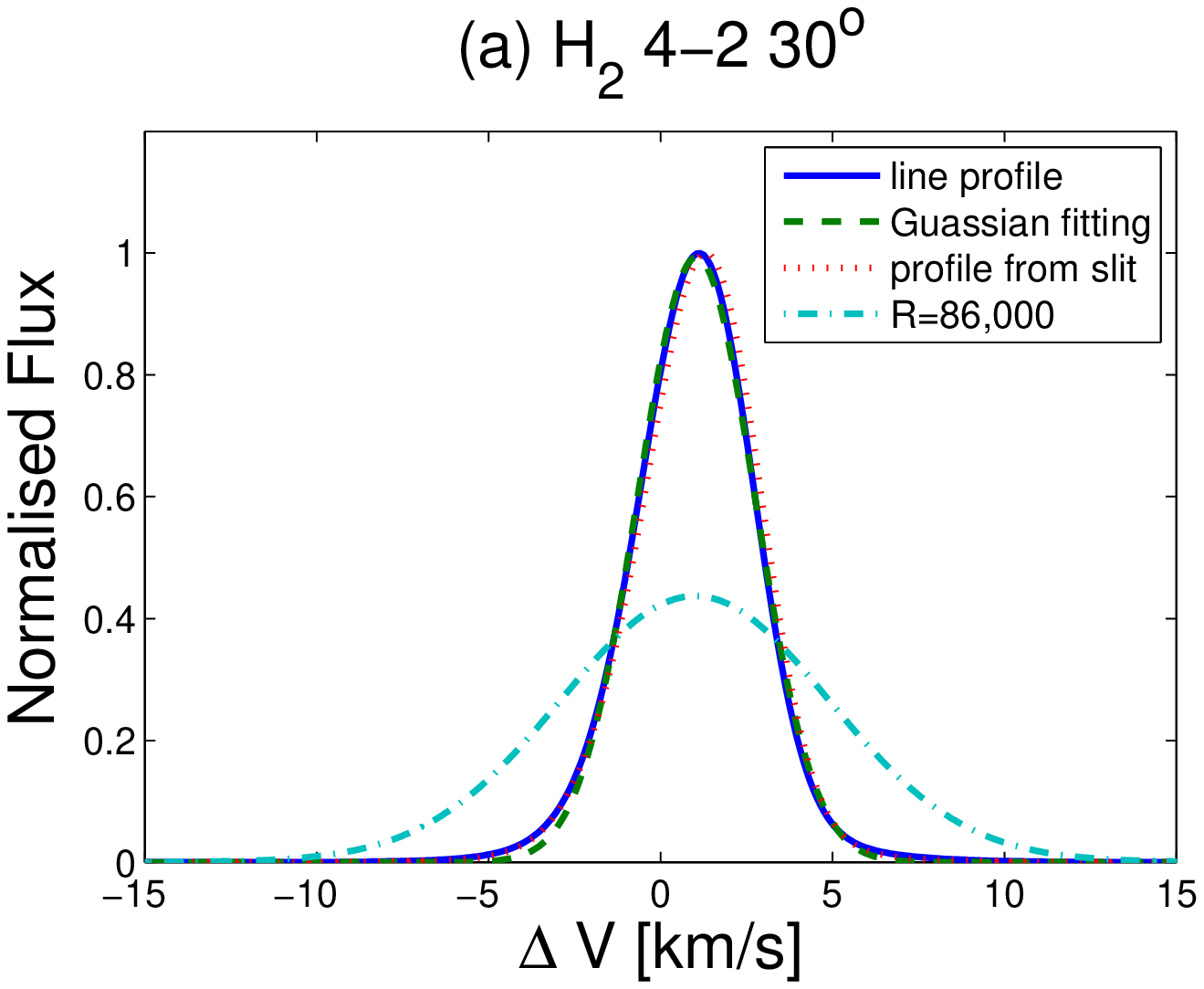}
\includegraphics[scale=.33]{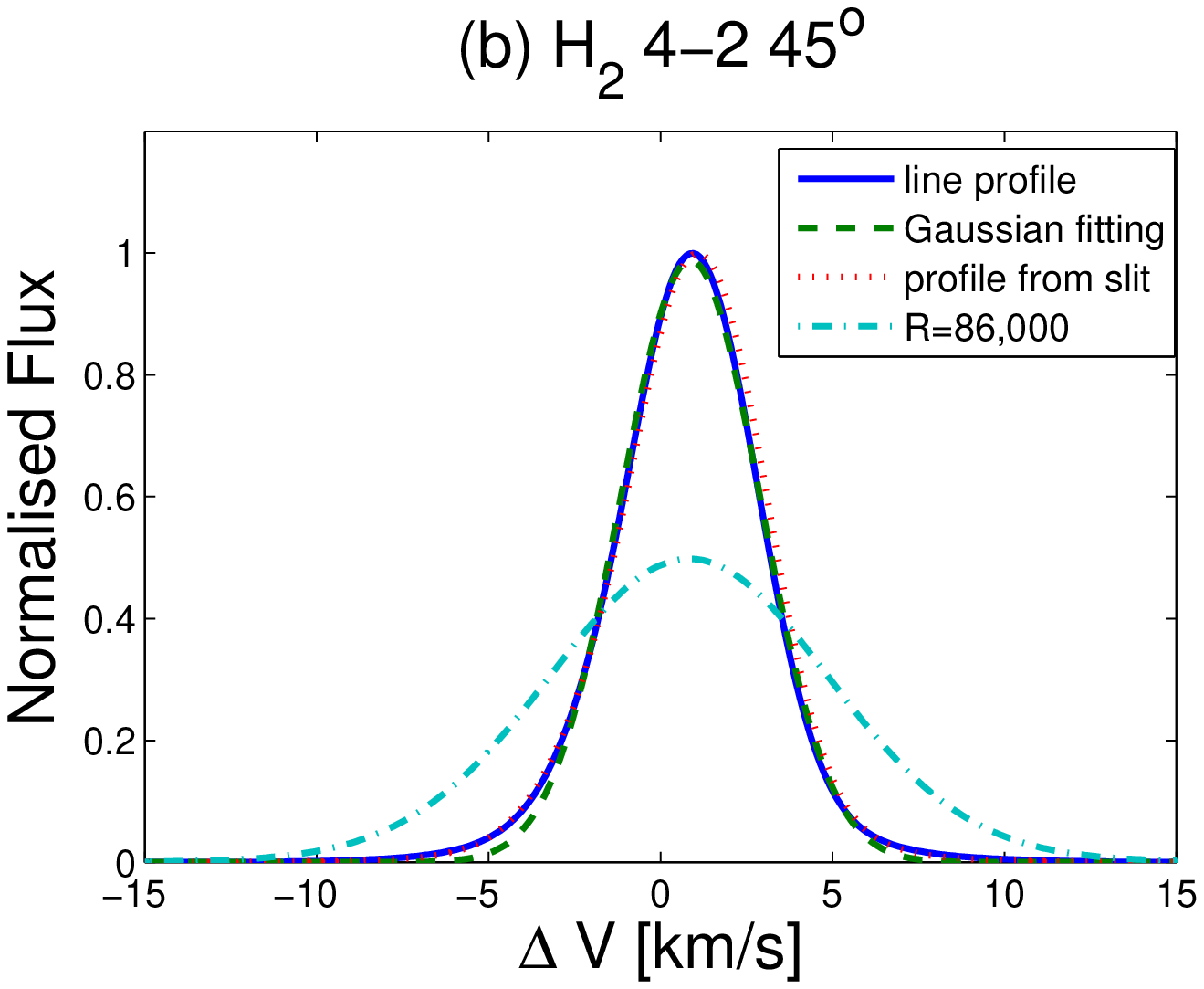}
\includegraphics[scale=.33]{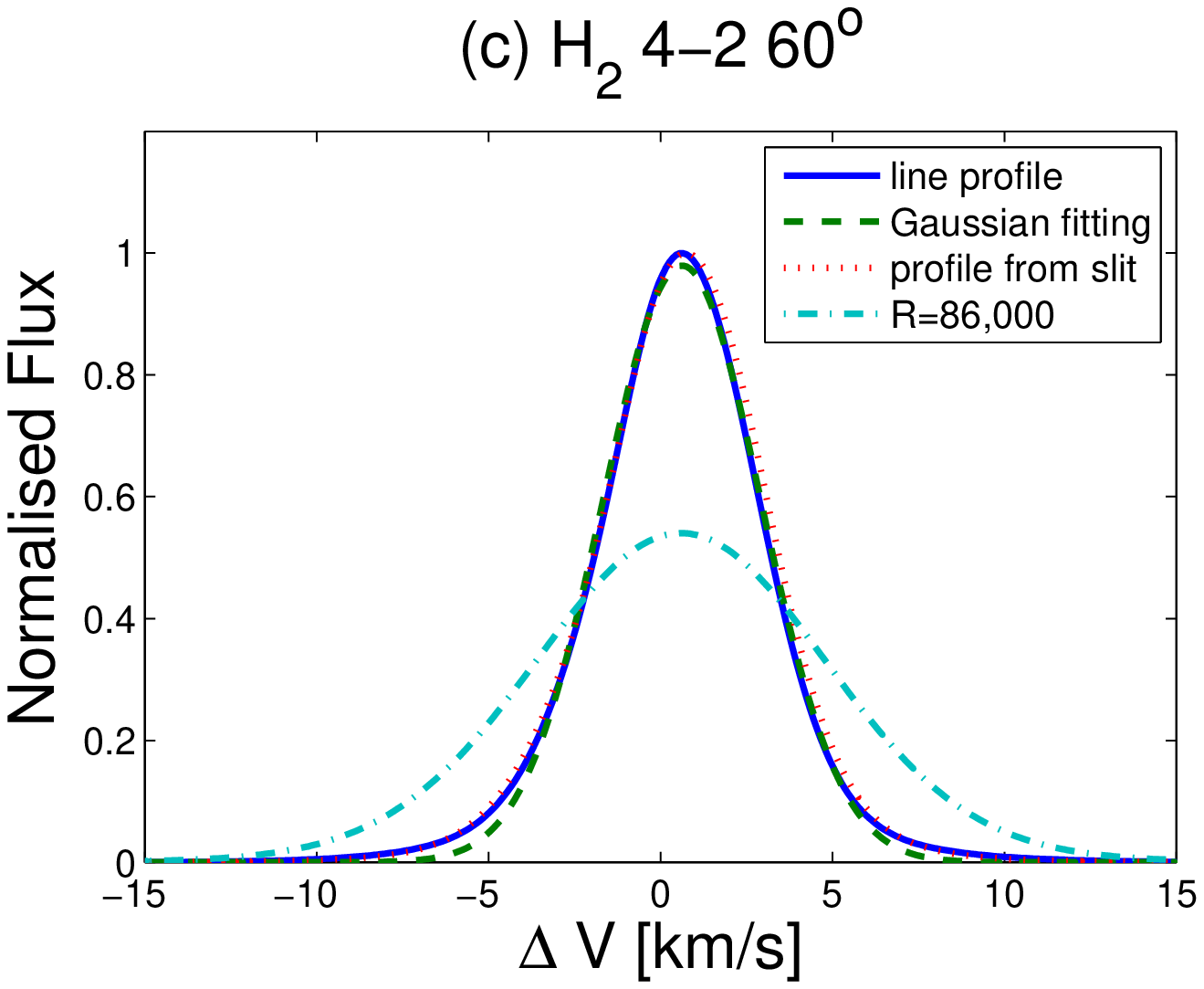}
\includegraphics[scale=.33]{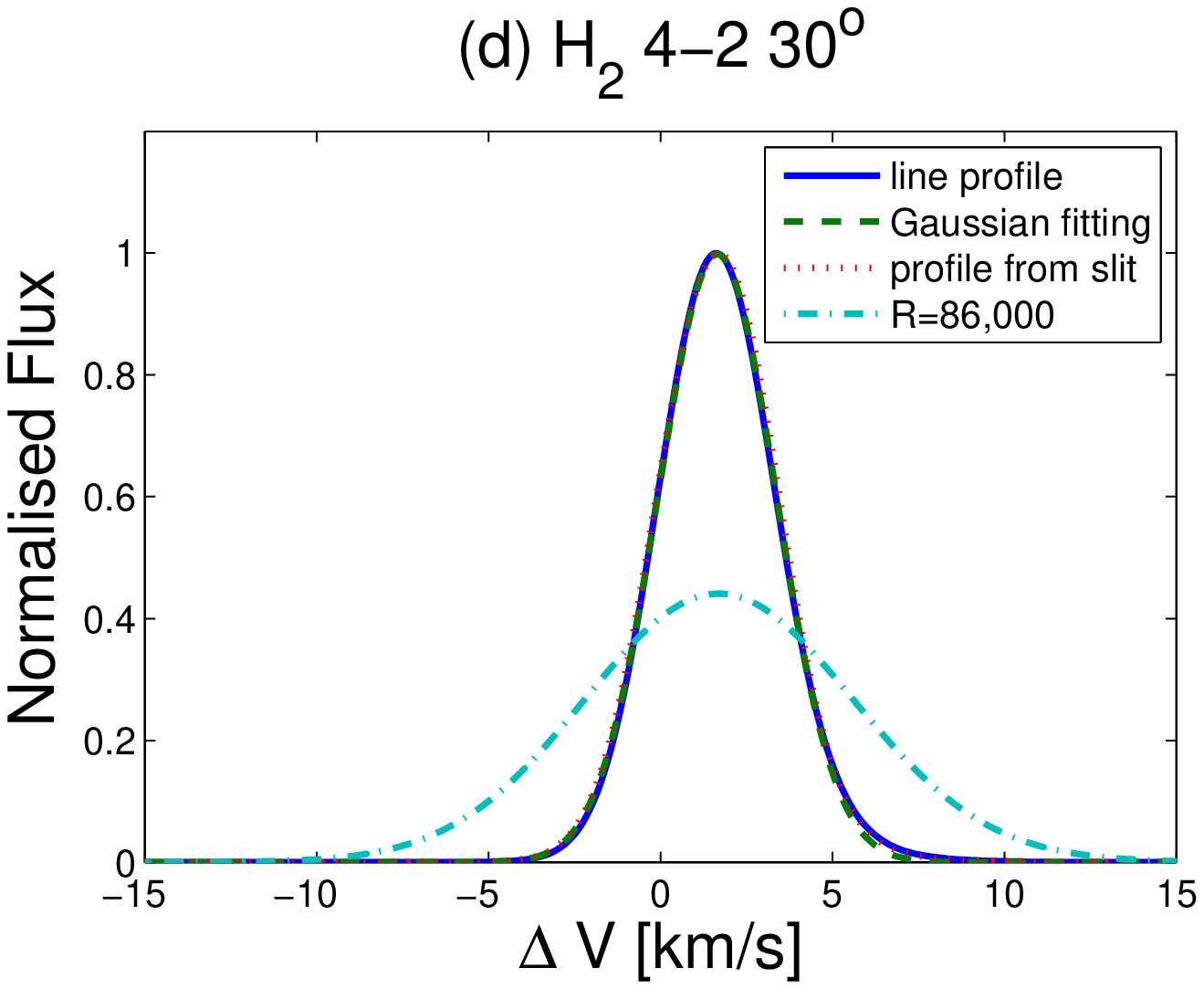}
\includegraphics[scale=.33]{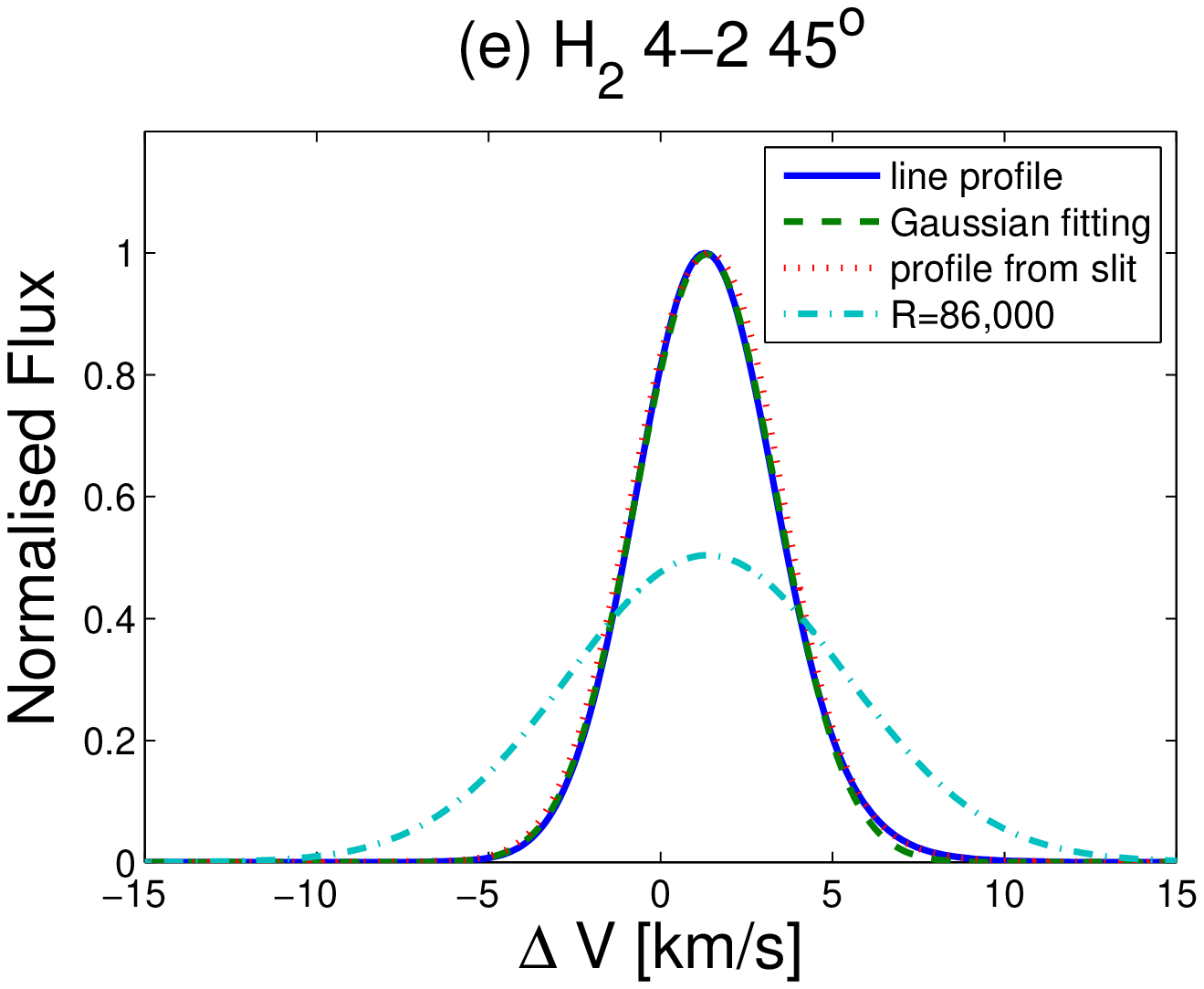}
\includegraphics[scale=.33]{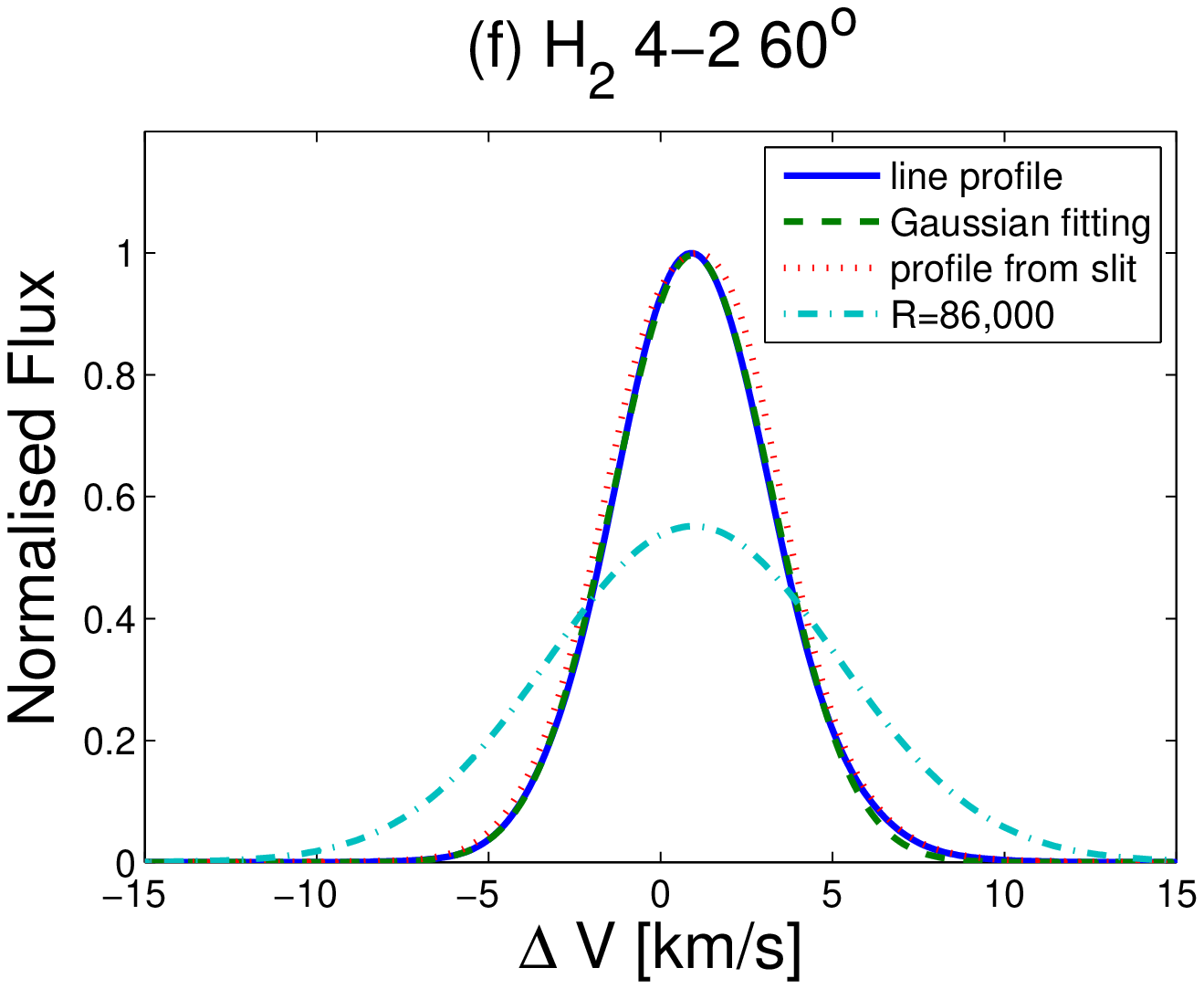}
\caption{Profiles of the $H_2~S(2)$ line (solid lines), the Gaussian fitting to the profiles (dashed lines) and the profiles at the resolution of $R=86,000$ (dot-dashed lines) at three inclination angles ($\theta=30^o$, $45^o$ and $60^o$). The $H_2~S(2)$ line profiles from the slit along the \textrm{symmetry axis} of the projected 2D image are also presented (dotted lines). The top panels are plotted for model C, and the bottom panels are shown for model D.}
\begin{flushleft}
\end{flushleft}
\label{fig_modcdfitH}
\end{figure}

The profiles of the $H_2~S(2)$ line are plotted in Figure \ref{fig_modcdfitH}, and their line properties are listed in Table \ref{tab_modcd}. The FWCVs of the $H_2$ line and the centers of the Gaussian fitting curves are $0.99,~0.81,~0.57~km~s^{-1}$ and $1.04,~0.87,~0.62~km~s^{-1}$ for \textrm{inclination angles} of $30^o,~45^o~\textrm{and}~60^o$ in model C, \textrm{respectively, and} those in model D are $1.73,~1.41,~1.00~km~s^{-1}$ and $1.65,~1.33,~0.94~km~s^{-1}$. The FWCV and the \textrm{line centers} are close to each other for the corresponding inclinations because the line profiles are very close to Gaussian. All values in model D are a little higher than the corresponding ones in model C. This is because the higher-mass star with larger ionization luminosity and stronger stellar wind in model D \textrm{leads to a faster moving} ionization front. The FWHMs of the fitted Gaussian in \textrm{models} C and D are $4.00,~4.76,~5.40~km~s^{-1}$ and $4.03,~4.79,~5.45~km~s^{-1}$ for \textrm{inclination angles} of $30^o,~45^o,~60^o$, respectively. It is apparent that the FWHM increases with \textrm{rising inclination angle}. 
It is not easy to distinguish \textrm{models} C and D from model A with the same stellar parameters by only using the FWCVs, FWHMs and the centers of the fitting curves of the $H_2~S(2)$ line. Particularly, the inclination angle causes more ambiguities. The FWCVs of the $H_2~S(2)$ line in model B are \textrm{obviously} higher than those in the other models. Although the profiles of the $H_2$ line are closer to a Gaussian shape in the \textrm{champagne-flow} models, the difference of the line profiles is not significant enough. In addition, the skewnesses of the $H_2~S(2)$ line profiles in \textrm{models} C and D are $0.01,~-0.03,~-0.03$ and $0.26,~0.20,~0.13$ for the inclinations of $30^o,~45^o$, and $60^o$, respectively. The skewnesses of the $H_2$ line in champagne flows are generally lower than in bow shocks. The $H_2$ line profiles from the slit along the \textrm{symmetry axis} of the projected 2D image for the three inclinations in \textrm{models} C and D are hard \textrm{to distinguish from} the ones for the whole regions.

\begin{figure}[!htp]
\centering
\includegraphics[scale=.33]{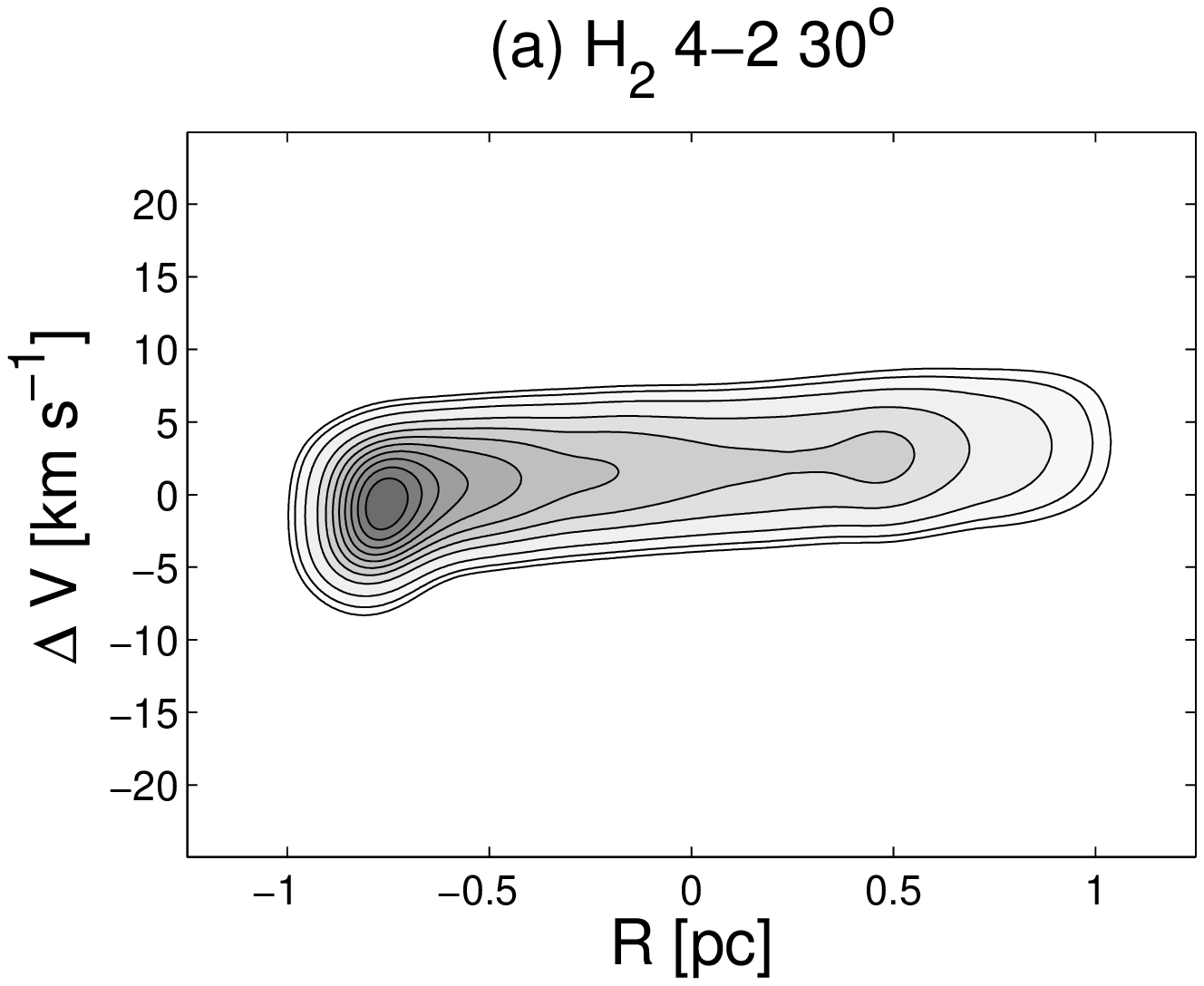}
\includegraphics[scale=.33]{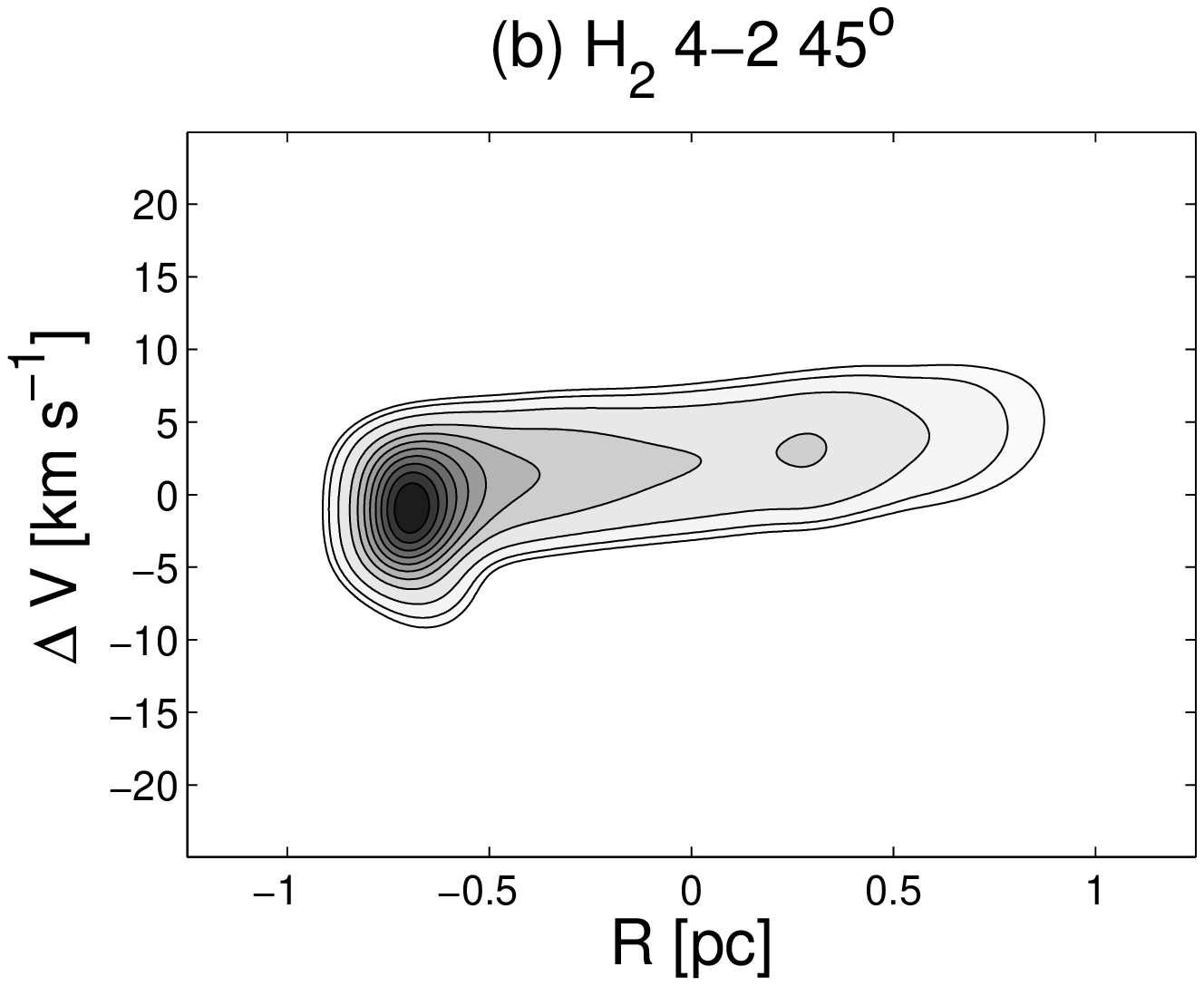}
\includegraphics[scale=.33]{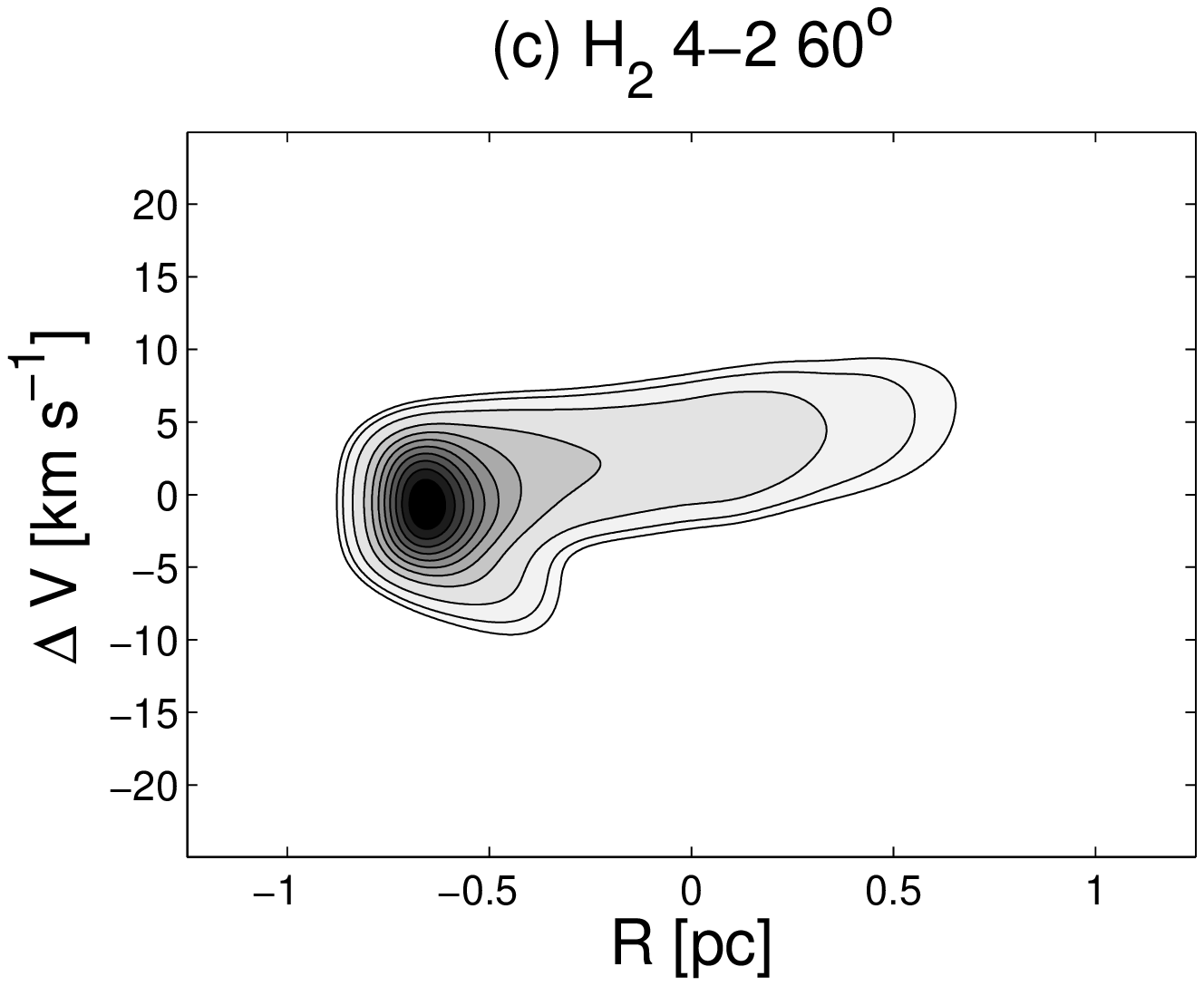}
\includegraphics[scale=.33]{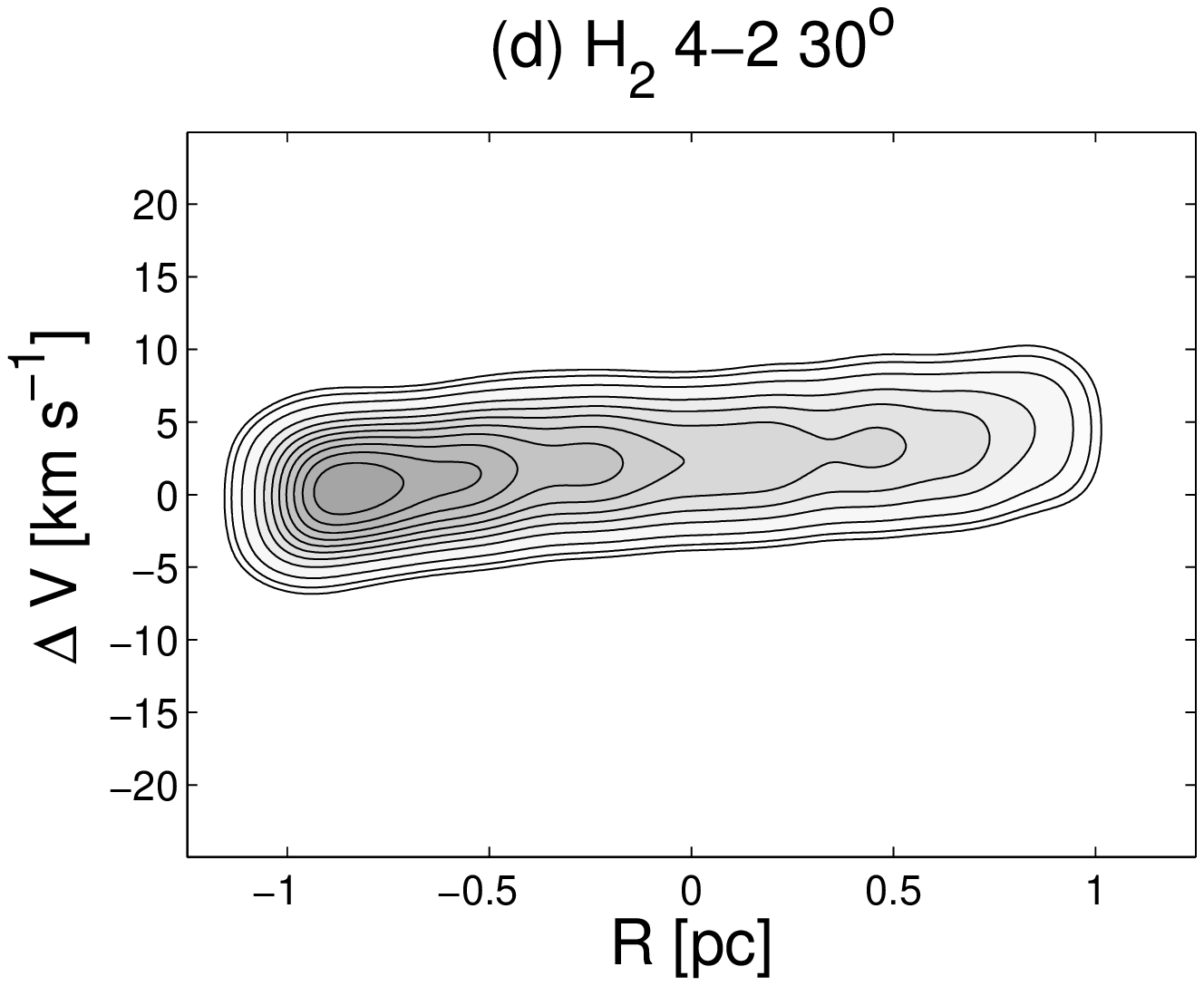}
\includegraphics[scale=.33]{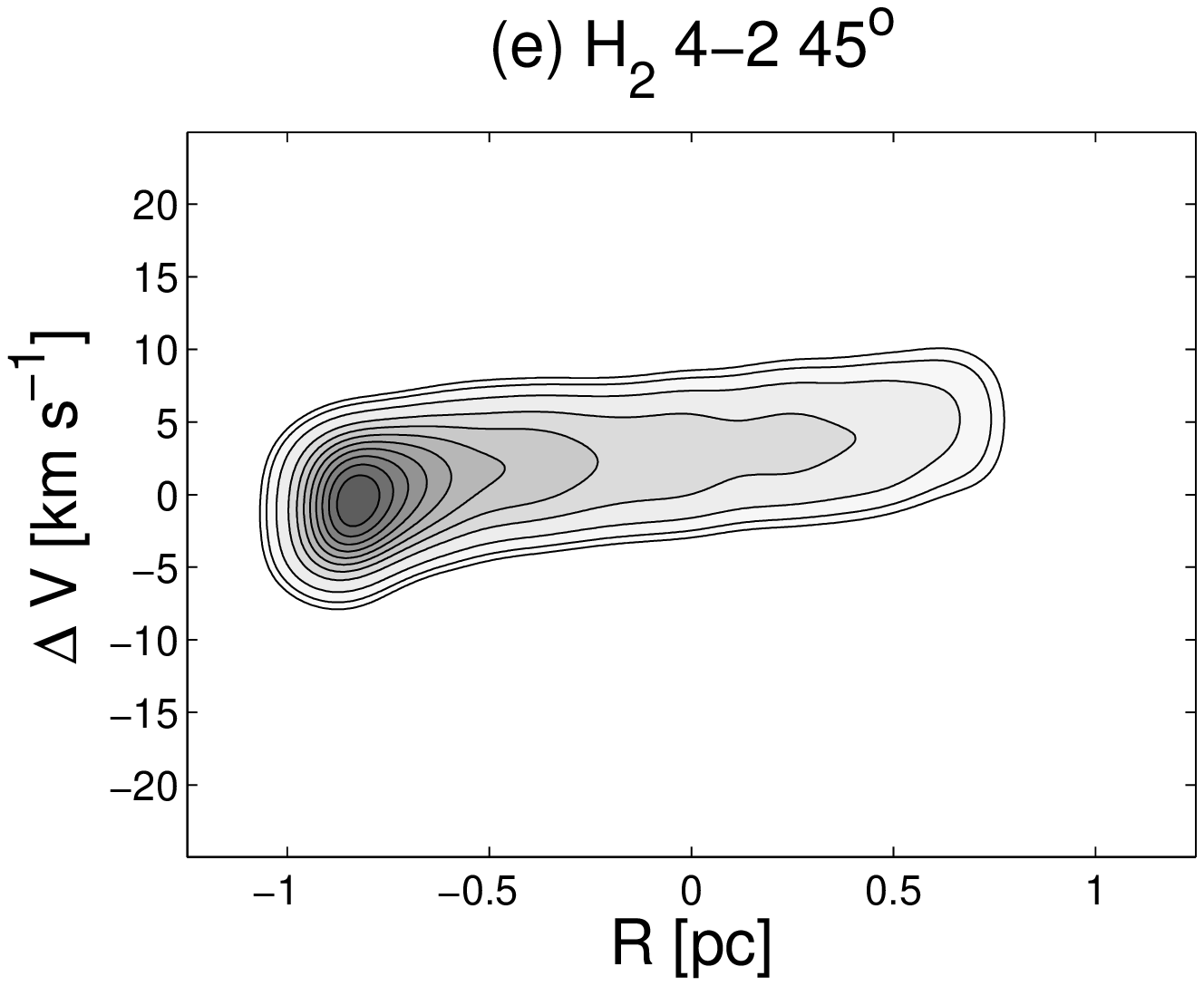}
\includegraphics[scale=.33]{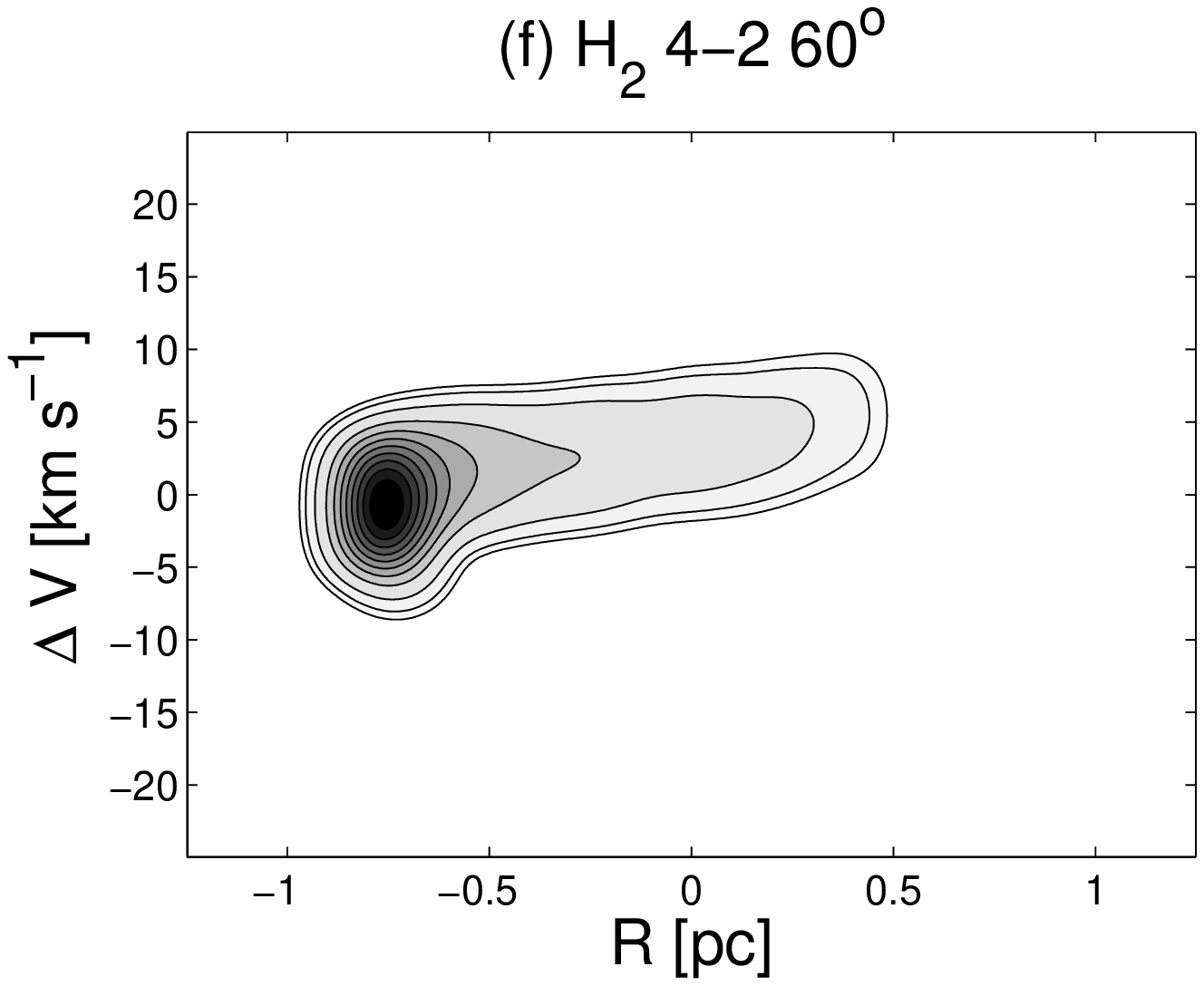}
\caption{The position-velocity diagrams of the $H_2$ 4-2 rotational transition line from the slit along the \textrm{symmetry axis} of the projected 2D image for three inclination angles ($\theta=30^o$, $45^o$ and $60^o$). The top panels are presented for model C, and the bottom ones are shown for model D. The contour levels are at $3,~5,~10,~20,~30,~40,~50,~60,~70,~80,~90\%$ of the emission peaks in each panel.}
\begin{flushleft}
\end{flushleft}
\label{fig_modcdpvHmol}
\end{figure}

The position-velocity diagrams of the $H_2~S(2)$ line from the slit along the \textrm{symmetry axis} of the projected 2D image for the inclination angles of $30^o,~45^o,~60^o$ in \textrm{models} C and D (Figure \ref{fig_modcdpvHmol}) are \textrm{obviously different} from those in \textrm{bow-shock} models. The axial components of the velocity of the molecular hydrogen in \textrm{models} C and D are not high. There is apparently no gas emission above $5~km~s^{-1}$ \textrm{in the diagrams}. 
The $H_2$ line emission is always close to the red-shifted side of the [Ne II] line emission at the same values of $R$. This is because \textrm{the $H_2$ dissociation front} falls into the dense shell almost in every direction in \textrm{models} C and D so that the $H_2$ line emission is all from the dense shell in the \textrm{champagne-flow} models and the distribution of the $H_2$ line emission is continuous. Both the [Ne II] line emission and the $H_2$ line emission \textrm{come from} the dense shell behind the shock front. Their dynamics is mainly due to the expansion of the H II region and the dense shell. The projected velocities of the \textrm{neutral gas in the shell} are typically more red-shifted than those of the nearby ionized gas. However, in \textrm{the bow-shock} models, \textrm{part of the $H_2$ line emission comes from} the region beyond the shock front. If the stellar velocity is not very high ($v_\ast\leq10~km~s^{-1}$), the DF of $H_2$ will also fall into the dense shell, and the $H_2$ \textrm{line emission} from the molecular gas close to the apex of the shell will reflect \textrm{in} the p-v diagrams. \textrm{On the other hand,} the number of $H_2$ line photons emitted from the region beyond the shock front is very small and can be neglected. 


\begin{table}[!htp]\footnotesize
\centering
\begin{tabular}{c|ccccc}
\hline
Model C Champagne Flow $21.9~M_\odot$  \\
\hline
Line & Inclination & FWCV & Skewness & Center of line & FWHM\\
 & & ($km~s^{-1}$) & & ($km~s^{-1}$) & ($km~s^{-1}$) \\
\hline
[Ne II] $12.81 \mu m$ & $30^o$ & -7.54 & -0.68 & -6.16 & 14.91 \\
 & $45^o$ & -6.16 & -0.70 & -4.99 & 12.91 \\
 & $60^o$ & -4.36 & -0.67 & -3.49 & 11.00 \\
\hline
$H_2~4-2$ & $30^o$ & 0.99 & 0.01 & 1.04 & 4.00 \\
 & $45^o$ & 0.81 & -0.03 & 0.87 & 4.76 \\
 & $60^o$ & 0.57 & -0.03 & 0.62 & 5.40 \\
\hline
Model D Champagne Flow $40.9~M_\odot$  \\
\hline
Line & Inclination & FWCV & Skewness & Center of line & FWHM \\
 & & ($km~s^{-1}$) & & ($km~s^{-1}$) & ($km~s^{-1}$) \\
\hline
[Ne II] $12.81 \mu m$ & $30^o$ & -5.42 & -0.76 & -4.23 & 13.43 \\
 & $45^o$ & -4.42 & -0.81 & -3.41 & 11.75 \\
 & $60^o$ & -3.13 & -0.79 & -2.36 & 10.40 \\
\hline
$H_2~4-2$ & $30^o$ & 1.73 & 0.26 & 1.65 & 4.03 \\
 & $45^o$ & 1.41 & 0.20 & 1.33 & 4.79 \\
 & $60^o$ & 1.00 & 0.13 & 0.94 & 5.45 \\
\hline
\end{tabular}
\caption{Flux weighted central velocities (FWCV), skewness, and the center and FWHMs of Gaussian fitting of the lines for different angles in \textrm{models} C and D\label{tab_modcd}}
\end{table}

\subsection{Comparison with observations}
\label{sect:comparison}

Two cometary H II regions in the DR 21 region were observed by \citet{cyg03} to investigate whether the H II regions are consistent with \textrm{bow-shock} models or \textrm{champagne-flow} models. The northern H II region can be explained by a \textrm{champagne-flow} model. And the observations of the southern H II region suggest a hybrid model including a stellar motion and a density gradient. \citet{imm14} presented deep Very Large Array $H66\alpha$ observations of the southern cometary H II regions in DR 21. The line profiles of $H66\alpha$ line from different positions in the H II regions are shown \textrm{separately}. The analysis of the observations shows that the line profiles mostly show two velocity components. The ionized gas near the cometary head is mainly blue-shifted (towards the cloud in the observations), and this is consistent with the results of our \textrm{bow-shock} models. The velocity of the ionized gas becomes increasingly red-shifted from the head to the cometary tail, and the ionized gas at the tail has a highly red-shifted velocity ($\sim30~km~s^{-1}$) relative to the molecular gas. Our \textrm{champagne-flow} models also show such kinematic features. Since, in the \textrm{bow-shock} models, the ionized gas velocity at the tail can not reach such a high value, and the ionized gas near the cometary head mainly flows downstream in our \textrm{champagne-flow} models, we agree that these features suggest the combined effects of a stellar wind, stellar motion (as in \textrm{bow-shock} models) and an ambient density gradient (as in \textrm{champagne-flow} models).

\section{Conclusions}
\label{sect:conclusion}

In this paper, we simulate the evolution of the cometary H II region with the PDR in \textrm{bow-shock} models and \textrm{champagne-flow} models.  The line profiles of the [Ne II] $12.81 \mu m$ line and the $H_2~S(2)$ line are presented, and the properties of these lines are studied. In \textrm{the champagne-flow} models, the ionized gas flows downstream into the low-density region due to the pressure gradient caused by the original density gradient, and the neutral gas in the dense shell enveloping the H II region moves slowly with the expansion of the ionized region. In \textrm{the bow-shock} models, the ionized gas in the head of the cometary H II region originally goes towards the molecular cloud with the moving star. But the ionized gas is gradually accelerated at the opposite direction because of the density gradient formed in \textrm{the ionized region}, and eventually flows toward the tail. The neutral dense shell \textrm{moves in the direction} of the stellar motion in \textrm{bow-shock} models. By using the [Ne II] $12.81~\mu m$ line, it is \textrm{possible} to distinguish a high stellar velocity \textrm{bow-shock} model from the \textrm{champagne-flow} models. If the center of the [Ne II] line profile from the head of the H II region has a doppler shift that indicates an ionized gas motion toward the molecular cloud, the cometary H II region should be caused by a bow shock. However, because the ionized gas is accelerated toward the tail, the line profiles from a low speed star bow shock might be confused with those from a champagne flow. In this case, it is possible to use the emission lines from the dense shell as additional criteria to discriminate between the bow-shock and \textrm{champagne-flow} scenarios. This is because the neutral gas in the shell is not influenced by the density gradient in the H II region. If the center of the $H_2~S(2)$ line profile from the molecular gas near the cometary head has a doppler shift larger than $3~km~s^{-1}$, it is likely a bow shock is present because such a high speed of the molecular gas can not be reached only by the expansion of the ionized gas unless the H II region is at very early times. The details of our models are summarized as follows:

1. In both the \textrm{bow-shock} and the \textrm{champagne-flow} models, the [Ne II] $12.81 \mu m$ line profiles are negative-skewed. The skewness of the [Ne II] line profiles increases from more negative values with the increasing inclination. However, in the \textrm{champagne-flow} models, because of the large opening angle, this trend is not obvious when the inclination is lower than $45^o$. The centers of the [Ne II] line profiles in the \textrm{champagne-flow} models are obviously blue-shifted ($FWCV<-3~km~s^{-1}$). In the \textrm{bow-shock} models, because of the pressure gradient in ionized regions mentioned above, the centers of the [Ne II] line could be blue-shifted when the \textrm{stellar velocity is} $v_\ast<15km~s^{-1}$. If the stellar velocity is high, the centers of the [Ne II] line are red-shifted. The FWHM of the Gaussian fitting of the [Ne II] line decreases with the increasing inclination in both the \textrm{bow-shock} models and the \textrm{champagne-flow} models.

2. For the $H_2~S(2)$ line, \textrm{we predict that} the line centers are red-shifted in the \textrm{bow-shock} models when the stellar velocity is high. But the center of the $H_2$ line is always less than $5~km~s^{-1}$. This is because the apex of the dense shell is fully dissociated by the FUV photons if the stellar velocity is higher than $10~km~s^{-1}$, and the emission is mostly from the sides of the dense shell and the undisturbed region where the gas has less red-shifted velocity. Because of the discontinuity of the velocity field of the molecular gas in shell, the $H_2~S(2)$ line profiles in the \textrm{bow-shock} models are far from Gaussian and often double-peaked. In the \textrm{champagne-flow} models, the line profiles are very close to the Gaussian shape. And the centers of the lines are lower than $2~km~s^{-1}$ and limited by the expansion of the H II region.

3. The position-velocity diagrams of the [Ne II] line from the slit along the \textrm{symmetry axis} of the projected 2D image show that the line tends to be blue-shifted in the \textrm{champagne-flow} models. In the \textrm{bow-shock} models with stellar velocity $v_\ast=15~km~s^{-1}$, the emission peaks of the [Ne II] line are all red-shifted. Comparing the p-v diagrams of the [Ne II] line and the $H_2~S(2)$ line, we find that the dynamics of the ionized gas is influenced by pressure gradients in the ionized region for both the \textrm{champagne-flow} models and the \textrm{bow-shock} models, but the $H_2$ line emission is mainly affected by the moving shell or the expansion of the ionized region. So the correlation between the kinematics of the ionized gas and that of the warm molecular gas is low.

4. In the \textrm{bow-shock} model, the IF and the DFs are at roughly constant distances from the star after a certain age. In the \textrm{champagne-flow} model, the distances from the star of the IF and DFs increase with time. The DFs gradually fall into the dense shell along the evolution of the champagne flows.

\begin{acknowledgements}
\textrm{Acknowledgements}: The work is partially supported by National Basic Research Program of China (973 program) No. 2012CB821805 and the Strategic Priority Research Program " The Emergence of Cosmological Structures" of the Chinese Academy of Sciences. Grant No. XDB09000000. The authors are also grateful to the supports by Doctoral Fund of Ministry of Education of China No. 20113402120018 and Natural Science Foundation of Anhui Province of China No. 1408085MA13. F.-Y. Zhu thanks Dr. X.-L. Deng for generous financial aids during the visit to Beijing Computational Science Research Center and suggestions on the hydrodynamical computation methods, and also thanks Dr. J.-X. Wang for financial supports.

\end{acknowledgements}

\section{Appendix}

\subsection{Expansion of an H II region in a uniform medium}

Two test cases are shown in Appendix to assess the reliability of our models. In the first test case, we simulated the expansion of the H II region in a uniform medium of \textrm{density $n=10000~cm^{-3}$ and initial temperature $T=10~K$}. The \textrm{effective temperature} of the star is assumed to be $40,000~K$. The photo luminosities of the ionizing radiation and dissociating radiation of the star are $10^{48.78}~s^{-1}$ and $10^{48.76}~s^{-1}$, respectively. \textrm{Stellar winds are not} included. \textrm{The grid has $250\times500$ uniform cells of size $0.005~pc$.} We ceased the evolution at $100,000~yr$, and compared the results of this case with the analytical solution in \citet{spi78} and the numerical results of the expansion of the H II region and the PDR in \citet{hos06}.

\begin{figure}[!htp]
\centering
\includegraphics[scale=.45]{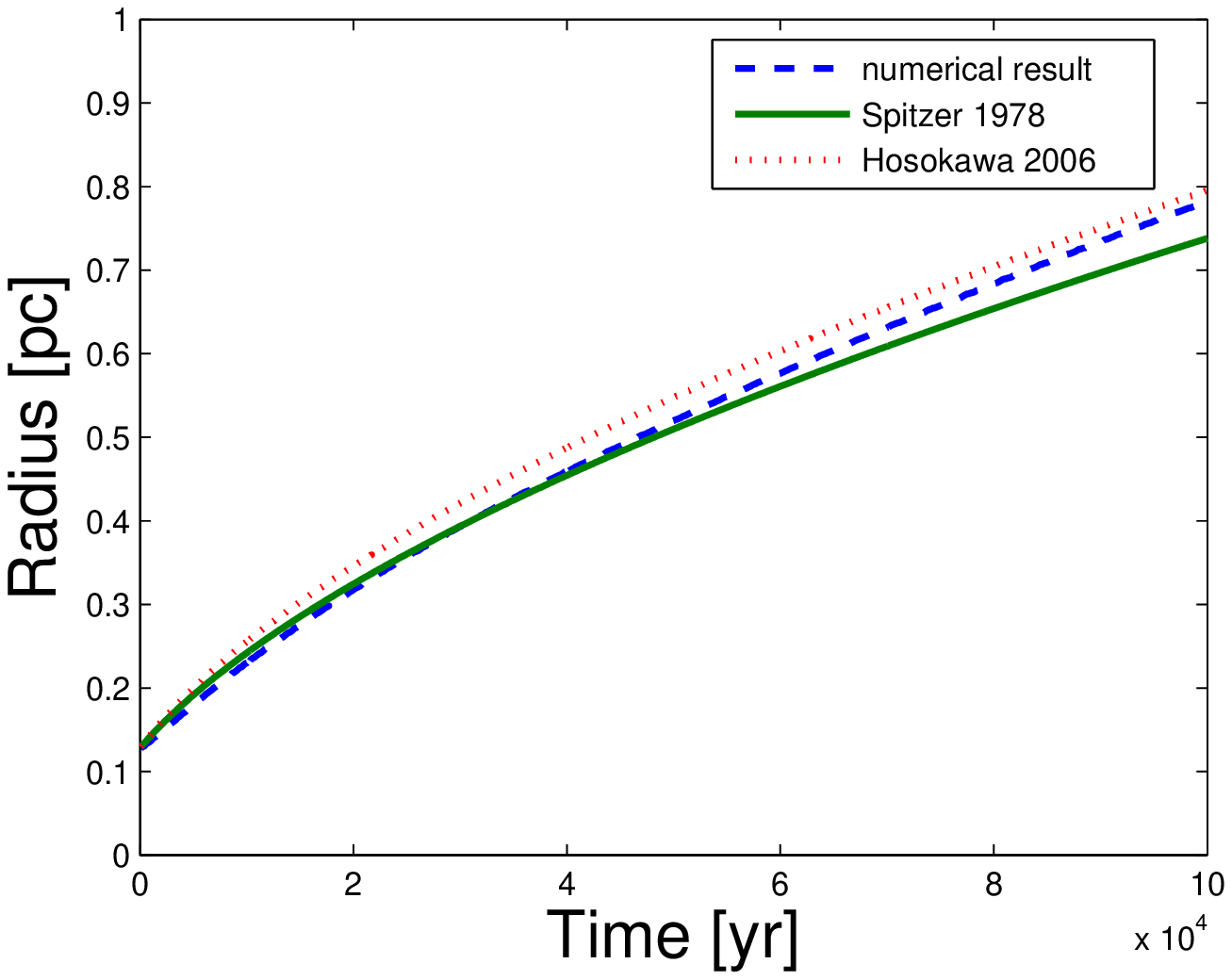}
\includegraphics[scale=.45]{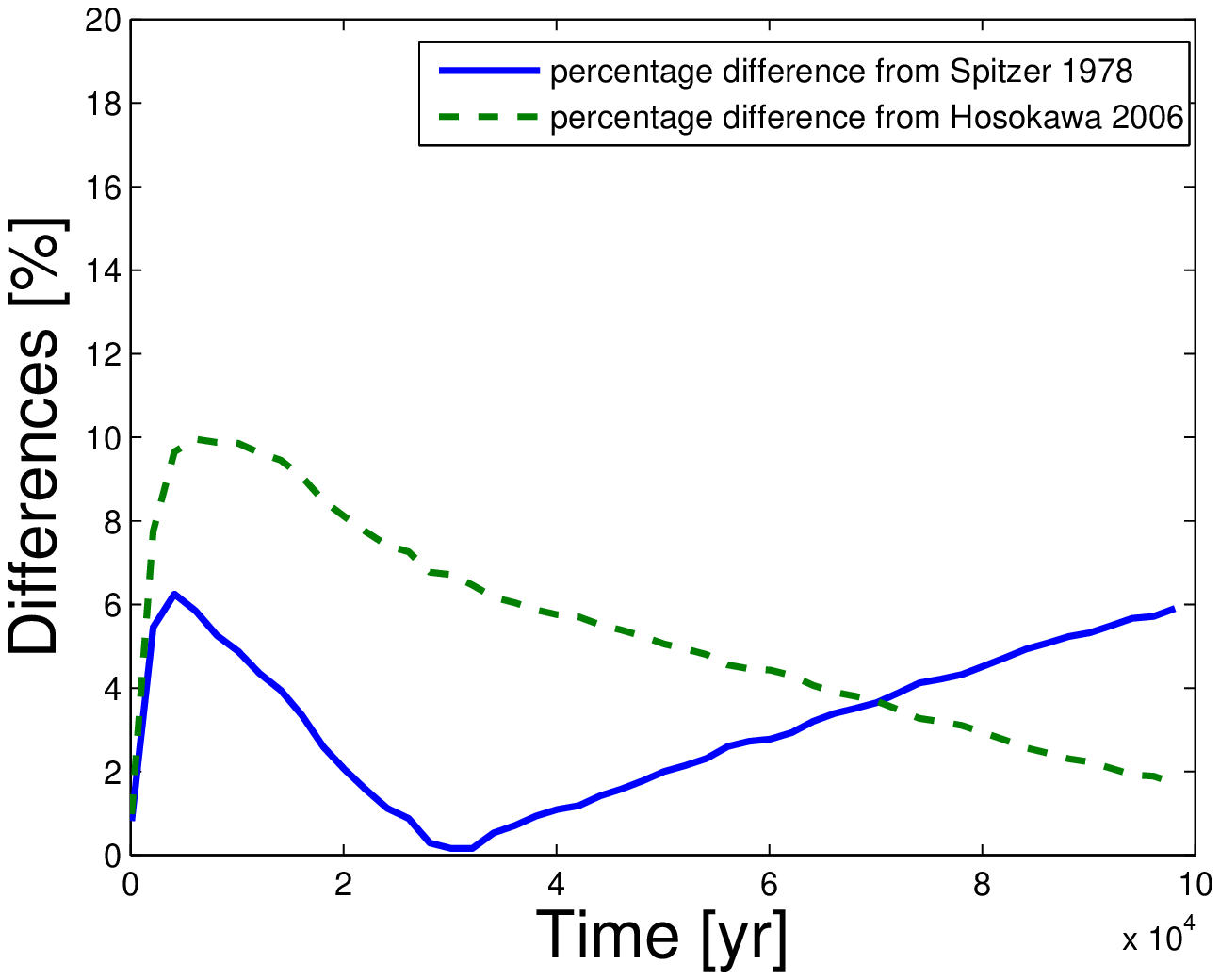}
\caption{Comparison between the numerical simulation and the analytical solutions of the time evolution of the radius of a spherical H II region in the first test case. The numerical and analytical solutions are shown in the left panel, and the percentage differences between the numerical result with analytical solutions from \citet{spi78} and \citet{hos06} are plotted in the right panel.}
\begin{flushleft}
\end{flushleft}
\label{fig_radius}
\end{figure}

In Figure \ref{fig_radius}, \textrm{the radial expansion of the numerical solution and the analytical solution of the spherical H II region are compared}. The analytical solution is given by \citet{spi78} as follows,

\begin{equation}
R_{IF}=R_0(1+\frac{7}{4}\frac{c_{II}t}{R_0})^{4/7}~~~,
\end{equation}
where $R_{IF}$ is the radius of the ionization front. $R_0$ is the radius of the theoretical initial St\"{o}mgren sphere. $c_{II}$ is the sound speed of the ionized gas, and $t$ is time. The percentage difference between the numerical solution and the analytical solution is also shown in Figure \ref{fig_radius}. An approximate formula derived from the equation of motion of the shell is provided by \citet{hos06} as follows,

\begin{equation}
R_{IF}=R_0(1+\frac{7}{4}\sqrt{\frac{4}{3}}\frac{c_{II}t}{R_0})^{4/7}~~~.
\end{equation}

This formula and the percentage difference between it and the result of the test case are also plotted in Figure \ref{fig_radius}. The numerical result is closer to the solution of Spitzer expansion at early times, and become closer to the formula provided by \citet{hos06} at late times. The percentage differences are always lower than $10\%$.

\begin{figure}[!htp]
\centering
\includegraphics[scale=.45]{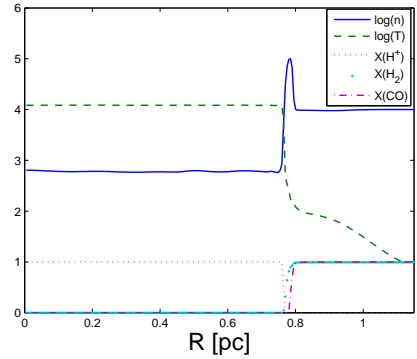}
\caption{The radial distributions of density, temperature, and fractions of $H^+$ ions, $H_2$ molecules and $CO$ molecules in the first test case at age of $100,000~yr$.}
\begin{flushleft}
\end{flushleft}
\label{fig_radial}
\end{figure}

The radial density and temperature distributions from the expansion of the spherical H II region in the first test case are plotted in Figure \ref{fig_radial}. These results are compared with the results in Fig. 3 from \citet{hos06}. Because of the same initial conditions, the results in our model are approximately consistent with those in \citet{hos06}. The temperature of ionized gas both in two models is $T\approx~10000~K$, and the radial temperature distribution in PDR is similar. The density distributions \textrm{in the ionized region} and PDR are also alike. But, since the resolution in the 1D model \citep{hos06} is higher than that in our 2D model, the results around the neutral dense shell in \textrm{two models differ slightly}. In \citet{hos06}, the temperature increases slightly near the H II region edge. However, this phenomenon does not appear in our result because the effect of spectral hardening is not considered. According to our calculations, this distinction will not lead to a significant difference.

\subsection{Evolution of an H II region with a moving star}

\begin{figure}[!htp]
\centering
\includegraphics[scale=.45]{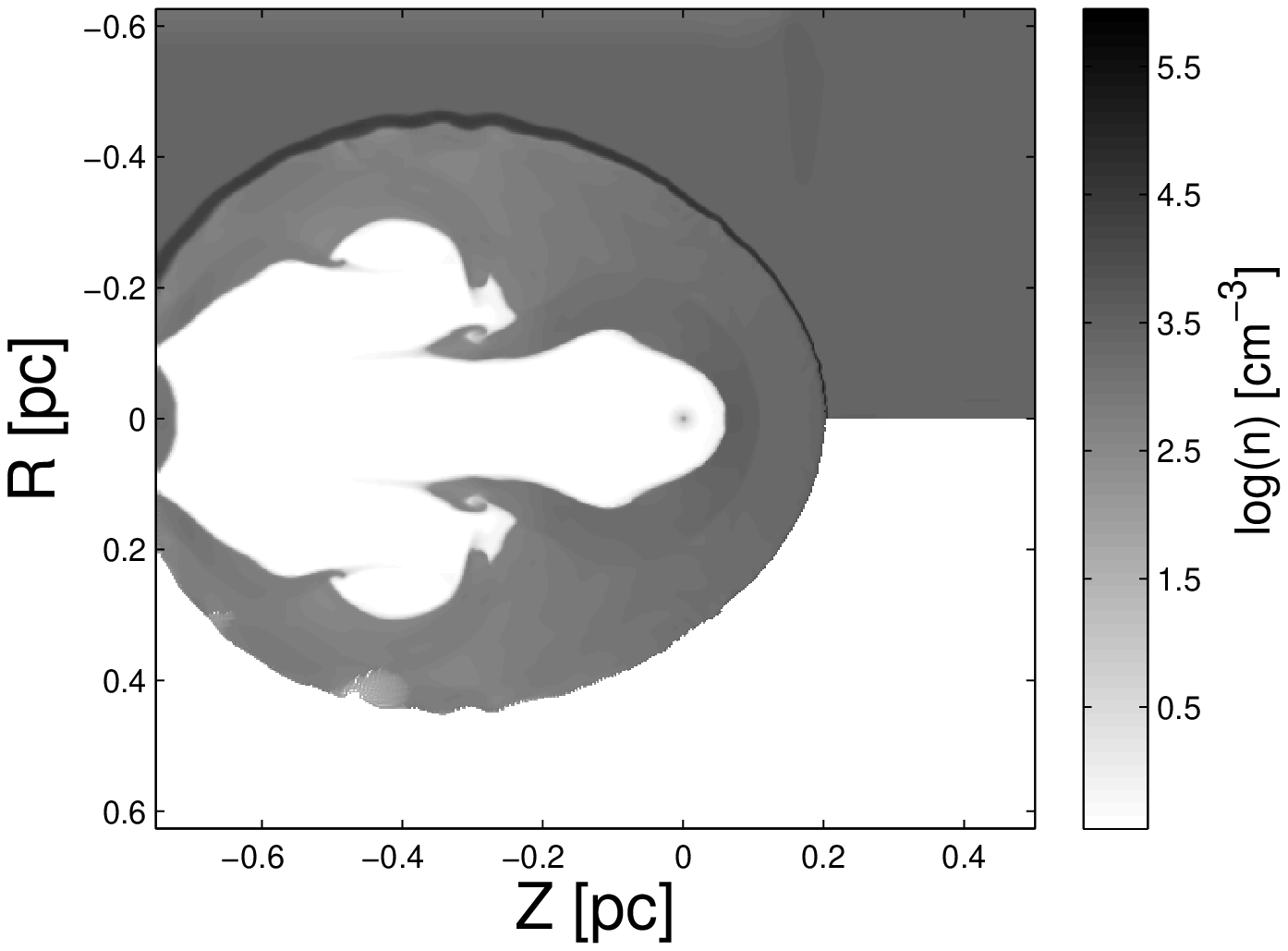}
\includegraphics[scale=.45]{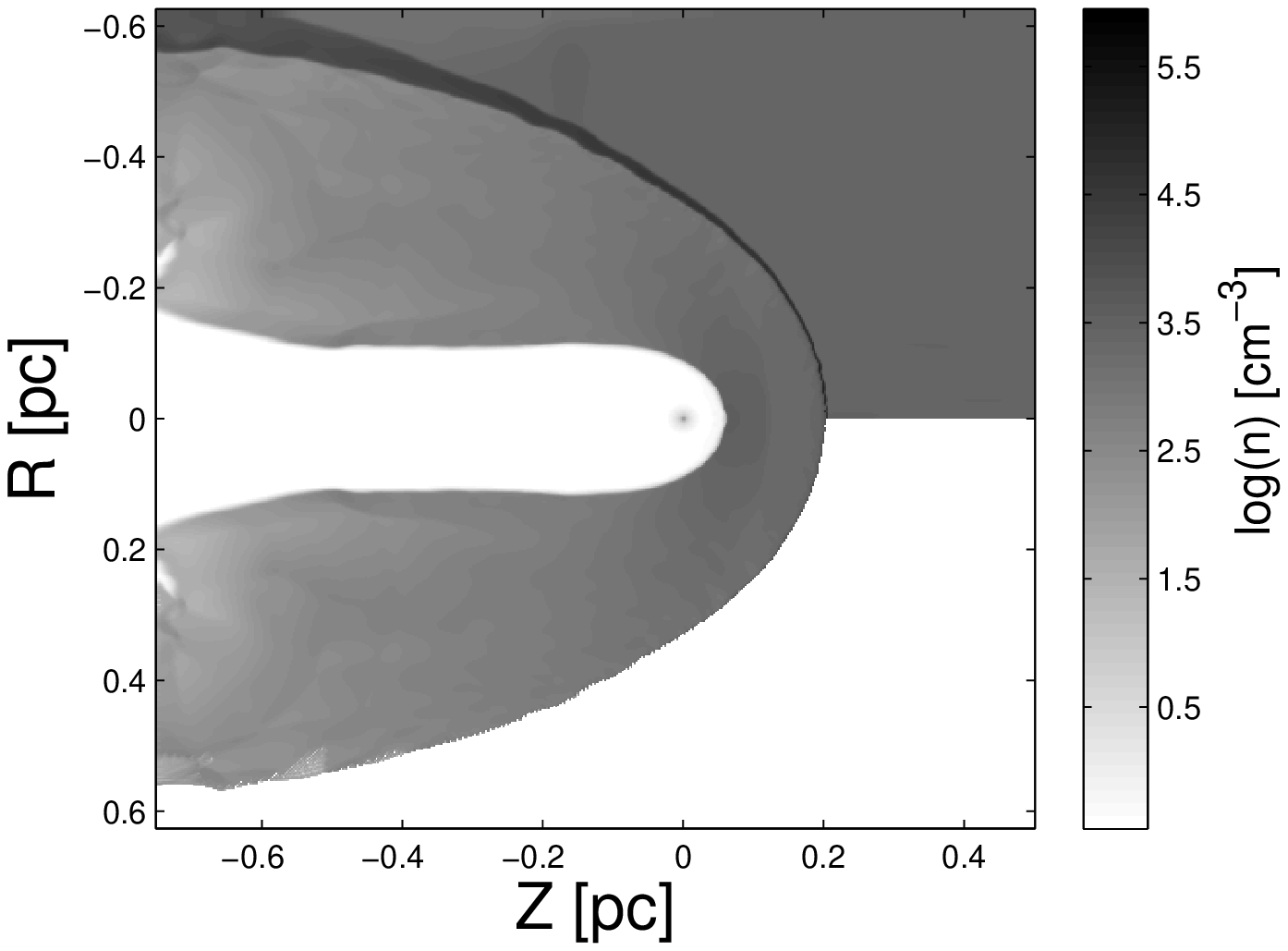}
\caption{The number density distributions of all materials and ionized hydrogen in the second test case at ages of $25,000~yr$ and $50,000~yr$.}
\begin{flushleft}
\end{flushleft}
\label{fig_ionizedbow}
\end{figure}

\textrm{A second test case, comparing a simulation of a bow shock with the models in \citet{mac15} is also studied}. In this case, the uniform density of the neutral medium is assumed to be $n=3000~cm^{-3}$, and the stellar velocity is $v_\ast=15~km~s^{-1}$. The other stellar parameters are same as those in \citet{mac15}. We also assume same minimum temperature $T_{min}=500~K$, and switch off calculating the chemical reactions in PDR because these reactions are not considered in \citet{mac15}. This simulation is ceased at $100,000~yr$. The density distributions of the total materials and the ionized hydrogen are presented in Figure \ref{fig_ionizedbow}.

\textrm{As in model V16 in \citet{mac15} our simulation shows that the supersonic motion of the star leads to a weak bow shock in the ionized region at $z=0.07~pc$}, the densities behind the bow shock are a little higher than those \textrm{ahead of} the bow shock, and the upstream radius of the H II region is approximately equal to $0.2~pc$. In Figure \ref{fig_ionizedbow}, the Kelvin-Helmholz instabilities are also shown, and these instabilities gradually flow to the tail of the H II region. There is also a density gradient \textrm{in the ionized region of our simulation, like in V16}. These features are similar in the two models. We find that the locations of the recombining ionized gas between our model and model V16 in \citet{mac15} are both at the sides of the H II region, but the location in our model is closer to the tail than that in V16. This could be due to the differences in the treatment of the radiative transfer and hydrodynamics and to grid resolution. 

After comparing the results in our test cases with previous works, we think the results from our models are reliable.

\clearpage



\end{document}